\documentclass{elsarticle}

\usepackage[english]{babel}
\usepackage[utf8]{inputenc}
\usepackage[T1]{fontenc}

\usepackage{lmodern}

\usepackage[hidelinks]{hyperref}
\hypersetup{pdfauthor=author}

\usepackage{lineno}

\usepackage{amsmath,amsfonts,amssymb,mathtools,mathrsfs,dsfont,mathabx}
\usepackage{accents}
\usepackage{listings}
\usepackage{upgreek}

\usepackage[left=3cm,right=3cm,top=2cm,bottom=3cm]{geometry}
\usepackage{parskip}

\usepackage{graphicx}
\usepackage{tikz}
\usetikzlibrary{arrows,positioning,automata,shadows,fit,shapes,calc,math,decorations.markings,decorations.pathreplacing}
\tikzset{mynode/.style={shape=rectangle,rounded corners,draw,align=center,top color=white,bottom color=white}}
\tikzset{->-/.style={decoration={markings,mark=at position #1 with {\arrow{>}}},postaction={decorate}}}
\usepackage{smartdiagram}
\usepackage[outdir=./]{epstopdf}

\usepackage{multirow}
\usepackage{makecell}
\usepackage{adjustbox}

\usepackage{enumerate}

\usepackage{subcaption}

\usepackage{doi}

\usepackage[toc,page]{appendix}

\usepackage[bottom]{footmisc}

\newcommand{\kb}{\boldsymbol{k}}

\newcommand{\lb}{\boldsymbol{\ell}}

\newcommand{\pb}{\boldsymbol{p}}

\newcommand{\ub}{\boldsymbol{u}}
\newcommand{\vb}{\boldsymbol{v}}
\newcommand{\xb}{\boldsymbol{x}}

\newcommand{\zb}{\boldsymbol{z}}

\newcommand{\Bb}{\boldsymbol{B}}

\newcommand{\Eb}{\boldsymbol{E}}

\newcommand{\Jb}{\boldsymbol{J}}

\newcommand{\fhat}{\widehat{f}}
\newcommand{\ghat}{\widehat{g}}

\newcommand{\Fhat}{\widehat{F}}

\newcommand{\Fcal}{\mathcal{F}}

\newcommand{\dd}{\textrm{d}}
\newcommand{\de}{\partial}
\newcommand{\Nabla}{\boldsymbol{\nabla}}
\newcommand{\eps}{\varepsilon}

\newcommand{\wh}{\widehat}

\newcommand{\Cbb}{\mathbb{C}}
\newcommand{\Nbb}{\mathbb{N}}
\newcommand{\Rbb}{\mathbb{R}}
\newcommand{\Zbb}{\mathbb{Z}}

\newcommand{\vbgal}{\vb_{\textrm{gal}}}
\newcommand{\vgal}{v_{\textrm{gal}}}

\newcommand{\ctxt}{\textrm{c}}
\newcommand{\htxt}{\textrm{h}}
\newcommand{\mtxt}{\textrm{m}}
\newcommand{\ntxt}{\textrm{n}}
\newcommand{\ptxt}{\textrm{p}}
\newcommand{\stxt}{\textrm{s}}
\newcommand{\Vtxt}{\textrm{V}}
\newcommand{\mm}{\text{mm}}
\newcommand{\nm}{\text{nm}}
\newcommand{\ps}{\text{ps}}
\newcommand{\fs}{\text{fs}}
\newcommand{\dx}{\Delta x}
\newcommand{\dy}{\Delta y}
\newcommand{\dz}{\Delta z}
\newcommand{\dt}{\Delta t}
\newcommand{\micron}{\upmu\mtxt}
\newcommand{\epsb}{\boldsymbol{\epsilon}}

\journal{Computer Physics Communications}

\begin{document}

\begin{frontmatter}

\title{A Hybrid Nodal-Staggered Pseudo-Spectral Electromagnetic Particle-In-Cell Method with Finite-Order Centering}

\author[lbl]{Edoardo Zoni\corref{coraut}}
\cortext[coraut]{Corresponding author}
\ead{ezoni@lbl.gov}

\author[lbl]{Remi Lehe}
\author[lbl]{Olga Shapoval}
\author[lbl]{Daniel Belkin\fnref{newaff}}
\fntext[newaff]{Current affiliation: Department of Physics, University of Illinois at Urbana-Champaign, Urbana, IL 61801, USA}
\author[cea]{\\ Neil Zaïm}
\author[cea]{Luca Fedeli}
\author[cea]{Henri Vincenti}
\author[lbl]{Jean-Luc Vay}

\address[lbl]{Lawrence Berkeley National Laboratory, Berkeley, CA 94720, USA}
\address[cea]{Université Paris-Saclay, CEA, CNRS, LIDYL, 91191 Gif-sur-Yvette, France}


\begin{abstract}
Electromagnetic particle-in-cell (PIC) codes are widely used to perform computer simulations of a variety of physical systems, including fusion plasmas, astrophysical plasmas, plasma wakefield particle accelerators, and secondary photon sources driven by ultra-intense lasers.
In a PIC code, Maxwell's equations are solved on a grid with a numerical method of choice.
This article focuses on pseudo-spectral analytical time-domain (PSATD) algorithms and presents a novel hybrid PSATD PIC scheme that combines the respective advantages of standard nodal and staggered methods.
The novelty of the hybrid scheme consists in using finite-order centering of grid quantities between nodal and staggered grids, in order to combine the solution of Maxwell's equations on a staggered grid with the deposition of charges and currents and the gathering of electromagnetic forces on a nodal grid.
The correctness and performance of the novel hybrid scheme are assessed by means of numerical tests that employ different classes of PSATD equations in a variety of physical scenarios, ranging from the modeling of electron-positron pair creation in vacuum to the simulation of laser-driven and particle beam-driven plasma wakefield acceleration.
It is shown that the novel hybrid scheme offers significant numerical and computational advantages, compared to purely nodal or staggered methods, for all the test cases presented.
\end{abstract}


\end{frontmatter}


\section{Introduction}
Electromagnetic particle-in-cell (PIC) codes \citep{HockneyEastwood1988,BirdsallLangdon1991} are widely used to perform computer simulations of a variety of physical systems, including turbulent plasmas in nuclear fusion devices \citep{Ethier2005,Wang2006,Ku2009,Bottino2010,Kraus2017}, relativistic astrophysical plasmas \citep{Kazimura1998,Silva2003,Spitkovsky2008,Keshet2009,Martins2009}, particle acceleration based on laser-plasma interactions \citep{Tajima1979,Gitomer1986,Gorbunov1987,Malka2009}, high-order harmonic sources based on laser-solid interactions and their applications \citep{Vincenti2014,Leblanc2017,Vincenti2019,Fedeli2021}.

In a PIC code, Maxwell's equations, which describe the dynamics and evolution of the electromagnetic fields, are solved on a grid and the plasma is modeled with a collection of macro-particles, each representing many real particles of the modeled system.
Macro-particles move according to the electromagnetic fields on the grid.
Charged macro-particles generate charge and current densities on the grid, which are used as sources for Maxwell's equations.
The finite-difference and pseudo-spectral algorithms are the common numerical methods of choice for the numerical solution of Maxwell's equations.

Finite-difference algorithms \citep{Yee1966,Cole1997,Cole2002,Taflove2005} typically approximate both spatial and time derivatives with finite differences (generally second-order), which usually lead to spurious numerical dispersion.
On the other hand, pseudo-spectral methods \citep{Haber1973,BunemanJCP1980,Dawson1983,Liu1997,Vay2013,Lehe2019} help mitigate such numerical artifacts by approximating spatial derivatives with high-order discrete expressions that use larger stencils of grid points~\citep{Vincenti2016,Jalas2017}.

This article focuses in particular on pseudo-spectral analytical time-domain (PSATD) algorithms~\citep{Haber1973,BunemanJCP1980,Vay2013}, which help mitigate the spurious numerical dispersion of finite-difference methods even further, by integrating Maxwell's equations in Fourier space analytically in time, instead of approximating time derivatives by finite differences.

More precisely, a novel PSATD PIC method is proposed that combines the respective advantages of standard nodal and staggered PIC algorithms. The novel scheme, which will be referred to as \emph{hybrid}, combines the solution of Maxwell's equations on a staggered grid with the deposition of charges and currents on a nodal grid as well as the gathering of electromagnetic forces from a nodal grid, using finite-order interpolation, based on the coefficients first introduced by Fornberg~\citep{Fornberg1990}, to center grid quantities between nodal and staggered grids.

The correctness and performance of the novel hybrid method are assessed by means of numerical tests that employ different classes of PSATD equations (standard PSATD~\citep{Haber1973,Vay2013}, standard Galilean PSATD~\citep{Lehe2016,Kirchen2020}, and averaged Galilean PSATD~\citep{Shapoval2021}), adapted to staggered grids, in a variety of physical scenarios, ranging from the modeling of electron-positron pair creation in vacuum to the simulation of laser-driven and particle beam-driven plasma wakefield acceleration.

This article is organized as follows.
Section~\ref{sec_motivations} presents the general idea and motivations for the novel hybrid method.
Section~\ref{sec_interpolation} describes how to perform finite-order interpolation between nodal and staggered grids by means of the Fornberg coefficients.
Section~\ref{sec_algo} presents the equations for three classes of PSATD schemes of interest, adapted to staggered grids.
Section~\ref{sec_tests} presents a variety of numerical tests, assessing the correctness and performance of the novel hybrid method.
Section~\ref{sec_conclusions} presents the conclusions of this work.
Finally, three appendices have been added to derive or illustrate in more detail some of the mathematical results presented in the article.

\section{Motivations}
\label{sec_motivations}
This section presents the general idea and motivations for the novel hybrid PSATD PIC method proposed in this article, starting with a brief review of the structure of a time step of a \emph{standard} PSATD PIC algorithm, illustrated by the following cycle:
\begin{center}
\begin{tikzpicture}[scale=1.]
\def \r {3cm}
\node at (90:\r/2) {\emph{standard} PSATD PIC cycle};
\node[mynode] at ({90}:\r)
{push particles by updating $\xb$, $\pb$};
\draw[<-, >=latex] ({0+8}:\r) arc ({0+8}:{90-30}:\r);
\node[mynode] at ({360}:\r)
{deposit $\rho$, $\Jb$ on \emph{nodal/Yee grid}};
\draw[<-, >=latex] ({270+38}:\r) arc ({270+38}:{360-8}:\r);
\node[mynode] at ({270}:\r)
{solve Maxwell's equations \\ on \emph{nodal/Yee grid} in Fourier space};
\draw[<-, >=latex] ({180+8}:\r) arc ({180+8}:{270-38}:\r);
\node[mynode] at ({180}:\r)
{gather $\Eb$, $\Bb$ from \emph{nodal/Yee grid}};
\draw[<-, >=latex] ({90+30}:\r) arc ({90+30}:{180-8}:\r);
\end{tikzpicture}
\end{center}
\vspace{1em}
Here, $\xb$ and $\pb$ denote the positions and momenta of the macro-particles, $\rho$ and $\Jb$ the charge and current densities generated by charged macro-particles, $\Eb$ and $\Bb$ the electromagnetic fields.
The deposition of $\rho$ and $\Jb$ usually includes a smoothing of the quantities using one pass (or more) of the bilinear filter \cite{BirdsallLangdon1991}. 
In general, the grid used for charge and current deposition, field gathering, and for the solution of Maxwell's equations, can be a nodal grid or a staggered Yee grid \citep{Yee1966,Cole1997,Cole2002}.
In the case of a nodal grid, all grid quantities (electromagnetic fields, charge and current densities) are evaluated at the cell nodes in each direction.
In the case of a staggered Yee grid, instead, the various grid quantities are evaluated at various cell nodes and centers, depending on the quantity itself and on the direction considered.
More precisely, the charge density $\rho$ is still evaluated at the cell nodes in each direction, while the positions of the electromagnetic fields $\Eb$ and $\Bb$ and the current density $\Jb$ are illustrated by the following three-dimensional schematics of a single cell:
\begin{center}
\begin{tikzpicture}[scale=.7]
\draw (0,0,0) -- (0,2,0) -- (0,2,2) -- (0,0,2) -- (0,0,0);
\draw (2,0,0) -- (2,0,2) -- (2,2,2) -- (2,2,0) -- (2,0,0);
\draw (0,0,0) -- (2,0,0);
\draw (0,2,0) -- (2,2,0);
\draw (2,0,2) -- (0,0,2);
\draw (2,2,2) -- (0,2,2);
\draw[->, >=latex] (2,0,0) -- (3,0,0);
\node[above] at (3,0,0) {$x$};
\draw[->, >=latex] (0,2,0) -- (0,3,0);
\node[left] at (0,3,0) {$y$};
\draw[->, >=latex] (0,0,2) -- (0,0,3);
\node[right] at (0.1,0,3) {$z$};
\node at (1.5,3,0) {$E^x$, $J^x$};
\node at (1,0,0) {\textbullet};
\node at (1,2,0) {\textbullet};
\node at (1,0,2) {\textbullet};
\node at (1,2,2) {\textbullet};
\end{tikzpicture}
\hspace{2em}
\begin{tikzpicture}[scale=.7]
\draw (0,0,0) -- (0,2,0) -- (0,2,2) -- (0,0,2) -- (0,0,0);
\draw (2,0,0) -- (2,0,2) -- (2,2,2) -- (2,2,0) -- (2,0,0);
\draw (0,0,0) -- (2,0,0);
\draw (0,2,0) -- (2,2,0);
\draw (2,0,2) -- (0,0,2);
\draw (2,2,2) -- (0,2,2);
\draw[->, >=latex] (2,0,0) -- (3,0,0);
\node[above] at (3,0,0) {$x$};
\draw[->, >=latex] (0,2,0) -- (0,3,0);
\node[left] at (0,3,0) {$y$};
\draw[->, >=latex] (0,0,2) -- (0,0,3);
\node[right] at (0.1,0,3) {$z$};
\node at (1.5,3,0) {$E^y$, $J^y$};
\node at (0,1,0) {\textbullet};
\node at (2,1,0) {\textbullet};
\node at (0,1,2) {\textbullet};
\node at (2,1,2) {\textbullet};
\end{tikzpicture}
\hspace{2em}
\begin{tikzpicture}[scale=.7]
\draw (0,0,0) -- (0,2,0) -- (0,2,2) -- (0,0,2) -- (0,0,0);
\draw (2,0,0) -- (2,0,2) -- (2,2,2) -- (2,2,0) -- (2,0,0);
\draw (0,0,0) -- (2,0,0);
\draw (0,2,0) -- (2,2,0);
\draw (2,0,2) -- (0,0,2);
\draw (2,2,2) -- (0,2,2);
\draw[->, >=latex] (2,0,0) -- (3,0,0);
\node[above] at (3,0,0) {$x$};
\draw[->, >=latex] (0,2,0) -- (0,3,0);
\node[left] at (0,3,0) {$y$};
\draw[->, >=latex] (0,0,2) -- (0,0,3);
\node[right] at (0.1,0,3) {$z$};
\node at (1.5,3,0) {$E^z$, $J^z$};
\node at (0,0,1) {\textbullet};
\node at (2,0,1) {\textbullet};
\node at (0,2,1) {\textbullet};
\node at (2,2,1) {\textbullet};
\end{tikzpicture}
\\
\begin{tikzpicture}[scale=.7]
\draw[fill=gray!10] (0,0,0) -- (0,2,0) -- (0,2,2) -- (0,0,2) -- (0,0,0);
\draw[fill=gray!10] (2,0,0) -- (2,0,2) -- (2,2,2) -- (2,2,0) -- (2,0,0);
\draw (0,0,0) -- (2,0,0);
\draw (0,2,0) -- (2,2,0);
\draw (2,0,2) -- (0,0,2);
\draw (2,2,2) -- (0,2,2);
\draw[->, >=latex] (2,0,0) -- (3,0,0);
\node[above] at (3,0,0) {$x$};
\draw[->, >=latex] (0,2,0) -- (0,3,0);
\node[left] at (0,3,0) {$y$};
\draw[->, >=latex] (0,0,2) -- (0,0,3);
\node[right] at (0.1,0,3) {$z$};
\node at (1.5,3,0) {$B^x$};
\node at (0,1,1) {\textbullet};
\node at (2,1,1) {\textbullet};
\end{tikzpicture}
\hspace{2em}
\begin{tikzpicture}[scale=.7]
\draw[fill=gray!10] (0,2,0) -- (2,2,0) -- (2,2,2) -- (0,2,2) -- (0,2,0);
\draw[fill=gray!10] (0,0,0) -- (2,0,0) -- (2,0,2) -- (0,0,2) -- (0,0,0);
\draw (0,0,0) -- (0,2,0);
\draw (2,0,0) -- (2,2,0);
\draw (0,0,2) -- (0,2,2);
\draw (2,0,2) -- (2,2,2);
\draw[->, >=latex] (2,0,0) -- (3,0,0);
\node[above] at (3,0,0) {$x$};
\draw[->, >=latex] (0,2,0) -- (0,3,0);
\node[left] at (0,3,0) {$y$};
\draw[->, >=latex] (0,0,2) -- (0,0,3);
\node[right] at (0.1,0,3) {$z$};
\node at (1.5,3,0) {$B^y$};
\node at (1,0,1) {\textbullet};
\node at (1,2,1) {\textbullet};
\end{tikzpicture}
\hspace{2em}
\begin{tikzpicture}[scale=.7]
\draw[fill=gray!10] (0,0,0) -- (2,0,0) -- (2,2,0) -- (0,2,0) -- (0,0,0);
\draw[fill=gray!10, fill opacity=0.5] (0,0,2) -- (2,0,2) -- (2,2,2) -- (0,2,2) -- (0,0,2);
\draw (0,0,0) -- (0,0,2);
\draw (0,2,0) -- (0,2,2);
\draw (2,0,0) -- (2,0,2);
\draw (2,2,0) -- (2,2,2);
\draw[->, >=latex] (2,0,0) -- (3,0,0);
\node[above] at (3,0,0) {$x$};
\draw[->, >=latex] (0,2,0) -- (0,3,0);
\node[left] at (0,3,0) {$y$};
\draw[->, >=latex] (0,0,2) -- (0,0,3);
\node[right] at (0.1,0,3) {$z$};
\node at (1.5,3,0) {$B^z$};
\node at (1,1,0) {\textbullet};
\node at (1,1,2) {\textbullet};
\end{tikzpicture}
\end{center}
\vspace{1em}

The \emph{hybrid} PSATD PIC method proposed in this article differs from the standard PSATD PIC algorithm summarized above and entails the main steps illustrated by the following cycle:
\vspace{1em}
\begin{center}
\begin{tikzpicture}[scale=1.]
\def \r {3cm}
\node at (90:\r/2) {\emph{hybrid} PSATD PIC cycle};
\node[mynode] at ({90}:\r)
{push particles by updating $\xb$, $\pb$};
\draw[<-, >=latex] ({0+8}:\r) arc ({0+8}:{90-30}:\r);
\node[mynode] at ({360}:\r)
{deposit $\rho$, $\Jb$ on \emph{nodal grid}};
\draw[<-, >=latex] ({330+12}:\r) arc ({330+12}:{360-8}:\r);
\node[mynode] at ({330}:\r)
{finite-order centering of $\Jb$ \\ from nodal grid to Yee grid};
\draw[<-, >=latex] ({270+36}:\r) arc ({270+36}:{330-14}:\r);
\node[mynode] at ({270}:\r)
{solve Maxwell's equations \\ on \emph{Yee grid} in Fourier space};
\draw[<-, >=latex] ({210+14}:\r) arc ({210+14}:{270-36}:\r);
\node[mynode] at ({210}:\r)
{finite-order centering of $\Eb$, $\Bb$ \\ from Yee grid to nodal grid};
\draw[<-, >=latex] ({180+8}:\r) arc ({180+8}:{210-12}:\r);
\node[mynode] at ({180}:\r)
{gather $\Eb$, $\Bb$ from \emph{nodal grid}};
\draw[<-, >=latex] ({90+30}:\r) arc ({90+30}:{180-8}:\r);
\end{tikzpicture}
\end{center}
\vspace{1em}

There are two main differences with respect to the standard PSATD PIC cycle.
First, charge and current densities are deposited on a nodal grid and the electromagnetic forces acting on the macro-particles are gathered from a nodal grid, while Maxwell's equations for the electromagnetic fields are solved on a staggered Yee grid. Secondly, finite-order interpolation based on the Fornberg coefficients~\citep{Fornberg1990} is used to center data between the nodal grid used for deposition and gathering and the staggered Yee grid used for the solution of Maxwell's equations.
The details of such finite-order centering are discussed in Section~\ref{sec_interpolation}.

Gathering the electromagnetic forces acting on the macro-particles from a nodal grid, after linear interpolation from the staggered Yee grid used to solve Maxwell's equations, has been employed in many standard finite-difference electromagnetic PIC simulation codes for decades, as well as described in the literature where it is referred to as the \emph{momentum-conserving gather}~\citep{BirdsallLangdon1991}.
The hybrid PIC cycle proposed in this article differs from this, in that (i) a nodal grid is used also for the deposition of charge and currents and (ii) interpolation of order typically higher than linear is employed to center data between nodal and staggered Yee grids.

The rationale behind the novel hybrid PIC method is summarized in the following few paragraphs.

On the one hand, solving Maxwell's equations on a staggered grid, instead of a nodal grid, presents several numerical and computational advantages (for example, more local stencils, lower levels of numerical dispersion, better stability at short wavelengths), which are discussed in more detail in this section.
On the other hand, purely staggered PIC schemes sometimes need much higher resolution in space or time to produce correct physical results, depending on the specific physics application under study.
This is true in particular for the modeling of relativistic plasmas, where staggering of quantities in space or time can lead to unacceptably large numerical errors from interpolation \cite{Vay2008,Lehe2014}.
An example where a purely staggered PIC scheme exhibits difficulties in producing correct physical results is given by the simulation of vacuum electron-positron pair creation illustrated in Section~\ref{sec_psatd_test}.
Another example occurs with physics applications that exhibit numerical Cherenkov instability (NCI) \citep{Godfrey1974,Godfrey1975,Martins2010,Vay2010,Vay2011,Xu2013,Godfrey2014}, such as in electromagnetic simulations of relativistic flowing plasmas, where fast particles may resonate unphysically with electromagnetic waves or aliases of matching phase velocity.
More precisely, with regard to one specific class of PSATD methods that have been shown to mitigate such instability, namely the Galilean algorithms \citep{Lehe2016,Kirchen2020,Shapoval2021}, it is observed that purely nodal PIC cycles mitigate the instability more effectively than purely staggered PIC cycles.
Examples of such behavior are illustrated in Sections~\ref{sec_gal_test}~and~\ref{sec_avg_test}, Figures~\ref{fig_gal_nodal_vs_hybrid}~and~\ref{fig_avg_nodal_vs_hybrid}, respectively, for two cases of laser-driven and particle beam-driven plasma wakefield acceleration.
A few heuristic arguments on the role of staggering and finite-order centering in relation to NCI mitigation are given in~\ref{appendix_3}.

The \emph{hybrid} PIC scheme proposed here represents an intermediate approach that combines the respective advantages of standard nodal and staggered PIC methods.
More precisely, the main idea is to construct a PIC cycle where, on the one hand, Maxwell's equations are solved on a staggered grid (in order to benefit from the locality of the stencils, lower levels of numerical dispersion, and better stability at short wavelengths, as discussed in more detail below) and, on the other hand, the resulting cycle is as close as possible to a fully nodal PIC cycle (in order to avoid numerical errors coming from low-order interpolation of grid quantities that are defined at different locations on the grid).
Hence the idea of depositing charges and currents on a nodal grid as well as gathering the electromagnetic forces acting on the macro-particles from a nodal grid, while keeping the solution of Maxwell's equations on a staggered grid.
Within this context, the finite-order interpolation of fields and currents represents a way to center grid quantities between the two sets of grids, with the aim of balancing numerical accuracy and locality by an appropriate choice of the finite order of interpolation, which can vary depending on the specific physics application under study.

As already mentioned, there are several numerical and computational advantages of solving Maxwell's equations on a staggered grid, instead of a nodal grid, which are detailed here:
\begin{enumerate}[(i)]
\item {\bf Staggered solvers usually exhibit less numerical dispersion than nodal solvers.} \\
Figure~\ref{fig_disprel_phasevel} shows the numerical dispersion relation and phase velocity in vacuum for the standard PSATD equations \citep{Haber1973,Vay2013}, namely equations \eqref{eq_update_B_with_rho_standard}-\eqref{eq_update_E_with_rho_standard} without sources, at finite spectral order~$16$ (quite typical for simulations of plasma wakefield acceleration) for nodal and staggered solvers, in a one-dimensional case.
The vacuum dispersion relation in this case is given by equation~\eqref{vacuum_disprel_psatd}, derived in~\ref{appendix_2}.
Figure~\ref{fig_disprel_phasevel} shows that in the case of the staggered equations, the dispersion relation is closer to linear and the phase velocity remains consequently closer to $c$, reducing the slowdown of high-frequency waves, which in the nodal case is so strong that it leads to standing waves at the Nyquist wavelength at any finite order, producing in turn undesirable effects \cite{Ohmura2010} that need to be suppressed.
\begin{figure}[!ht]
\centering
\includegraphics[width=0.49\linewidth,trim={0.5cm 0 1cm 0},clip]{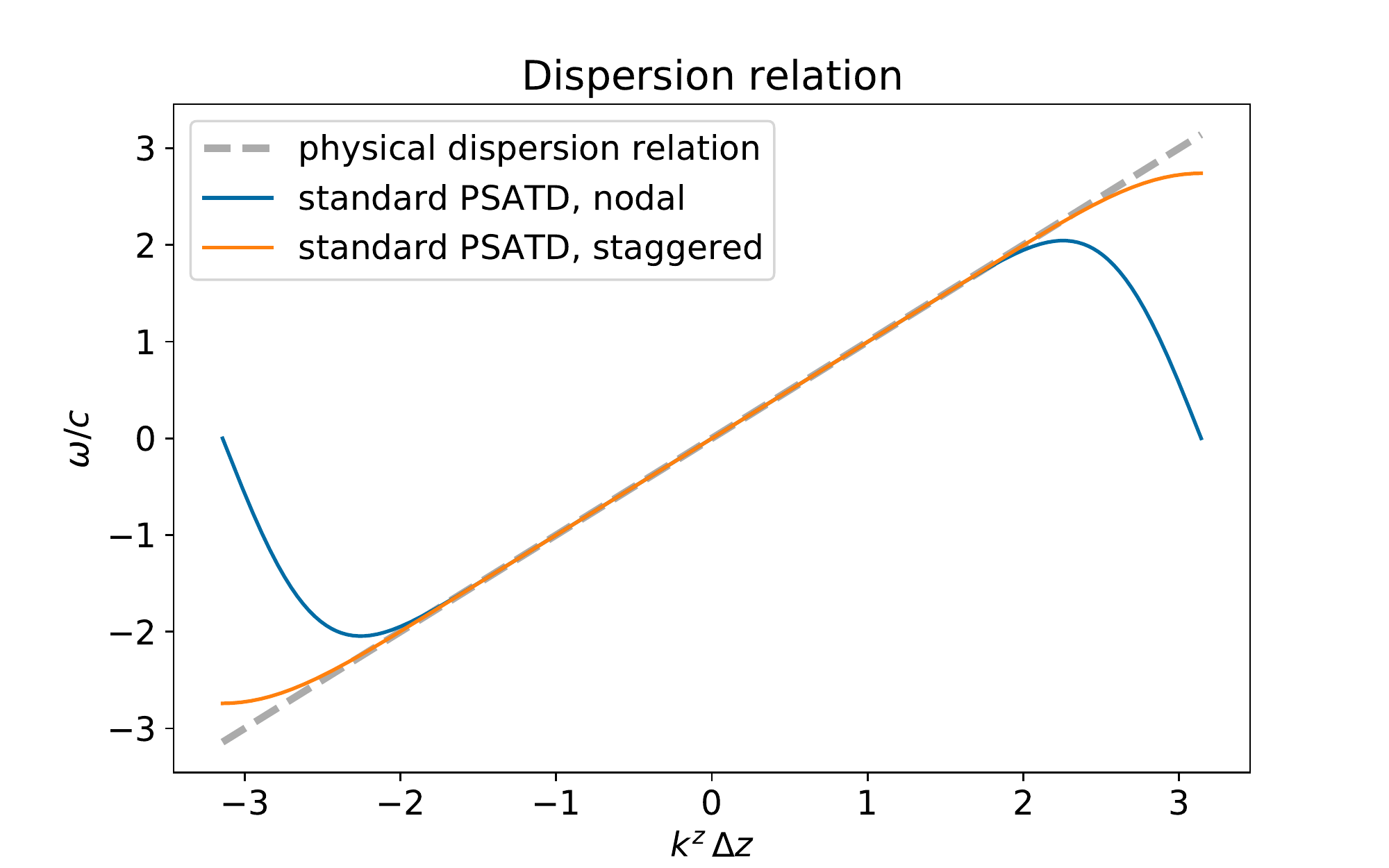}
\hfill
\includegraphics[width=0.49\linewidth,trim={0.5cm 0 1cm 0},clip]{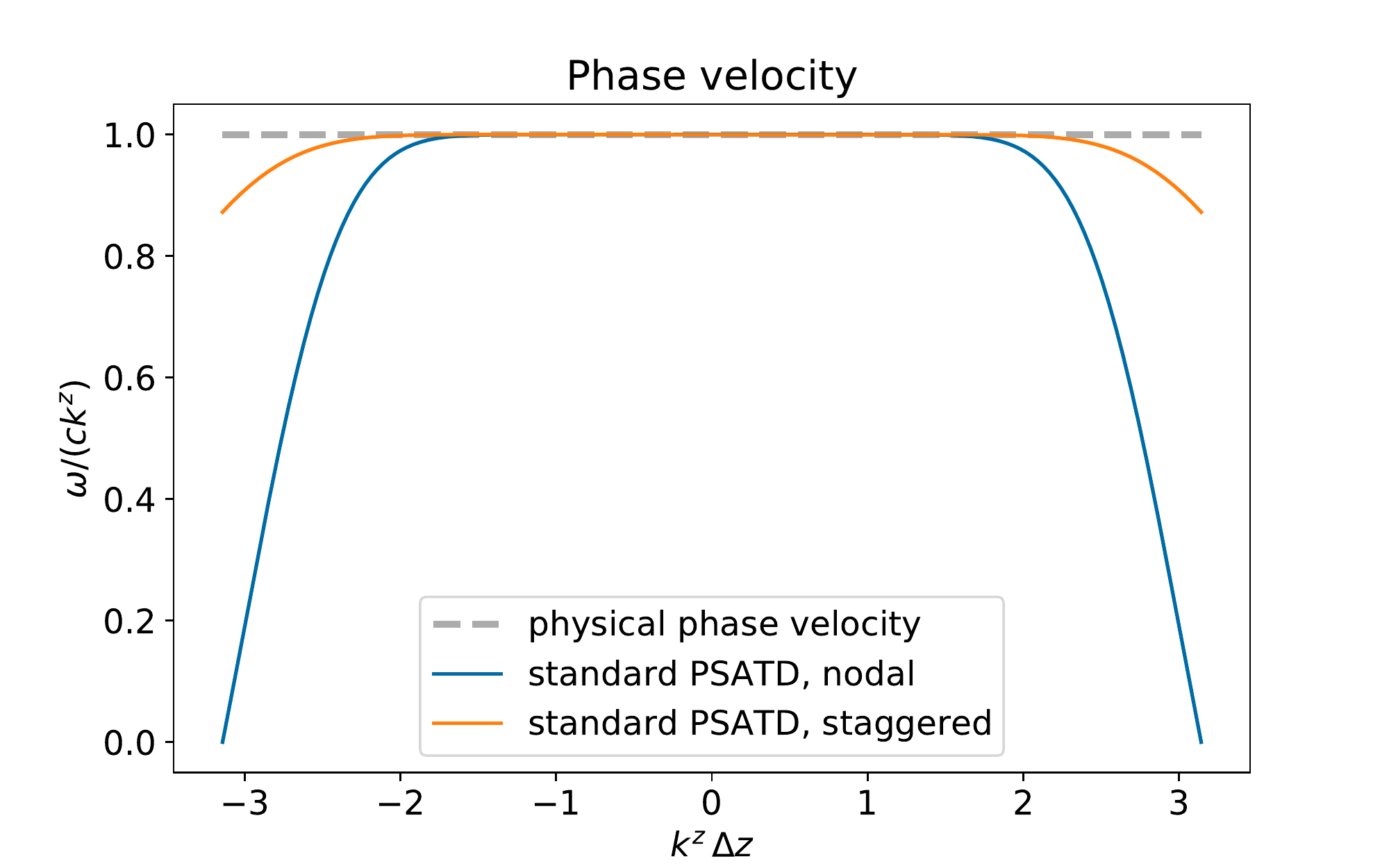}
\caption{\textbf{Numerical Dispersion}. Numerical dispersion relation and phase velocity in vacuum for the standard PSATD equations at finite spectral order $16$ for nodal and staggered solvers (solid lines), in a one-dimensional case. With the staggered equations, the dispersion relation is closer to linear and the phase velocity remains consequently closer to $c$, reducing the slowdown of high-frequency waves, which is instead relatively strong in the nodal case.}
\label{fig_disprel_phasevel}
\end{figure}
\item {\bf Staggered solvers offer a more local stencil than nodal solvers, at a given finite order.}
This can be illustrated by measuring the extent of the stencil of a given term in Maxwell's equations in Fourier space \citep{Jalas2017,Kirchen2020}.
For example, for a given quantity $\ghat$ in Fourier space (which can be, for instance, a coefficient in the update equations for $\Eb$ and~$\Bb$), a measure of its stencil extent along a given direction, say $x$, can be computed as
\begin{linenomath}
\begin{equation}
\label{eq_stencil_extent}
\Gamma_{\ghat}(x) = \bigg|\frac{1}{N_y N_z} \sum_{k_y} \sum_{k_z} \, [\Fcal^{-1}_x(\ghat)](x,k_y,k_z)\bigg| \,, 
\end{equation}
\end{linenomath}
and similarly for the stencil extents along $y$ and $z$, by cyclic permutation.
In other words, the inverse Fourier transform of $\ghat$ along the given axis is computed and the result is averaged over the remaining axes in Fourier space.
An example of such stencils is shown in Figure~\ref{fig_stencil_pulse} for a two-dimensional case with $512$ cells in each direction and $c \, \dt = \dx = \dz \approx 0.39 \, \micron$, at finite spectral order 64 (quite typical for simulations of high-order harmonic sources) as well as infinite spectral order.
More precisely, Figure~\ref{fig_stencil_pulse} shows the stencil along $x$ of the coefficient $C$ appearing in the standard PSATD equations~\eqref{eq_update_B_with_rho_standard}-\eqref{eq_update_E_with_rho_standard}, computed as prescribed in~\eqref{eq_stencil_extent}.
The idea is to look at how quickly such stencils fall off to machine precision, with respect to their extension in units of grid cells, and identify consequently the number of cells after which the stencils will be truncated, again with the aim of balancing numerical accuracy and locality.
In practice, when the computational domain is decomposed in parallel subdomains, the number of ghost cells used to exchange fields between neighboring subdomains is chosen based on the extent of such stencils. 
In the limit case of infinite spectral order, nodal and staggered solvers produce the same result: the stencil extends over the entire grid and the evolution of the fields on the grid is not local.
Other examples of such stencils are shown in Sections~\ref{sec_gal_test}~and~\ref{sec_avg_test}, Figures~\ref{fig_stencil_lwfa}~and~\ref{fig_stencil_pwfa}, respectively, for two cases of laser-driven and particle beam-driven plasma wakefield acceleration.
In general, the greater locality of the finite-order stencil offered by staggered solvers results in shorter runtimes and smaller computational costs overall, thanks to the fact that the number of ghost cells used to exchange fields between neighboring subdomains is smaller.
\begin{figure}[!ht]
\centering
\includegraphics[width=0.8\linewidth,trim={0 0 0 0},clip]{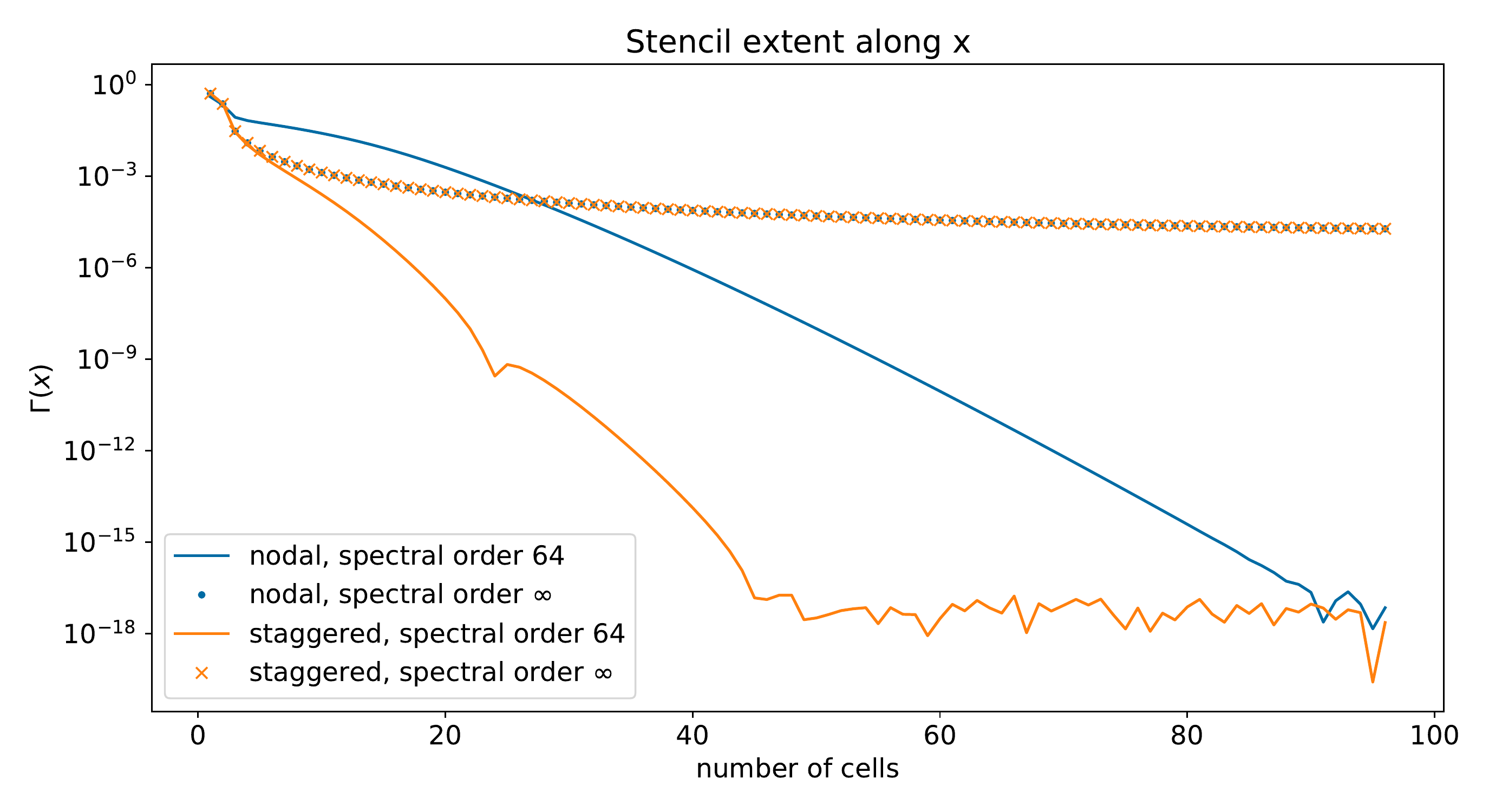}
\caption{\textbf{PSATD Stencils}. Stencil extent along $x$ of the coefficient $C$ appearing in the standard PSATD equations~\eqref{eq_update_B_with_rho_standard}-\eqref{eq_update_E_with_rho_standard}, computed as prescribed in~\eqref{eq_stencil_extent}, for a two-dimensional case with $512$ cells in each direction and $c \, \dt = \dx = \dz \approx 0.39 \, \micron$. At finite spectral order, staggered solvers offer a more local stencil than nodal solvers. In the limit case of infinite spectral order, nodal and staggered solvers produce the same result: the stencil extends over the entire grid and the evolution of the fields on the grid is not local.}
\label{fig_stencil_pulse}
\end{figure}
\item {\bf 
Staggered solvers exhibit better behavior than nodal solvers at short wavelengths.}
Another feature of staggered solvers is that they exhibit better behavior than nodal solvers at short wavelengths, in particular at the Nyquist cutoff~\citep{Ohmura2010}.
Figure~\ref{fig_pulse_nodal_vs_stagg} shows an example of this phenomenon for a two-dimensional rectangular electric pulse (with unitary amplitude), initialized at the center of a two-dimensional periodic domain.
More precisely, a snapshot of the component $E^y$ of the electric field is shown after $200$ iterations, with $c \, \dt = \dx = \dz \approx 0.39 \, \micron$, with a nodal solver (left) and a staggered solver (right).
Both cases used spectral order 64 and 8 ghost cells along $(x,z)$ (first row) or infinite spectral order and again 8 ghost cells along $(x,z)$ (second row).
The domain was decomposed in 256 subdomains, with $32 \times 32$ cells per subdomain.
The nodal solver exhibits much stronger short-wavelength noise than the staggered solver, both at finite spectral order and at infinite spectral order.
Consequently, the nodal solver leads to a strong non-physical growth over time of the electromagnetic field energy $W = \frac{1}{2} \sum_{\text{cells}} (\epsilon_0 |\Eb|^2 + |\Bb|^2/\mu_0) \, \Delta V$, which instead grows much more slowly in the staggered case, as shown in Figure~\ref{fig_pulse_nodal_vs_stagg}.
In conclusion, in addition to offering a more local stencil, staggered solvers are inherently more stable than nodal solvers.
The fundamental reasons for this and the optimization of the number of ghost cells for a given spectral order are being studied and will be reported in future work.
\begin{figure}[!ht]
\centering
\includegraphics[width=0.42\linewidth,trim={0 1cm 0.5cm 1cm}, clip]{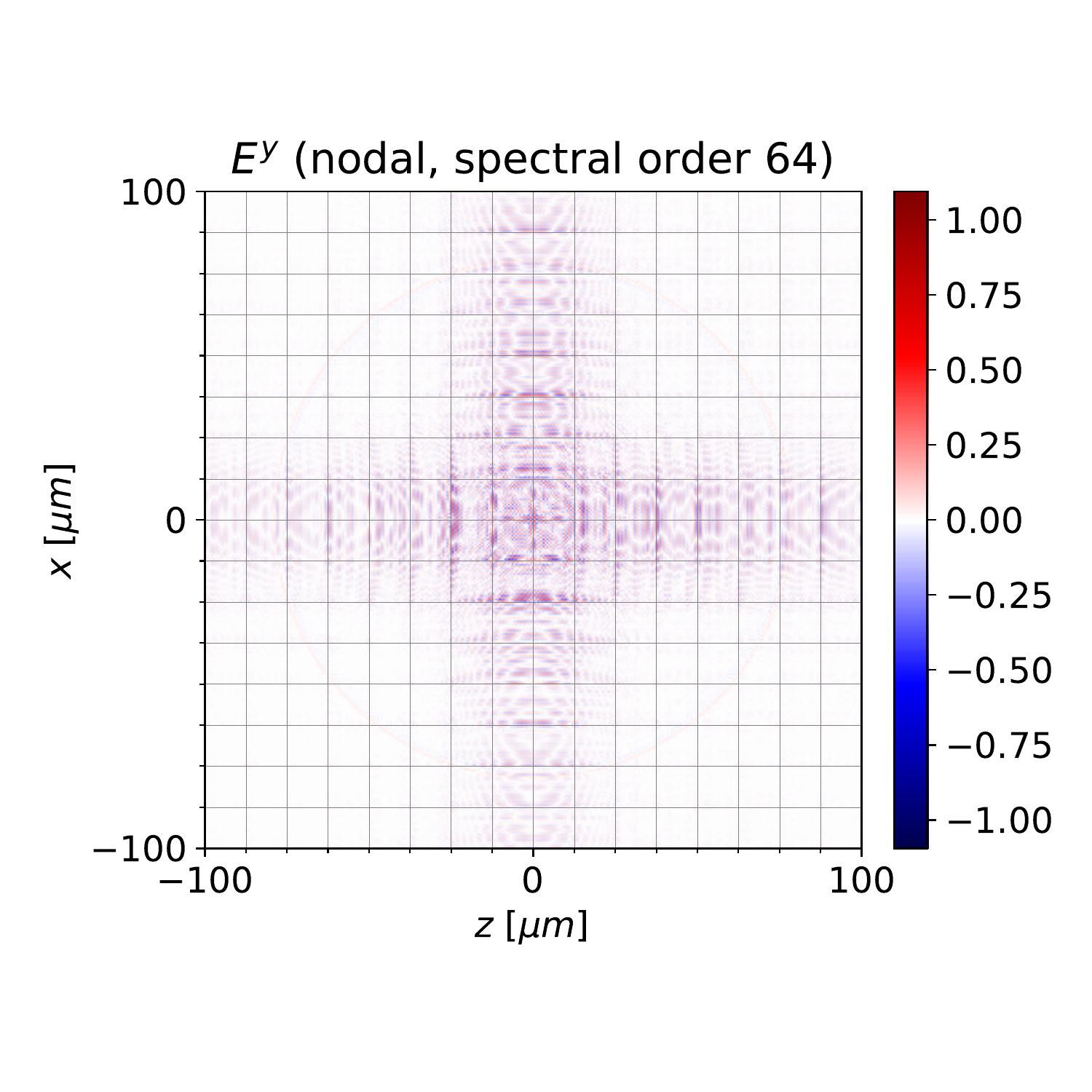}
\hspace{0.5cm}
\includegraphics[width=0.42\linewidth,trim={0 1cm 0 0.5cm}, clip]{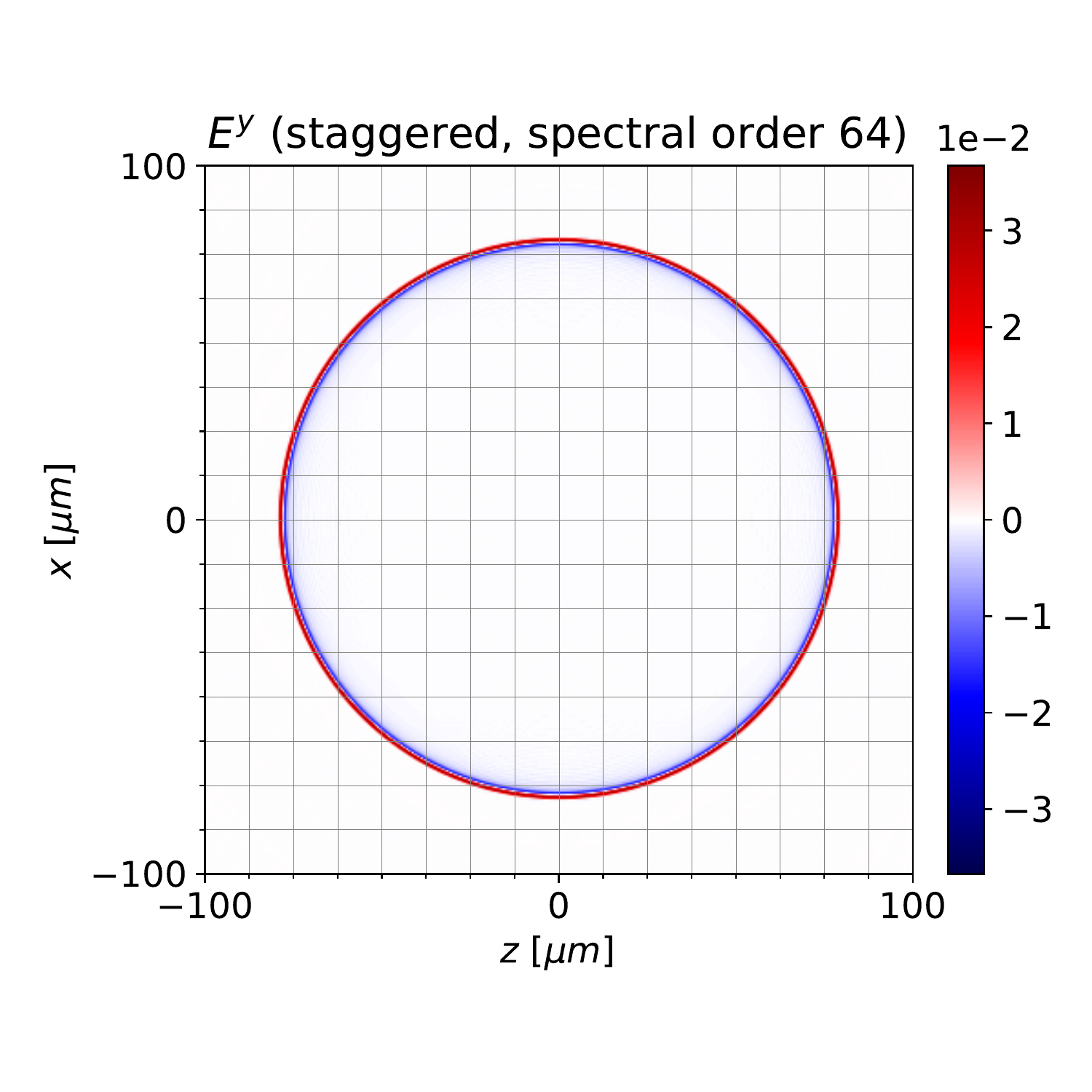}
\includegraphics[width=0.42\linewidth,trim={0 1cm 0 0.5cm}, clip]{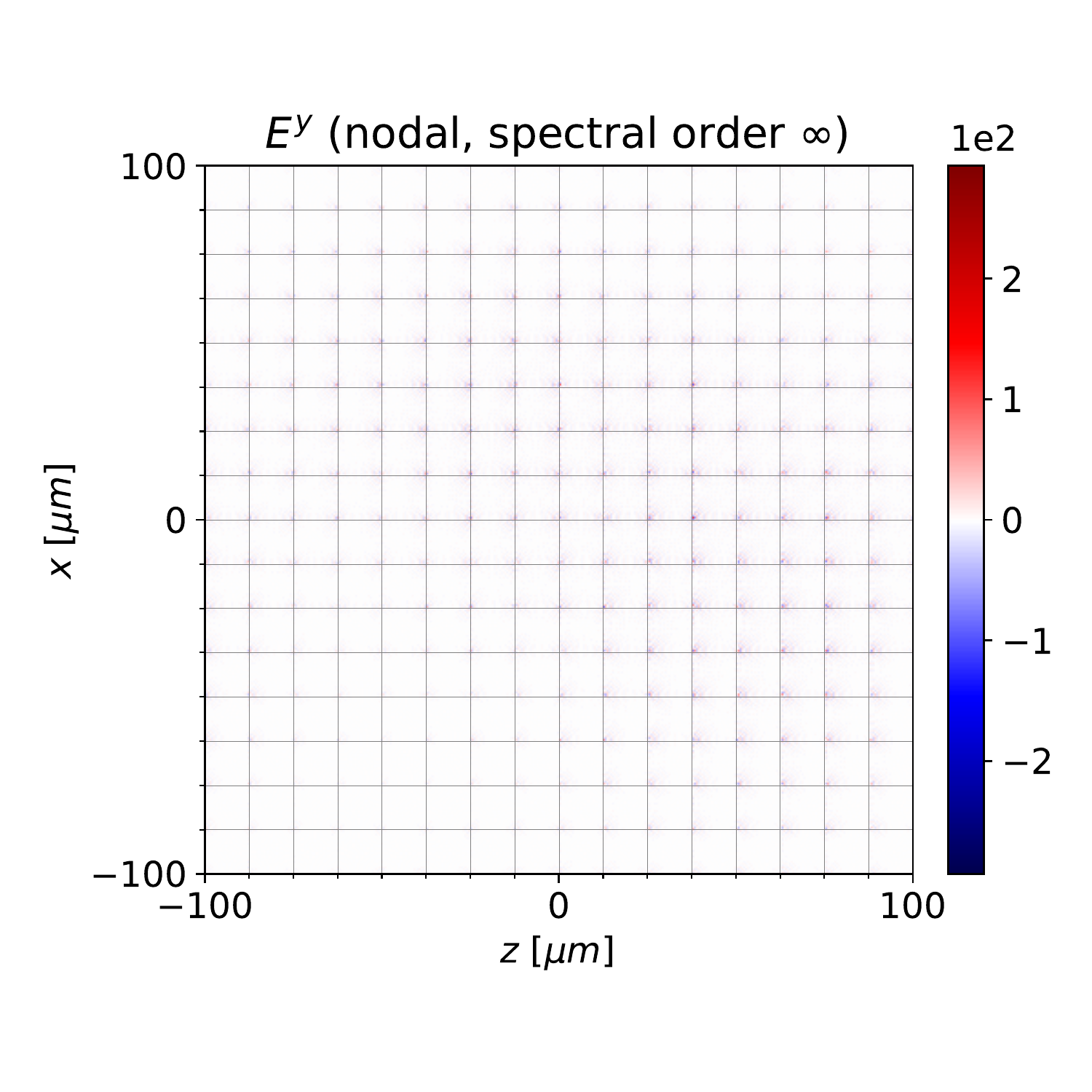}
\hspace{0.5cm}
\includegraphics[width=0.42\linewidth,trim={0 1cm 0 0.5cm}, clip]{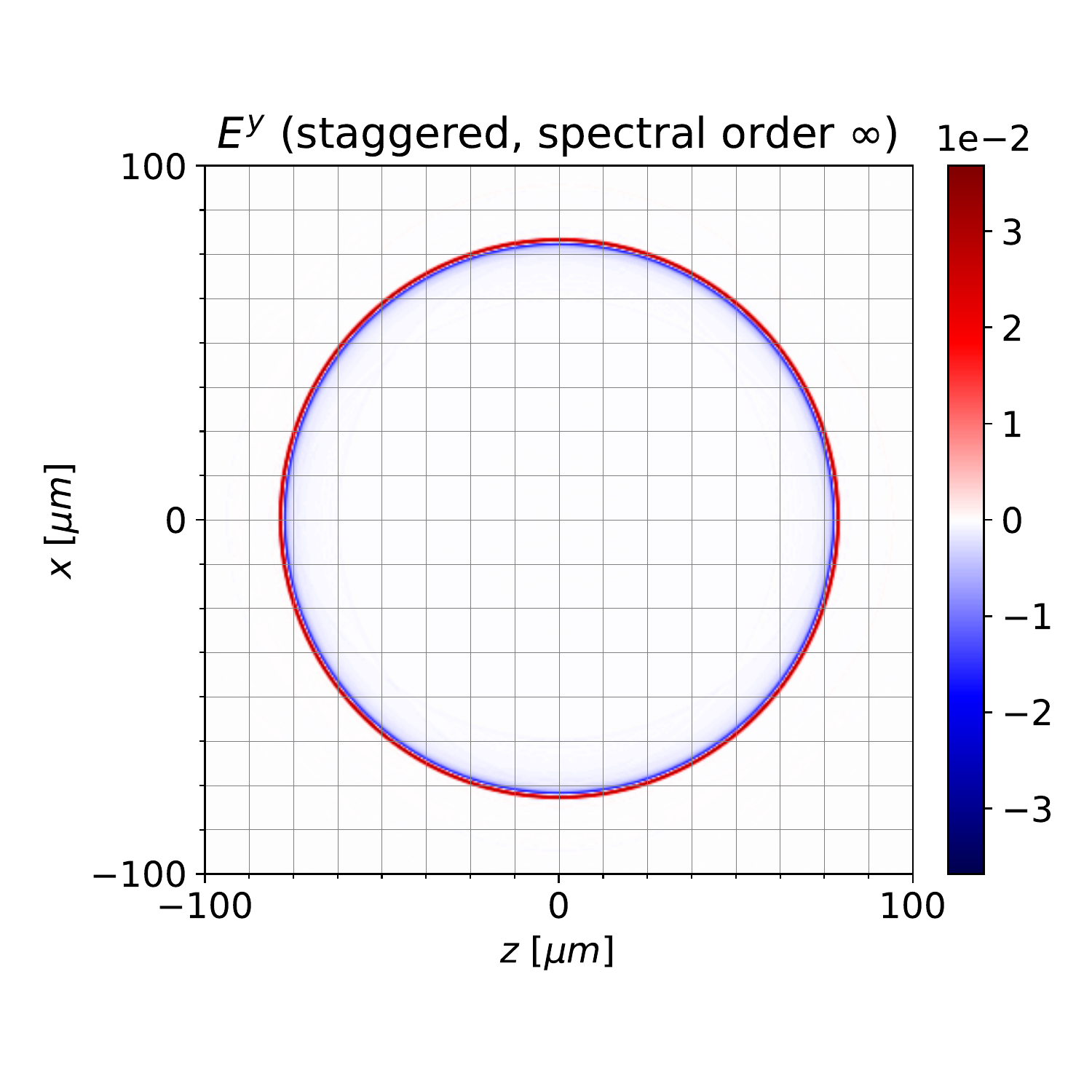}
\includegraphics[width=0.8\linewidth,trim={0 0 0 0},clip]{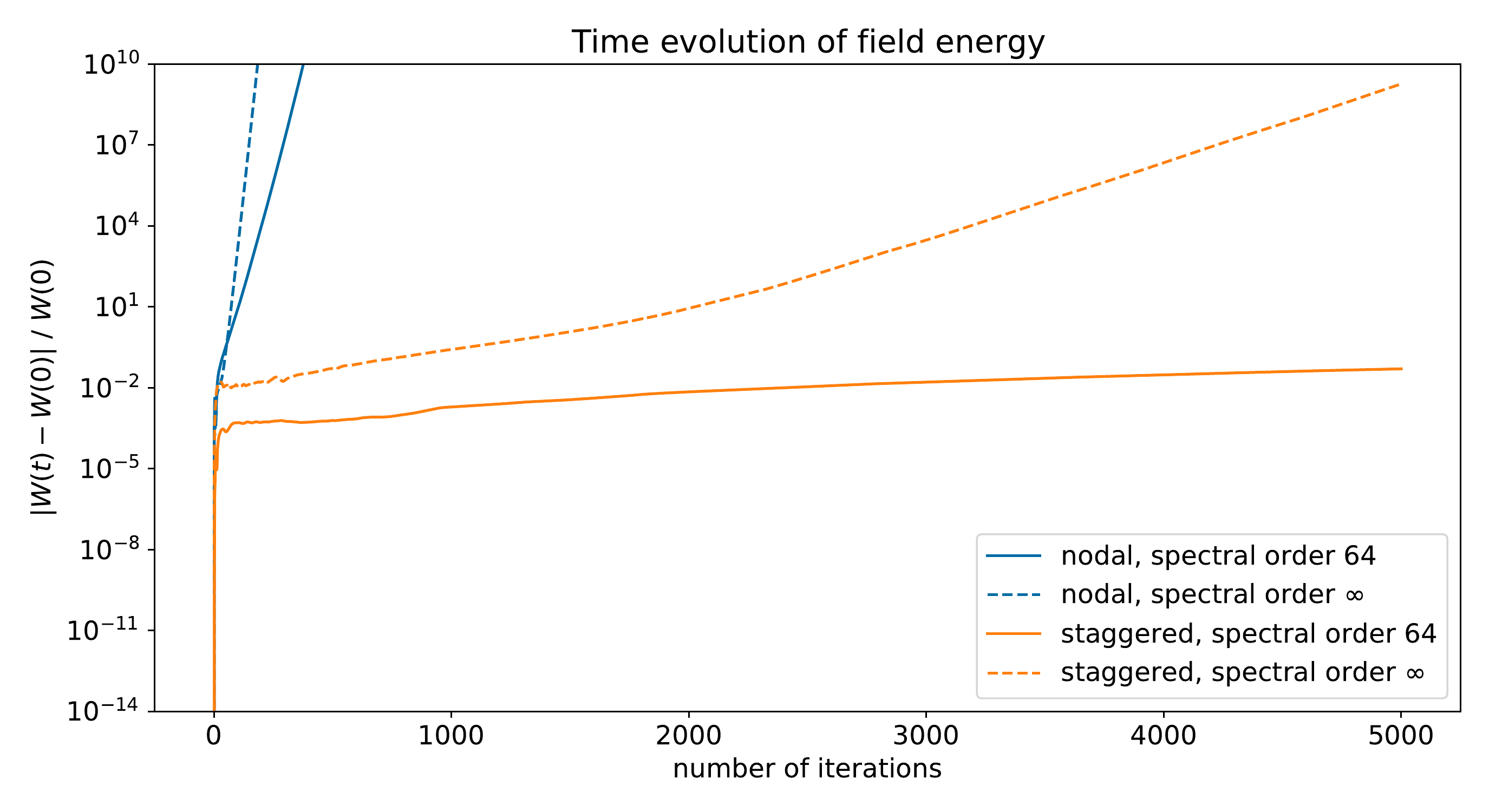}
\caption{\textbf{Nyquist Noise}. Example of short-wavelength noise with a nodal Maxwell solver (left) and a staggered Maxwell solver (right) for a two-dimensional rectangular electric pulse (with  unitary amplitude) initialized at the center of a two-dimensional periodic domain, for parallel runs using domain decomposition with 256 subdomains. 
The nodal solver exhibits much stronger short-wavelength noise than the staggered solver and leads to a strong non-physical growth over time of the electromagnetic field energy $W$, which instead grows much more slowly in the staggered case.}
\label{fig_pulse_nodal_vs_stagg}
\end{figure}
\end{enumerate}

\section{Finite-order Centering with Fornberg Coefficients}
\label{sec_interpolation}
As mentioned in the introduction, PSATD algorithms help mitigate the spurious numerical dispersion of finite-difference algorithms by approximating spatial derivatives with high-order discrete expressions that use large stencils of grid points and by integrating Maxwell's equations, in Fourier space, analytically in time, instead of approximating time derivatives by finite differences.

This section shows that the coefficients originally introduced by Fornberg~\citep{Fornberg1990} for the high-order approximation of spatial derivatives can be employed also to perform finite-order centering of fields and currents between nodal and staggered grids, within the context of the hybrid PSATD PIC scheme outlined in Section~\ref{sec_motivations}.

We first review how spatial derivatives can be approximated with high-order expressions using the Fornberg coefficients.
Since the goal is to show how the same coefficients can be used for finite-order centering of grid quantities from a staggered grid to a nodal grid (or vice versa), we consider here the case of staggered finite differences applied to a function $f: \Lambda \ni x \mapsto f(x) \in \Rbb$, evaluated on the cell centers of a one-dimensional domain $\Lambda := \{ x_{j+\frac 12} := j \Delta x + \Delta x / 2, j \in \Zbb\}$, for a given cell size $\Delta x \in \Rbb$.
``Staggered finite differences'' means that we are interested in providing an approximation of the derivative of $f$ on a cell node (rather than a cell center), say $x_j$.
In this case, the approximation of $\dd f / \dd x$ in $x_j$ at order $2m$, for $\Nbb \ni m > 0$, reads
\begin{linenomath}
\begin{equation}
\label{staggered_finite_difference}
\left(\frac{\dd f}{\dd x}\right)_j = \sum_{n=1}^{m} \alpha_{m,n}^\stxt \frac{f_{j+n-1/2} - f_{j-n+1/2}}{(2n-1) \, \dx} + O(\Delta x^{2m+1}) \,,
\end{equation}
\end{linenomath}
where $\alpha_{m,n}^\stxt$ denote the staggered Fornberg coefficients
\begin{linenomath}
\begin{equation}
\alpha_{m,n}^\stxt := (-1)^{n+1} \left[\frac{(2m)!}{2^{2m} m!}\right]^2 \frac{4}{(2n-1) (m-n)! \, (m+n-1)!} \,.
\end{equation}
\end{linenomath}
Here and in the following, $f_\ell := f(\ell \Delta x)$, for a given integer or half-integer index~$\ell$.
The following one-dimensional schematic helps understand the geometric meaning of the indices used in \eqref{staggered_finite_difference}:
\begin{center}
\begin{tikzpicture}[scale=1.]
\draw (-1,1)--(7,1);
\draw (-1,0)--(7,0);
\foreach \x in {-1,...,7} \node at (\x,1) {$\vert$};
\foreach \x in {-1,...,7} \node at (\x,0) {$\vert$};
\node at (0+0.5,1+0.5) {$j-n+\frac 12$};
\node at (5+0.5,1+0.5) {$j+n-\frac 12$};
\node at (3,0-0.5) {$j$};
\draw[->-=.5] (0.5,1)--(3,0);
\draw[->-=.5] (5+0.5,1)--(3,0);
%
\foreach \x in {-1,...,6} \node at (\x+0.5,1) {\color{blue}{x}};
\foreach \x in {-1,...,7} \node at (\x,0) {\color{red}{\small\textbullet}};
\end{tikzpicture}
\end{center}

A Taylor expansion of the right-hand side of \eqref{staggered_finite_difference} around $x_j = j \Delta x$ yields
\begin{linenomath}
\begin{equation}
\begin{split}
\left(\frac{\dd f}{\dd x}\right)_j & = \sum_{k=1}^{2m+1} \left(\frac{\dd^k f}{\dd x^k}\right)_j \frac{\Delta x^{k-1}}{k!} \sum_{n=1}^{m} \alpha_{m,n}^\stxt \frac{(n-1/2)^k-(-n+1/2)^k}{2n-1} + O(\Delta x^{2m+1}) \\[5pt]
& = \sum_{k=0}^{m} \left(\frac{\dd^{2k+1} f}{\dd x^{2k+1}}\right)_j \frac{\Delta x^{2k}}{(2k+1)!} \sum_{n=1}^{m} \alpha_{m,n}^\stxt (n-1/2)^{2k} + O(\Delta x^{2m+1}) \,,
\end{split}
\end{equation}
\end{linenomath}
where even values of $k$ canceled out in the first line and the first partial sum was re-indexed over the index $k$. This implies
\begin{linenomath}
\begin{equation}
\label{eq_Fornberg_property}
\sum_{n=1}^{m} \alpha_{m,n}^\stxt (n-1/2)^{2k} =
\begin{dcases}
1 & k = 0 \,, \\
0 & k = 1, \dots, m \,.
\end{dcases}
\end{equation}
\end{linenomath}

Thanks to this property of the Fornberg coefficients, the same coefficients $\alpha_{m,n}^\stxt$ can also be used  to perform finite-order interpolation. More precisely, the function $f$ can be interpolated in $x_j$ via
\begin{linenomath}
\begin{equation}
\label{staggered_finite_sum}
f_j = \sum_{n=1}^{m} \alpha_{m,n}^\stxt \frac{f_{j+n-1/2} + f_{j-n+1/2}}{2} + O(\Delta x^{2m+2}) \,.
\end{equation}
\end{linenomath}
This can be shown by performing a Taylor expansion of the right-hand side of \eqref{staggered_finite_sum} around $x_j = j \Delta x$,
\begin{linenomath}
\begin{equation}
\begin{split}
\sum_{n=1}^{m} & \alpha_{m,n}^\stxt \frac{f_{j+n-1/2} + f_{j-n+1/2}}{2} \\
& = f_j \sum_{n=1}^{m} \alpha_{m,n}^\stxt
+ \sum_{k=1}^{2m+1} \left(\frac{\dd^k f}{\dd x^k}\right)_j \frac{\Delta x^{k}}{k!} \sum_{n=1}^{m} \alpha_{m,n}^\stxt \frac{(n-1/2)^k+(-n+1/2)^k}{2} + O(\Delta x^{2m+2}) \\[5pt]
& = f_j \sum_{n=1}^{m} \alpha_{m,n}^\stxt
+ \sum_{k=1}^{m} \left(\frac{\dd^{2k} f}{\dd x^{2k}}\right)_j \frac{\Delta x^{2k}}{(2k)!} \sum_{n=1}^{m} \alpha_{m,n}^\stxt (n-1/2)^{2k} + O(\Delta x^{2m+2}) 
= f_j + O(\Delta x^{2m+2}) \,,
\end{split}
\end{equation}
\end{linenomath}
which proves \eqref{staggered_finite_sum}, thanks to the property \eqref{eq_Fornberg_property}.

\begin{figure}[!ht]
\centering
\includegraphics[width=\linewidth,trim={3cm 0 3cm 1cm},clip]{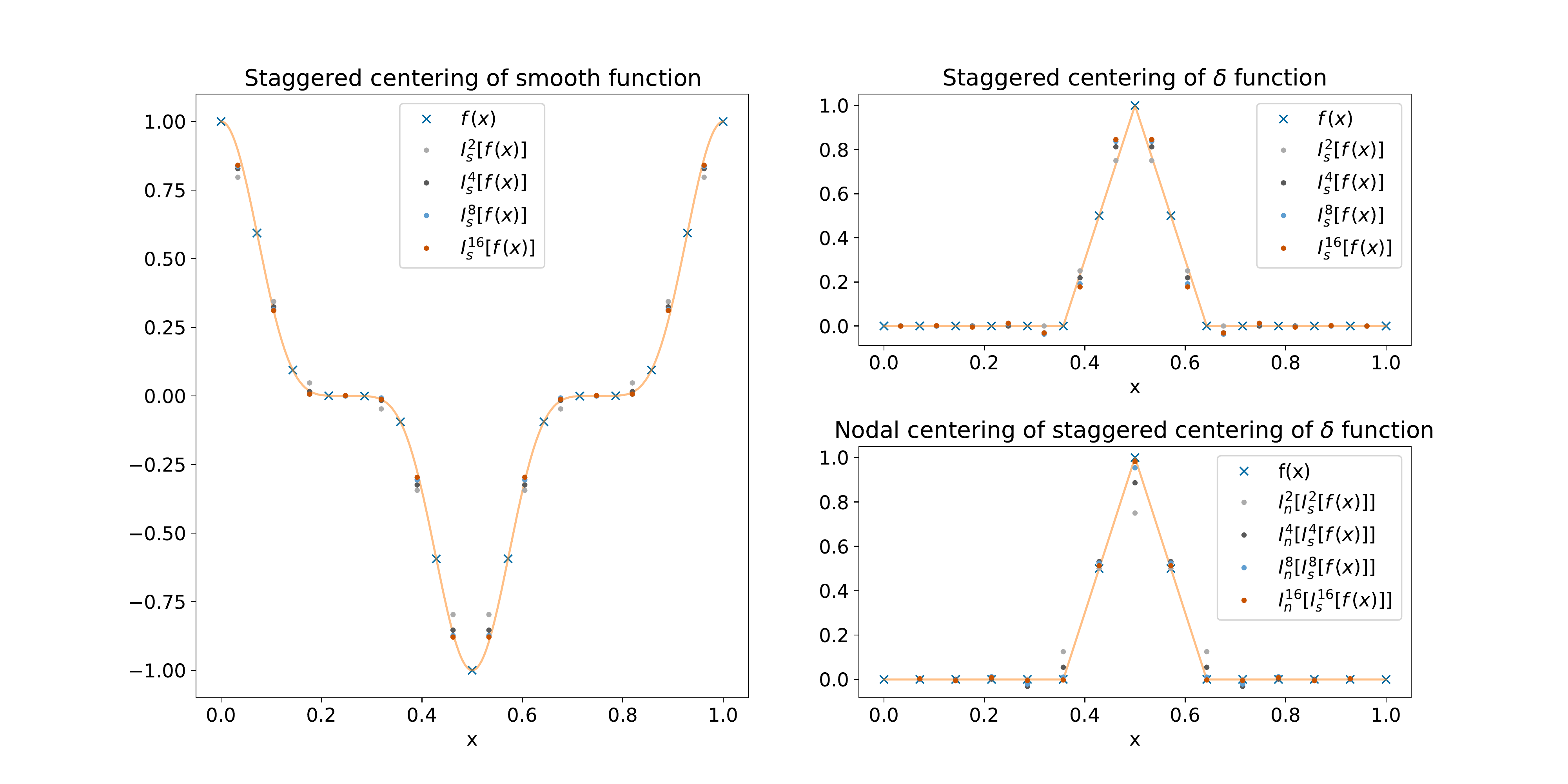}
\caption{\textbf{Finite-order Centering}. One-dimensional interpolation of a smooth function $f(x) := \cos^5(4 \pi x)$ (left) as well as a delta function (right). The delta function on the right is chosen to be a Kronecker pulse extending over a single cell, followed by one pass of binomial filter, representing a unit of charge, current or field on the grid. The crosses represent the discrete values of $f$ used to compute the interpolated values, represented by the colored bullets. The interpolation is performed using the Fornberg coefficients as prescribed in~\eqref{staggered_finite_sum}, at increasing interpolation orders $2m = 2,4,8,16$. $I^{2m}_\ntxt[f]$ and $I^{2m}_\stxt[f]$ denote the centering of $f$ at order $2m$ to a nodal and staggered grid, respectively.}
\label{fig_interpolation}
\end{figure}

Equation \eqref{staggered_finite_sum} is the type of finite-order method that is used for the centering of fields and currents between nodal and staggered grids, within the context of the hybrid PSATD PIC scheme outlined in Section~\eqref{sec_motivations}.
Figure~\ref{fig_interpolation} shows an example of one-dimensional interpolation of a smooth function $f(x) := \cos^5(4 \pi x)$ (left) as well as a delta function (right), where the interpolated values converge to the exact values by increasing the interpolation order $2m$.
The delta function on the right is chosen to be a Kronecker pulse extending over a single cell, followed by one pass of binomial filter, representing a unit of charge, current or field on the grid. 
$I^{2m}_\ntxt[f]$ and $I^{2m}_\stxt[f]$ denote the centering of $f$ at order $2m$ to a nodal and staggered grid, respectively.
It is remarkable that the centering to a nodal grid of data that were centered to a staggered grid recovers the original nodal signal as the order of interpolation goes to infinity.

\section{PSATD Algorithms of Interest on Staggered Grids}
\label{sec_algo}
This section summarizes the equations for the update of the electromagnetic fields in Fourier space, for three PSATD algorithms of interest that are considered in this paper to test the novel hybrid scheme.
These are:
\begin{itemize}
\item the standard PSATD PIC algorithm \citep{Haber1973,Vay2013,Godfrey2014};
\item the standard Galilean PSATD PIC algorithm \citep{Lehe2016,Kirchen2020};
\item the averaged Galilean PSATD PIC algorithm (for large time steps) \citep{Shapoval2021}.
\end{itemize}
The derivation of these solvers, previously performed only for the nodal case, is extended to the staggered case,
to be used with purely staggered or hybrid PIC cycles.
In particular, the derivation of the Galilean equations required extra care because of the presence of the Galilean coordinate transformation for which the new update equations on staggered grids cannot be obtained trivially by replacing all nodal quantities with staggered quantities in the old update equations valid on nodal grids~\citep{Lehe2016,Kirchen2020,Shapoval2021}.

A thorough mathematical derivation is presented in \ref{appendix_1} for the case of the standard Galilean PSATD algorithm.
The case of the standard PSATD algorithm can be derived trivially as a limit of the standard Galilean PSATD results with zero Galilean velocity.
Similarly, the case of the averaged Galilean PSATD algorithm requires simply to perform an additional averaging in time of the standard Galilean PSATD results, as described in~\citep{Shapoval2021}.

\subsection{Notation}
Common notations are introduced that will be used from here on, starting with the modified wave numbers used to express finite-order spatial derivatives in Fourier space.
Considering a one-dimensional domain with cell size $\dx$ for which $k^x$ denotes the wave numbers of the corresponding dual grid in Fourier space, 
centered and staggered finite differences at order $2m$ are expressed in Fourier space by means of centered and staggered modified wave numbers defined~as
\begin{linenomath}
\begin{subequations}
\begin{align}
\label{centered_modified_wave_number}
& [k^x]_\ctxt := \sum_{n=1}^{m} \alpha_{m,n}^\ctxt \frac{\sin(k \, n \, \dx)}{n \, \dx} \,, \\[5pt]
\label{staggered_modified_wave_number}
& [k^x]_\stxt := \sum_{n=1}^{m} \alpha_{m,n}^\stxt \frac{\sin(k \, (n-1/2) \, \dx)}{(n-1/2) \, \dx} \,,
\end{align}
\end{subequations}
\end{linenomath}
where $\alpha_{m,n}^\ctxt$ and $\alpha_{m,n}^\stxt$ denote the centered and staggered Fornberg coefficients \citep{Fornberg1990}
\begin{linenomath}
\begin{subequations}
\begin{align}
\label{eq_Fornberg_centered}
& \alpha_{m,n}^\ctxt := (-1)^{n+1} \frac{2 (m!)^2}{(m-n)! \, (m+n)!} \,, \\[5pt]
\label{eq_Fornberg_staggered}
& \alpha_{m,n}^\stxt := (-1)^{n+1} \left[\frac{(2m)!}{2^{2m} m!}\right]^2 \frac{4}{(2n-1) (m-n)! \, (m+n-1)!} \,.
\end{align}
\end{subequations}
\end{linenomath}

For the Galilean PSATD and averaged Galilean PSATD algorithms, we denote the Galilean velocity by $\vbgal$ and define the additional quantities
$\Omega_\ctxt := \vbgal \cdot [\kb]_\ctxt$,
$\omega_\ctxt := c \, [k]_\ctxt$ and
$\omega_\stxt := c \, [k]_\stxt$,
where the centered and staggered modified wave vectors $[\kb]_\ctxt$ and $[\kb]_\stxt$ are defined as vectors with components defined as in \eqref{centered_modified_wave_number}-\eqref{staggered_modified_wave_number}, and $[k]_\ctxt$ and $[k]_\stxt$ denote their magnitudes, respectively.
We also define the additional quantities
$C := \cos(\omega_\stxt \, \dt)$,
$S := \sin(\omega_\stxt \, \dt)$,
$\theta_\ctxt := e^{i \Omega_\ctxt \dt / 2}$ and
$\theta_\ctxt^* := e^{- i \Omega_\ctxt \dt / 2}$,
as well as the coefficient
\begin{linenomath}
\begin{equation}
\label{eq_chi1}
\chi_1 := \frac{\omega_\ctxt^2}{\omega_\stxt^2 - \Omega_\ctxt^2} \left(\theta_\ctxt^* - \theta_\ctxt \, C + i \, \Omega_\ctxt \, \theta_\ctxt \, \frac{S}{\omega_\stxt} \right) \,.
\end{equation}
\end{linenomath}

In the case of the standard PSATD algorithm, $\vbgal = 0$, $\Omega_\ctxt = 0$, $\theta_\ctxt = \theta_\ctxt^* = 1$, and $\displaystyle \lim_{\Omega_\ctxt \to 0} \chi_1 = \left(1 - C\right) \omega_\ctxt^2 / \omega_\stxt^2$, assuming $\omega_\stxt \neq 0$.


\subsection{Standard PSATD Algorithm}
\label{sec_psatd}
In the case of the standard PSATD algorithm \citep{Haber1973,Vay2013,Godfrey2014}, Faraday's and Ampère-Maxwell's equations in physical space read
\begin{linenomath}
\begin{subequations}
\begin{align}
\label{eq_Faraday}
\frac{\de \Bb}{\de t} & = - \Nabla \times \Eb \,, \\[5pt]
\label{eq_AmpereMaxwell}
\frac{1}{c^2} \frac{\de \Eb}{\de t} & = \Nabla \times \Bb - \mu_0 \Jb \,.
\end{align}
\end{subequations}
\end{linenomath}
Their expressions in Fourier space read
\begin{linenomath}
\begin{subequations}
\begin{align}
\label{eq_Faraday_Fourier_main}
\frac{\de \wh\Bb}{\de t} & = - i \, [\kb]_\stxt \times \wh\Eb \,, \\[5pt]
\label{eq_AmpereMaxwell_Fourier_main}
\frac{1}{c^2} \frac{\de \wh{\Eb}}{\de t} & = i \, [\kb]_\stxt \times \wh{\Bb} - \mu_0 \wh{\Jb} \,.
\end{align}
\end{subequations}
\end{linenomath}

By integrating \eqref{eq_Faraday_Fourier_main}-\eqref{eq_AmpereMaxwell_Fourier_main} analytically in time, along with the continuity equation, the update equations for the electromagnetic fields $\wh\Eb$ and $\wh\Bb$ from time $n \Delta t$ to time $(n+1) \Delta t$, read
\begin{linenomath}
\begin{subequations}
\begin{align}
\label{eq_update_B_with_rho_standard}
\wh\Bb^{n+1} & = C \wh\Bb^{n}
- i \, \frac{S}{\omega_\stxt} \, [\kb]_{\stxt} \times \wh\Eb^{n}
+ i \, X_1 \, [\kb]_{\stxt} \times \wh\Jb^{n+\frac 12} \,, \\[5pt]
\label{eq_update_E_with_rho_standard}
\wh\Eb^{n+1} & = C \wh\Eb^{n}
+ i \, c^2 \, \frac{S}{\omega_\stxt} \, [\kb]_{\stxt} \times \wh\Bb^{n}
+ X_4 \, \wh\Jb^{n+\frac 12}
+ i \, \left(X_3 \, \wh\rho^{\, n} - X_2 \, \wh\rho^{\, n+1}\right) \, [\kb]_{\stxt} \,,
\end{align}
\end{subequations}
\end{linenomath}
where the coefficients $X_1$, $X_2$, $X_3$ and $X_4$ are defined as
\begin{linenomath}
\begin{equation}
X_1 := \frac{1 - C}{\eps_0 \, \omega_\stxt^2} \,, \quad
X_2 := \frac{c^2}{\eps_0 \, \omega_\stxt^2} \left(1 - \frac{S}{\omega_\stxt \, \dt}\right) \,, \quad
X_3 := \frac{c^2}{\eps_0 \, \omega_\stxt^2} \left(C - \frac{S}{\omega_\stxt \, \dt}\right) \,, \quad
X_4 := - \frac{S}{\eps_0 \, \omega_\stxt} \,.
\end{equation}
\end{linenomath}

The update equations \eqref{eq_update_B_with_rho_standard}-\eqref{eq_update_E_with_rho_standard} contain quantities related only to the staggered modified wave vectors, as one might intuitively expect, and they can be obtained also by trivially replacing standard wave numbers with (staggered) modified wave numbers in the update equations valid at infinite spectral order~\citep{Vay2013}.

\subsection{Standard Galilean PSATD Algorithm}
\label{sec_gal}
In the case of the standard Galilean PSATD algorithm \citep{Lehe2016,Kirchen2020}, Faraday's and Ampère-Maxwell's equations in physical space read
\begin{linenomath}
\begin{subequations}
\begin{align}
\label{eq_Faraday_galilean}
\left(\frac{\de}{\de t} - \vbgal \cdot \Nabla\right) \Bb & = - \Nabla \times \Eb \,, \\[5pt]
\label{eq_AmpereMaxwell_galilean}
\frac{1}{c^2}\left(\frac{\de}{\de t} - \vbgal \cdot \Nabla\right) \Eb & = \Nabla \times \Bb - \mu_0 \Jb \,.
\end{align}
\end{subequations}
\end{linenomath}
Their expressions in Fourier space read
\begin{linenomath}
\begin{subequations}
\begin{align}
\label{eq_Faraday_Fourier_galilean}
\left(\frac{\de}{\de t} - i \, \Omega_\ctxt\right) \wh\Bb
& = - i \, [\kb]_\stxt \times \wh\Eb \,, \\[5pt]
\label{eq_AmpereMaxwell_Fourier_galilean}
\frac{1}{c^2}\left(\frac{\de}{\de t} - i \, \Omega_\ctxt\right) \wh{\Eb}
& = i \, [\kb]_\stxt \times \wh{\Bb} - \mu_0 \wh{\Jb} \,.
\end{align}
\end{subequations}
\end{linenomath}
While a thorough mathematical derivation of~\eqref{eq_Faraday_Fourier_galilean}-\eqref{eq_AmpereMaxwell_Fourier_galilean}, along with the results shown in the following, is presented in \ref{appendix_1}, it is important to note that, because of the presence of the Galilean coordinate transformation, a new term involving derivatives of the electromagnetic fields appears on the left hand side of both equations. This, in turns, results in having quantities related to both centered and staggered modified wave vectors, because the finite differences acting on $\Eb$ and $\Bb$ need to be defined differently, according to the spatial staggering of the two fields.

By integrating \eqref{eq_Faraday_Fourier_galilean}-\eqref{eq_AmpereMaxwell_Fourier_galilean} analytically in time, along with the continuity equation, the update equations for the electromagnetic fields $\wh\Eb$ and $\wh\Bb$ in Fourier space, from time $n \Delta t$ to time $(n+1) \Delta t$, read
\begin{linenomath}
\begin{subequations}
\begin{align}
\label{eq_update_B_with_rho_galilean}
\wh\Bb^{n+1} & = \theta_\ctxt^2 C \wh\Bb^{n}
- i \, \theta_\ctxt^2 \frac{S}{\omega_\stxt} \, [\kb]_{\stxt} \times \wh\Eb^{n}
+ i \, X_1 \, [\kb]_{\stxt} \times \wh\Jb^{n+\frac 12} \,, \\[5pt]
\label{eq_update_E_with_rho_galilean}
\wh\Eb^{n+1} & = \theta_\ctxt^2 C \wh\Eb^{n}
+ i \, c^2 \, \theta_\ctxt^2 \frac{S}{\omega_\stxt} \, [\kb]_{\stxt} \times \wh\Bb^{n}
+ X_4 \, \wh\Jb^{n+\frac 12}
+ i \, \left(\theta_\ctxt^2 X_3 \, \wh\rho^{\, n} - X_2 \, \wh\rho^{\, n+1}\right) \, [\kb]_{\stxt} \,,
\end{align}
\end{subequations}
\end{linenomath}
where the coefficients $X_1$, $X_2$, $X_3$ and $X_4$ are defined as
\begin{linenomath}
\begin{subequations}
\begin{align}
& X_1 := \frac{\theta_\ctxt \, \chi_1}{\eps_0 \, \omega_\ctxt^2} \,, \\[5pt]
& X_2 := \frac{c^2}{\theta_\ctxt^* - \theta_\ctxt}
\left(\theta_\ctxt^* X_1 - \theta_\ctxt \frac{1 - C}{\eps_0 \, \omega_\stxt^2}\right) \,, \\[5pt]
& X_3 := \frac{c^2}{\theta_\ctxt^* - \theta_\ctxt}
\left(\theta_\ctxt^* X_1 - \theta_\ctxt^* \frac{1 - C}{\eps_0 \, \omega_\stxt^2}\right) \,, \\[5pt]
& X_4 := i \, \Omega_\ctxt \, X_1 - \frac{\theta_\ctxt^2}{\eps_0} \frac{S}{\omega_\stxt} \,.
\end{align}
\end{subequations}
\end{linenomath}

Unlike the standard PSATD algorithm, the update equations \eqref{eq_update_B_with_rho_galilean}-\eqref{eq_update_E_with_rho_galilean} contain quantities related to both centered and staggered modified wave vectors.
Moreover, the update equations \eqref{eq_update_B_with_rho_standard}-\eqref{eq_update_E_with_rho_standard} for the standard PSATD algorithm can be obtained by taking the limit of \eqref{eq_update_B_with_rho_galilean}-\eqref{eq_update_E_with_rho_galilean} for $\Omega_\ctxt \to 0$.
As expected for consistency,  \eqref{eq_update_B_with_rho_galilean}-\eqref{eq_update_E_with_rho_galilean} correspond to equations (4a)-(4b) of \citep{Kirchen2020} when considering the purely nodal case where all staggered quantities are replaced by the corresponding centered quantities.

\subsection{Averaged Galilean PSATD Algorithm}
\label{sec_avg}
In the case of the averaged Galilean PSATD algorithm \citep{Shapoval2021}, the update equations for the electromagnetic fields $\wh\Eb$ and $\wh\Bb$ in Fourier space, from time $n \Delta t$ to time $(n+1) \Delta t$, are the same as for the Galilean PSATD algorithm, namely \eqref{eq_update_B_with_rho_galilean}-\eqref{eq_update_E_with_rho_galilean}.

Moreover, the update equations for the averaged electromagnetic fields $\langle \wh\Eb \rangle$ and $\langle \wh\Bb \rangle$ in Fourier space, from time $n \Delta t$ to time $(n+1) \Delta t$, are obtained by performing an additional averaging in time, as described in~\citep{Shapoval2021}, which yields
\begin{linenomath}
\begin{subequations}
\begin{align}
\label{eq_update_B_with_rho_averaged}
\langle \wh\Bb \rangle^{n+1} & = \Psi_1 \wh\Bb^{n} + i \, \Psi_2 \, [\kb]_{\stxt} \times \wh\Eb^{n} + i \, Y_1 \, [\kb]_{\stxt} \times \wh\Jb^{n+\frac 12} \,, \\[5pt]
\label{eq_update_E_with_rho_averaged}
\langle \wh\Eb \rangle^{n+1} & = \Psi_1 \wh\Eb^{n} - i \, c^2 \, \Psi_2 \, [\kb]_{\stxt} \times \wh\Bb^{n} + Y_4 \, \wh\Jb^{n+\frac 12} + \left(Y_3 \, \wh\rho^{\, n} + Y_2 \, \wh\rho^{\, n+1}\right) \, [\kb]_{\stxt} \,,
\end{align}
\end{subequations}
\end{linenomath}
where the coefficients $\Psi_1$ and $\Psi_2$ are defined as
\begin{linenomath}
\begin{subequations}
\begin{align}
& \Psi_1 := \frac{\theta_\ctxt^3 \, (\omega_\stxt \, S_3 + i \, \Omega_\ctxt \, C_3) - \theta_\ctxt \, (\omega_\stxt \, S_1 + i \, \Omega_\ctxt \, C_1)}{(\omega_\stxt^2 - \Omega_\ctxt^2) \, \dt} \,, \\
& \Psi_2 := \frac{\theta_\ctxt^3 \, (C_3 - i \, \Omega_\ctxt \, S_3/\omega_\stxt) - \theta_\ctxt \, (C_1 - i \, \Omega_\ctxt \, S_1/\omega_\stxt)}{(\omega_\stxt^2 - \Omega_\ctxt^2) \, \dt} \,,
\end{align}
\end{subequations}
\end{linenomath}
with $C_m = \cos(m \, \omega_\stxt \, \dt / 2)$ and $S_m = \sin(m \, \omega_\stxt \, \dt / 2)$, for $m = 1,2,3$, and the coefficients $Y_1$, $Y_2$, $Y_3$ and $Y_4$ are defined as
\begin{linenomath}
\begin{subequations}
\begin{align}
& Y_1 := \frac{1 - \Psi_1 - i \, \Omega_\ctxt \, \Psi_2}{\eps_0 \, (\omega_\stxt^2 - \Omega_\ctxt^2)} \,, \\[5pt]
& Y_2 := i \, c^2 \, \frac{\eps_0 \, \omega_\stxt^2 \, Y_1 - \Psi_3 + \Psi_1}{\eps_0 \, \omega_\stxt^2 \, (\theta_\ctxt^2 - 1)} \,, \\[5pt]
& Y_3 := i \, c^2 \, \frac{\Psi_3 - \Psi_1 - \eps_0 \, \theta_\ctxt^2 \, \omega_\stxt^2 \, Y_1}{\eps_0 \, \omega_\stxt^2 \, (\theta_\ctxt^2 - 1)} \,, \\[5pt]
& Y_4 := i \, \Omega_\ctxt \, Y_1 + \frac{\Psi_2}{\eps_0} \,,
\end{align}
\end{subequations}
\end{linenomath}
with $\Psi_3 := - i \, \theta_\ctxt \, (\theta_\ctxt^2 - 1) / (\Omega_\ctxt \, \dt)$.

As for the standard Galilean PSATD algorithm, all update equations contain quantities related to both centered and staggered modified wave vectors. As expected here again for consistency, \eqref{eq_update_B_with_rho_averaged}-\eqref{eq_update_E_with_rho_averaged} correspond to equations (10)-(11) of \citep{Shapoval2021}, when considering the purely nodal case where all staggered quantities are replaced by the corresponding centered quantities.

\section{Numerical Tests}
\label{sec_tests}
This section presents various physics applications to test the novel hybrid scheme, with the different PSATD PIC algorithms described in Section~\ref{sec_algo}. All simulations and results have been performed and obtained with the open-source electromagnetic PIC code WarpX \citep{Vay2018,Vay2020,VayPoP2021}.

\subsection{Standard PSATD Algorithm: Vacuum Electron-Positron Pair Creation}
\label{sec_psatd_test}

A first example where it is advantageous to use the novel hybrid scheme presented here is the modeling of electron-positron pair creation in vacuum. This effect, known as the Schwinger effect~\citep{Sauter1931,Heisenberg1936,Schwinger1951,Bulanov2010}, is among the most fundamental predictions of strong-field quantum electrodynamics. Observing the Schwinger process experimentally would shed light on the profound properties of the quantum vacuum, and is a major scientific goal for several research fields, including quantum field theory, high-energy astrophysics, and the design of future particle colliders. The Schwinger effect is expected to occur for electric fields approaching the Schwinger field, $E_S = 1.32 \times 10^{18}$ V/m, which is more than 3 orders of magnitude greater than the most intense fields produced by femtosecond lasers to date~\citep{Yoon2021}.

It was recently proposed~\citep{Vincenti2019} that the Schwinger field $E_S$ could be approached by focusing a multi-Petawatt laser pulse on a so-called plasma mirror, i.e., a solid-density plasma with a sharp density gradient on its front surface. As the laser field is reflected by the plasma, it is periodically compressed in time by the relativistic oscillation of the plasma surface (induced by the laser itself). As a result, the reflected field is emitted as a train of attosecond pulses. In the frequency domain, this corresponds to the generation of high-harmonics by Doppler upshift. Since the reflected harmonic beam is made of higher frequency components than the incident laser, it can be focused to much tighter focal spots. In fact, the focusing of the harmonics does not even require additional optical elements and can be achieved through the curvature of the plasma mirror, which is induced by the incident laser itself. The combination of temporal compression and tighter focusing results in multiple orders of magnitude intensity gains and, under optimal conditions, could be sufficient to bridge the gap towards the Schwinger limit~\citep{Vincenti2021}.

Accurate modelling of vacuum electron-positron pair creation is crucial to the design of future experiments at the Schwinger limit. In particular, it is important to determine the exact intensity thresholds for pair creation, the number of pairs generated as a function of the harmonic beam parameters, and the new physics that is expected to come at play in these extreme regimes.

We have therefore performed two-dimensional PIC simulations of the generation of Schwinger pairs at the focus of a very intense harmonic beam. For these test simulations, we have considered an idealized harmonic beam with a fundamental wavelength $\lambda$ = 800 nm. Its spectrum contains more than 100 harmonic orders and has been obtained from a one-dimensional PIC simulation of a 20 fs, 10$^{22}$~W/cm$^2$ laser impinging with 55$^{\circ}$ incidence on a plasma mirror~\citep{Vincenti2021}. Each of these harmonics has a Gaussian transverse spatial profile and is focused down to diffraction limit ($w_{0,n} = \lambda / n$, where $w_{0,n}$ is the beam waist at focus of the harmonic of order $n$)~\citep{SainteMarie2021}. Finally, the peak intensity of the harmonic beam has been manually set to approximately 10 times the Schwinger limit. Although this value is currently unrealistic, it is convenient for test purposes because it leads to the generation of a very high number of pairs, which smooths out statistical fluctuations between different simulations. In the test case, the harmonic beam is injected by an antenna 5 $\micron$ before focus. It propagates in the $z$ direction and its magnetic field is directed towards the $y$ direction, which is perpendicular to the simulation plane. Since this harmonic beam contains about 100 harmonic orders, it can be subject to strong numerical dispersion and it is therefore absolutely necessary to use a spectral Maxwell solver to mitigate this effect \cite{Blaclard2017a}. We use here the standard PSATD algorithm, with stencils of order 16 and 8 ghost cells in each direction, both with and without the new hybrid scheme.

The Schwinger process is implemented in the PIC code WarpX through the PICSAR library~\citep{Fedeli2021_2} and is enabled in the simulations. However, in order to separate the effect of pair creation in vacuum from further self-consistent effects, the generated electrons and positrons do not deposit neither their charge nor their current on the grid for this test case. Therefore, they do not influence the propagation of the harmonic beam in vacuum and only serve as probes of the Schwinger effect. Finally, in order to convert the Schwinger pair production rate (which is a number of pairs per unit volume per unit time) into an estimated number of particles generated per cell per time step, a transverse cell size of 20 nm, which corresponds to the typical transverse size of the harmonic beam at focus, is used.

To understand the results of these numerical tests, it is instructive to consider the most important features of the Schwinger pair production rate. This rate only depends on the invariants of the electromagnetic field $\mathcal{F} := \left( \mathbf{E}^2 - c^2 \mathbf{B}^2 \right) / E_S^2$ and $\mathcal{G} := c \, \mathbf{E} \cdot \mathbf{B} / E_S^2$ (which are normalized here by the Schwinger field). Close to the pair generation threshold, $\mathcal{F}$ is the most important of the two invariants, and it is in fact the only one that is non-zero in the two-dimensional simulations presented here, in which the magnetic field is perpendicular to the simulation plane. The pair production rate is extremely sensitive to small changes of $\mathcal{F}$ and becomes significant when $\mathcal{F}$ is positive and reaches the percent level. This can occur for a very strong electrostatic field or for two counterpropagative plane waves (near the nodes of the magnetic field). On the other hand, for a single plane wave, the $\mathbf{E}$ and $\mathbf{B}$ fields have the same amplitude everywhere, which means that $\mathcal{F}$, and thus the pair production rate, is zero regardless of the field amplitude. Our test case, with a single harmonic beam, is closer to that of a single plane wave. However, due to the tight focusing of the harmonics, the $\mathbf{E}$ and $\mathbf{B}$ fields do not have the same amplitude everywhere and pair creation can still occur in regions with stronger $\mathbf{E}$ field. Yet, since the invariant $\mathcal{F}$ is computed by subtracting two numbers that are very close in amplitude, great care must be taken to avoid numerical errors. In particular, it is absolutely necessary that all field components used in the calculation of $\mathcal{F}$ be located at the same points in the grid as the difference is then otherwise easily dominated by interpolation errors.

A natural solution to achieve this could be to solve Maxwell's equations on a nodal grid. However, as shown in Figure~\ref{fig_pulse_nodal_vs_stagg}, the nodal solver is subject to strong noise at the Nyquist wavelength. While this noise does not significantly change the amplitude of the fields, it can heavily affect the amplitude of the invariant $\mathcal{F}$. This is particularly true when the noise at the Nyquist wavelength is counterpropagative with the harmonic beam -- which is similar to the case of two counterpropagative plane waves. As a result, it is observed that using the nodal solver results in spurious Schwinger pair creations, which can increase the total number of generated pairs by several orders of magnitude, or lead to pair creation at intensities where there should be none. This issue effectively makes it infeasible to solve Maxwell's equation on a nodal grid in simulations that include the Schwinger effect.

Therefore, it is necessary to solve Maxwell's equations on a staggered Yee grid and then interpolate all field components on the cell nodes to compute the invariant $\mathcal{F}$. This precisely corresponds to the novel hybrid scheme presented in this article. In order to evaluate the impact of interpolation errors in the computation of the invariant, two convergence scans of the test case presented in this section were performed, with field centering of order 2 and 8, respectively. The resolutions used in the convergence scan are given in Table~\ref{table_qed_convergence_scan}. In each case, the peak intensity is adjusted so that the same total field energy is used in all simulations.

\begin{table}[h]
\centering
\renewcommand{\arraystretch}{1.5}
\setlength{\tabcolsep}{10pt}
\begin{tabular}{c|c|c}
\textbf{Simulation \#} &
\textbf{Longitudinal resolution} $\Delta z$ &
\textbf{Transverse resolution} $\Delta x$ \\
\hline
1 &
$\lambda$ / 328$\phantom{0}$ &
$\lambda$ / 60$\phantom{0}$ \\
\hline
2 &
$\lambda$ / 655$\phantom{0}$ &
$\lambda$ / 120 \\
\hline
3 &
$\lambda$ / 1311 &
$\lambda$ / 241 \\
\hline
4 &
$\lambda$ / 2621 &
$\lambda$ / 482 \\
\hline
5 &
$\lambda$ / 5243 &
$\lambda$ / 482 \\
\end{tabular}
\caption{Spatial resolutions used in the vacuum pair creation convergence scans. Each row corresponds to a simulation of the convergence scan. $\Delta x$ and $\Delta z$ denote the cell sizes in the transverse and longitudinal directions, respectively. Note that $\Delta x$ has been set to the same value in the last two simulations to save computational time. This should not affect the results presented here since it was observed that $\Delta z$ has, by far, the most significant impact on the calculation of the invariant. In all cases, the time step satisfies $c \, \Delta t = \Delta z$.}
\label{table_qed_convergence_scan}
\end{table}

\begin{figure}[!ht]
\centering
\includegraphics[width=1.\linewidth,trim={0 0 0 0},clip]{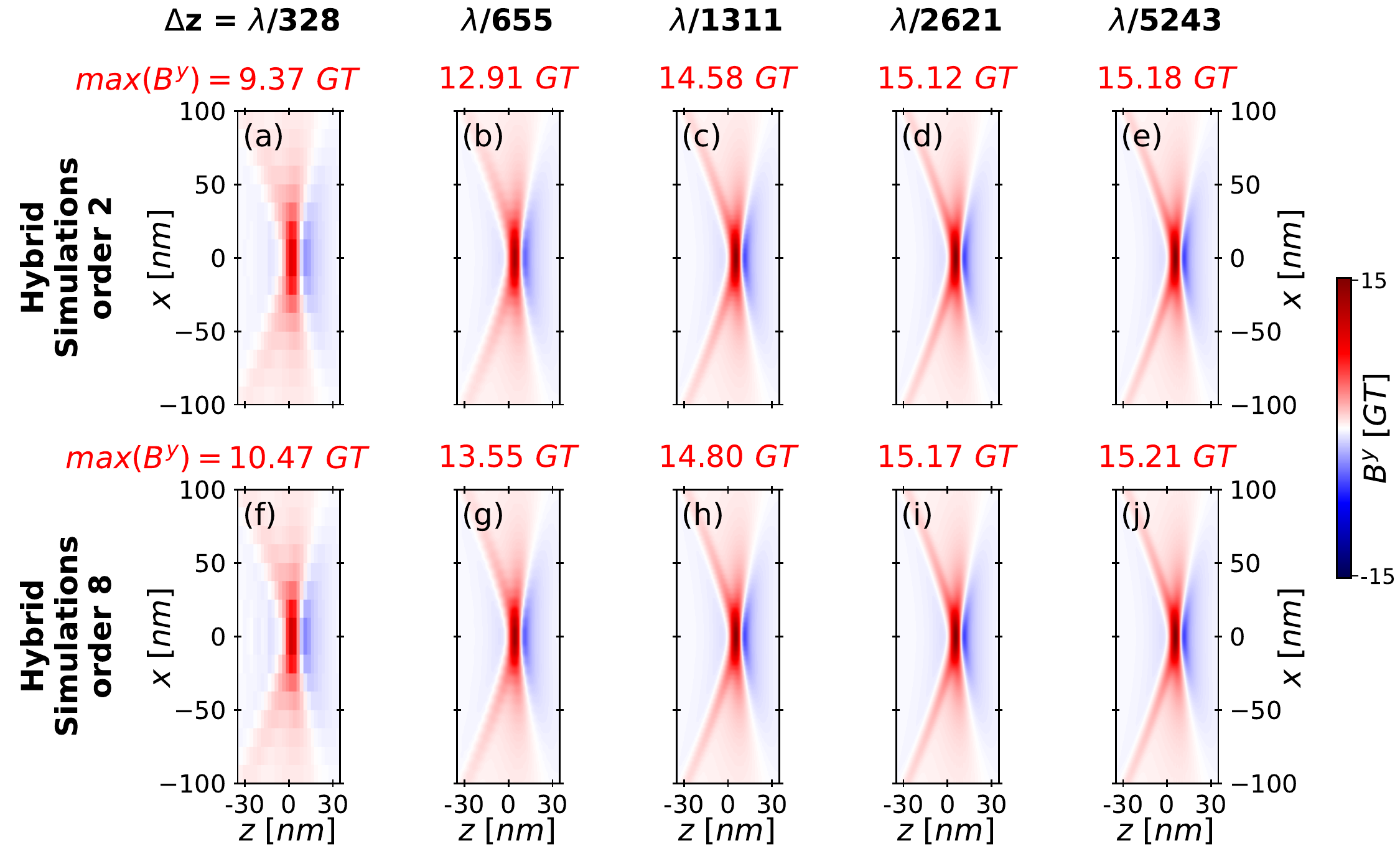}
\caption{\textbf{Magnetic Field}. Snapshots of the $B^y$ field at the focus of the most intense attosecond pulse with field centering of order 2 (a)-(e) or order 8 (f)-(j). The longitudinal resolution is given in the top row and the peak absolute value of the $B^y$ field is written in red above each snapshot.}
\label{fig_snapshot_Bfield_qed}
\end{figure}

Figure~\ref{fig_snapshot_Bfield_qed} shows snapshots of the component $B^y$ of the magnetic field at the focus of the most intense attosecond pulse, for all simulations of the convergence scan. The attosecond pulse looks similar in all cases. Yet, the peak amplitude of the electromagnetic field substantially increases with resolution. This is likely due to the combination of two factors: (i) the peak of the attosecond pulse is better resolved in time and space at higher resolution, and (ii) the highest harmonic orders are absent or do not propagate well at the lowest resolutions. It is expected that this effect will tend to increase the number of pairs generated when increasing the resolution.

\begin{figure}[!ht]
\centering
\includegraphics[width=\linewidth,trim={0 0 0 0},clip]{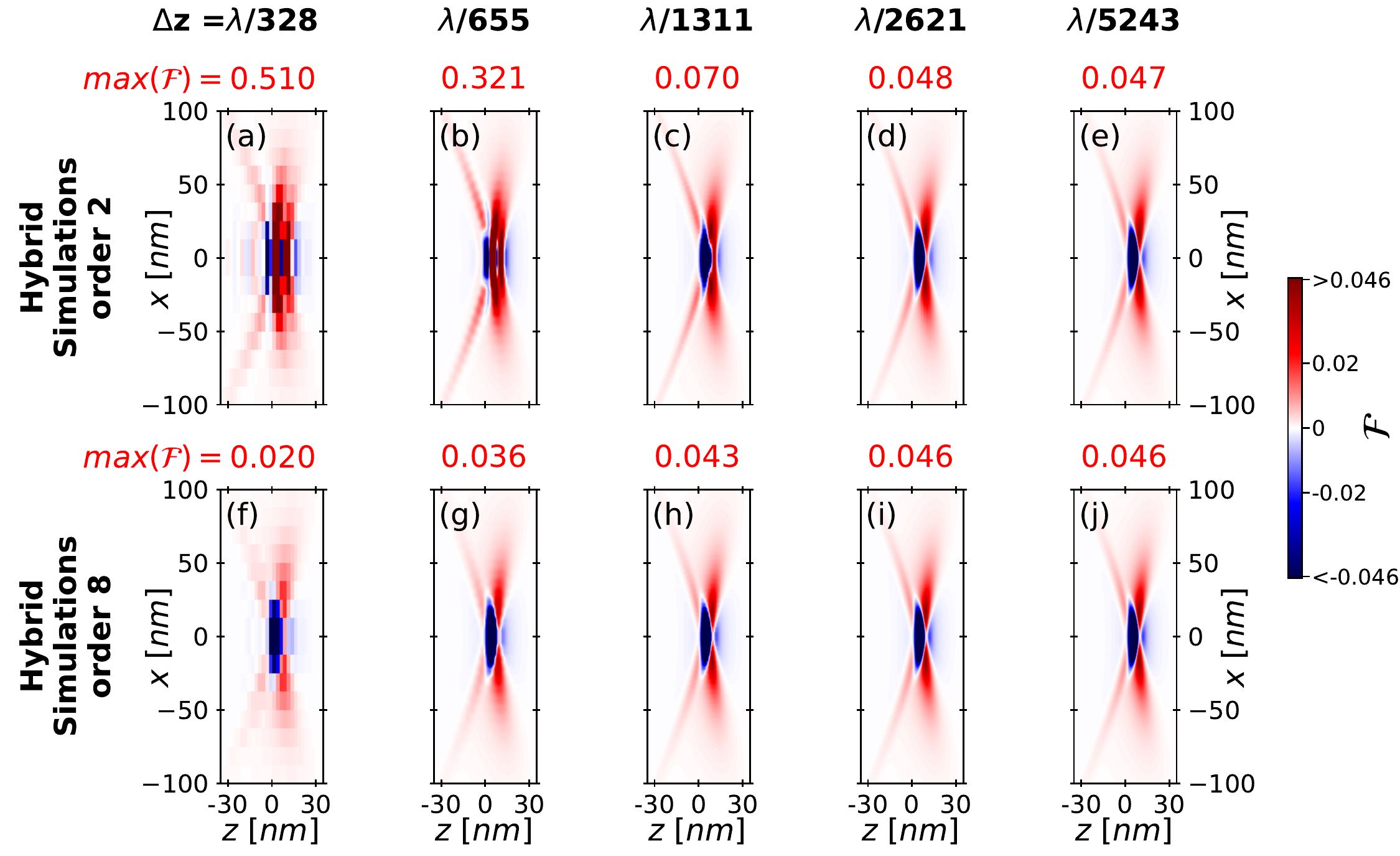}
\caption{\textbf{Field Invariant}. Snapshots of the invariant $\mathcal{F}$ at the focus of the most intense attosecond pulse with field centering of order 2 (a)-(e) or order 8 (f)-(j). The longitudinal resolution is given in the top row and the peak positive value of the invariant $\mathcal{F}$ is written in red above each snapshot.}
\label{fig_snapshot_invariant_qed}
\end{figure}

\begin{figure}[!ht]
\centering
\includegraphics[width=\linewidth,trim={0 0 0 0},clip]{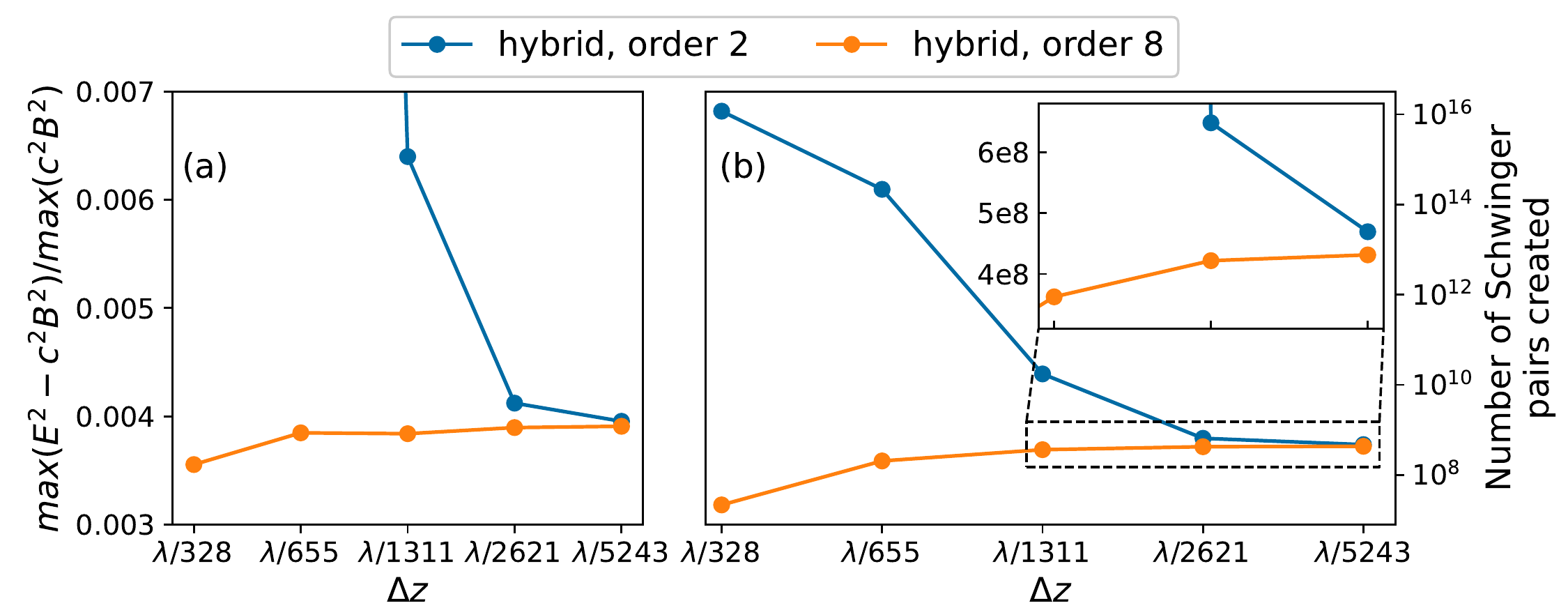}
\caption{\textbf{Convergence}. (a) Normalized ratio between the peak invariant $\mathcal{F}$ and the peak intensity as a function of spatial resolution. (b) Total number of Schwinger pairs generated in the simulations as a function of spatial resolution. The upper-right inset is a zoom over the rectangle marked by the dashed lines (note the change from logarithmic to linear scale in the $y$-axis).}
\label{fig_convergence_scan_qed}
\end{figure}

Figure~\ref{fig_snapshot_invariant_qed} shows snapshots of the invariant $\mathcal{F}$ at the same time and position as in Figure~\ref{fig_snapshot_Bfield_qed}. With field centering of order 2, the spatial shape of the invariant changes with resolution, until it starts converging from $\Delta z = \lambda / 2621$. This is because, until that resolution, the computation of the invariant is dominated by errors in the interpolation of the different field components on the nodes. Consequently, the peak values of $\mathcal{F}$ are much higher at lower resolution, even though the peak intensity is smaller at lower resolution. With field centering of order 8, the behavior is radically different: the spatial shape of the invariant is the same regardless of resolution, meaning that the computation of the invariant is never affected by interpolation errors. This time, the peak values attained by the invariant $\mathcal{F}$ increase with resolution, which is expected since the harmonic beam intensity itself increases with resolution.
To cancel out these intensity variations, Figure~\ref{fig_convergence_scan_qed}(a) shows the ratio between the peak invariant $\mathcal{F}$ and the peak intensity. This ratio is virtually constant for all resolutions with field centering of order 8. This result shows that the computation of the invariant is always accurate when using high-order field centering, and that the changes in the invariant amplitude are simply driven here by the peak intensity variations.

Figure~\ref{fig_convergence_scan_qed}(b) shows the total number of Schwinger pairs as a function of resolution. The general trends are the following: with field centering of order 2, interpolation errors in the computation of $\mathcal{F}$ lead to non-physical pair creation at lower resolution, which can quickly increase the number of pairs by several orders of magnitude. With field centering of order 8, the number of pairs increases with resolution, following the increase of the harmonic beam intensity. To obtain a correct order of magnitude for the number of Schwinger pairs, a longitudinal resolution $\dz = \lambda / 655$ is sufficient with field centering of order 8, whereas a longitudinal resolution $\dz = \lambda / 2621$ is required with field centering of order 2. Even at the highest resolution ($\dz = \lambda / 5243$), field centering of order 2 still results in a $9\%$ overestimation of the number of pairs created, approximately. For comparison, field centering of order 8 is already more accurate for $\Delta z = \lambda / 2621$, with a $2\%$ underestimation of the number of generated pairs, approximately.

\begin{figure}[!ht]
\centering
\includegraphics[width=0.5\linewidth,trim={0 0 0 0},clip]{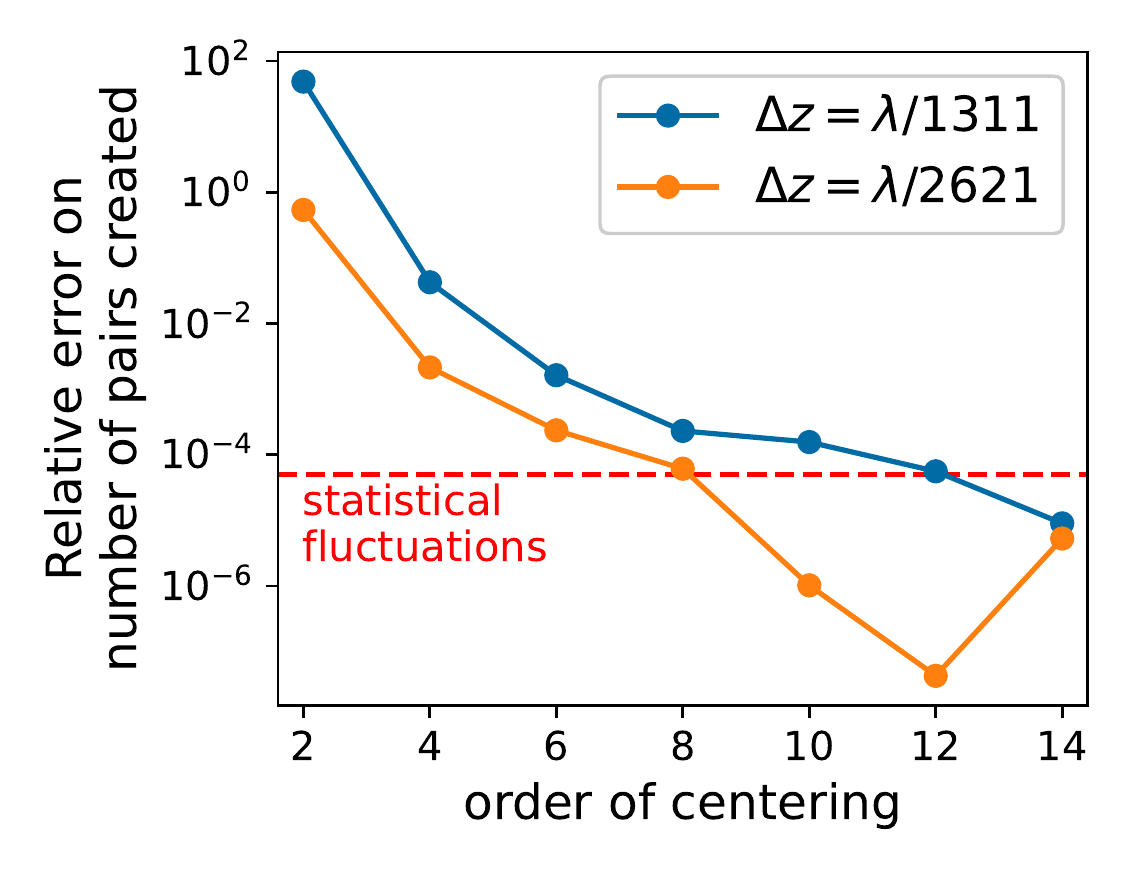}
\caption{\textbf{Effect of Centering Order}. Relative difference between the number of pairs created with a given field centering order and the number of pairs created with field centering of order 16, for a longitudinal resolution of $\dz = \lambda / 1311$ (blue curve) and $\dz = \lambda / 2621$ (orange curve). The dashed-red line shows the standard deviation of the number of generated Scwhinger pairs for identical simulations, estimated from the number of pairs obtained with $\dz = \lambda / 2621$ and field centering of order 16.}
\label{fig_centering_order_qed}
\end{figure}

Figure~\ref{fig_centering_order_qed} shows that increasing the field centering order results in a steady decrease of spurious pair generation coming from interpolation errors, up until the point where non-physical pair creation becomes comparable to the statistical fluctuations inherent to the Schwinger process. In our case, this typically occurs near field centering of order 8. Table~\ref{table_qed_runtimes} shows the runtimes of the simulations shown in Figure~\ref{fig_centering_order_qed}. The runtimes are mainly driven by the resolution and are also moderately affected by the amount of pair creation (which results in slightly load imbalanced simulations). On the other hand, the choice of the field centering order appears to have no sizeable effect on the simulation runtimes. These results indicate that any field centering order greater or equal to 8 (and compatible with the number of ghost cells used with domain decomposition) is perfectly appropriate for this test case.

\begin{table}[h]
\centering
\renewcommand{\arraystretch}{1.5}
\setlength{\tabcolsep}{7pt}
\begin{adjustbox}{center}
\begin{tabular}{c|c|c|c|c|c|c|c|c}
\makecell{\textbf{Simulation}\\ \textbf{Runtime}} &
\makecell{hybrid, \\ order 2} &
\makecell{hybrid, \\ order 4} &
\makecell{hybrid, \\ order 6} &
\makecell{hybrid, \\ order 8} &
\makecell{hybrid, \\ order 10} &
\makecell{hybrid, \\ order 12} &
\makecell{hybrid, \\ order 14} &
\makecell{hybrid, \\ order 16} \\
\hline
\makecell{$\dz = \lambda / 1311$ \\ (32 Summit nodes)} &
$634 \, \stxt$ &
$610 \, \stxt$ &
$602 \, \stxt$ &
$596 \, \stxt$ &
$604 \, \stxt$ &
$604 \, \stxt$ &
$604 \, \stxt$ &
$600 \, \stxt$ \\
\hline
\makecell{$\dz = \lambda / 2621$ \\ (128 Summit nodes)} &
$1111 \, \stxt$ &
$1024 \, \stxt$ &
$1022 \, \stxt$ &
$1048 \, \stxt$ &
$1046 \, \stxt$ &
$1046 \, \stxt$ &
$1052 \, \stxt$ &
$1049 \, \stxt$
\end{tabular}
\end{adjustbox}
\caption{Runtimes of the vacuum pair creation test case as a function of the field centering order for $\dz = \lambda / 1311$ and $\dz = \lambda / 2621$.}
\label{table_qed_runtimes}
\end{table}

Overall, the novel hybrid scheme has proven very useful for simulations of the Schwinger effect. Compared to a fully nodal scheme, it is much less sensitive to noise at the Nyquist frequency, which suppresses severe non-physical pair creation. Moreover, the ability to use high-order field centering entirely removes spurious pair creation coming from interpolation errors in the computation of the invariant $\mathcal{F}$. This feature significantly accelerates convergence and allows to reduce the resolution by a factor 3 to 4 in simulations of vacuum pair generation.

\subsection{Standard Galilean PSATD Algorithm: Laser-Driven Plasma Wakefield Acceleration}
\label{sec_gal_test}
This section presents a test of the novel hybrid scheme with the standard Galilean PSATD algorithm in a Lorentz-boosted frame \citep{Vay2007,Vay2010,Vaypop2011}, on the numerical simulation of laser-driven plasma wakefield acceleration (LWFA)~\cite{Esareyrmp09}.
A laser beam propagating through an under-dense plasma displaces electrons, creating a plasma wakefield that produces very high electric fields, which can be used to accelerate a short charged particle beam to high energy.
Plasma wakefield acceleration represents a novel accelerator technology, alternative to traditional particle accelerators (where the accelerating fields are produced by radio-frequency electromagnetic waves shaped by metallic cavities), and holds the promise of smaller and cheaper particle accelerators, making these machines more accessible for uses in many fields of science and technology, ranging from fundamental physics to medicine, security, and industrial applications.

This section reports on simulations of a laser propagating through a column of pre-ionized plasma that were performed on a three-dimensional computational domain, parametrized by the Cartesian coordinates \mbox{$(x,y,z) \in [-200 \, \micron, 200 \, \micron] \times [-200 \, \micron, 200 \, \micron] \times [-160 \, \micron, 0 \, \micron]$}.

The plasma is made of electrons and protons, injected in the simulation with 1 particle per cell in each direction. The plasma transverse density profile is parabolic, with a flat longitudinal profile terminated by cosine-like ramps at each end. The density for both electrons and protons reads $n(x,y,z) = n_0 \, n(x,y) \, n(z)$, where
\begin{linenomath}
\begin{subequations}
\begin{align}
& n(x,y) = 1 + 4 \, \frac{x^2 + y^2}{k_p^2 \, R_c^4} \,, \\[5pt]
& n(z) =
\begin{dcases}
\frac 12 \, \left[1 - \cos\left(\frac{\pi z}{L_{+}}\right)\right] & 0 \leq z < L_{+} \,, \\ 
1 & L_{+} \leq z < L_{+} + L_{\ptxt} \,, \\
\frac 12 \, \left[1 + \cos\left(\frac{\pi(z - L_{+} - L_{\ptxt})}{L_{-}}\right)\right] & L_{+} + L_{\ptxt} \leq z < L_{+} + L_{\ptxt} + L_{-} \,,
\end{dcases}
\end{align}
\end{subequations}
\end{linenomath}
with $n_0 = 1.7 \times 10^{23} \, \mtxt^{-3}$, $k_p = (q/c) \sqrt{n_0 / (m \, \eps_0)}$, $R_c = 40 \, \micron$, $L_{+} = 20 \, \mm$, $L_{-} = 3 \, \mm$, and $L_{\ptxt} = 0.297 \, \mtxt$. Both electrons and protons are injected assuming zero momentum (cold plasma) in the laboratory frame.

The laser propagates in the longitudinal direction after injection using a virtual antenna \cite{Vaypop2011} located at $x = y = 0$ and \mbox{$z = -1.0 \, \nm$} in the laboratory frame, and it is polarized in the $y$ direction. The peak amplitude of the laser field is $E_{\text{max}} \approx 6.82 \times 10^{12} \, \Vtxt/\mtxt$. The peak intensity is reached at $t \approx 0.14 \, \ps$ and the laser pulse has a duration of $\tau \approx 73.4 \, \fs$. The laser profile is Gaussian along both the transverse and longitudinal directions, with a transverse waist $w = 50 \, \micron$. The laser wavelength is $\lambda = 0.8 \, \micron$ and the distance between the antenna and the focal plane is $\delta = 8.75 \, \mm$ in the laboratory frame.

\begin{figure}[!ht]
\centering
\includegraphics[width=0.49\linewidth,trim={0 0 0 0},clip]{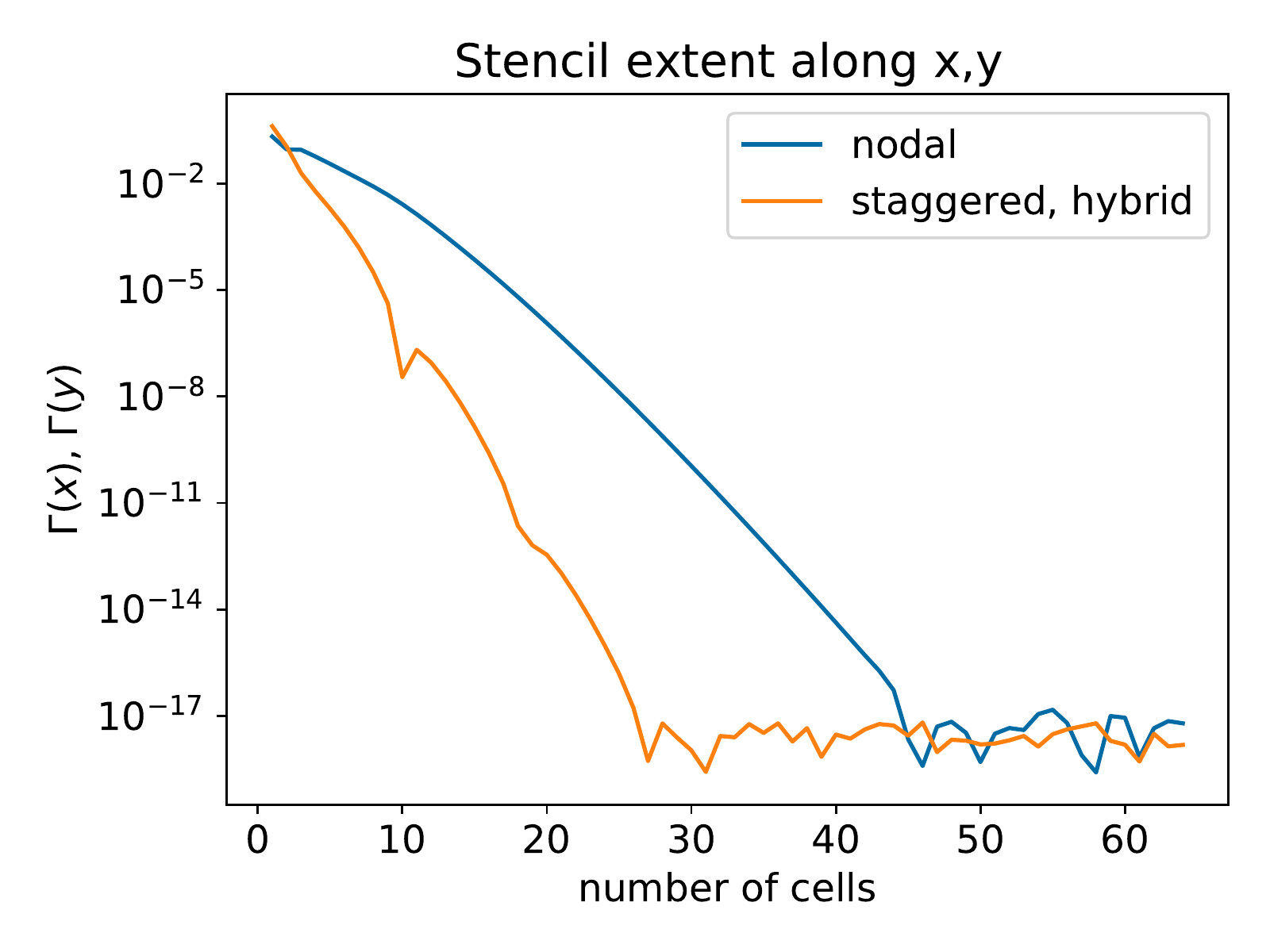}
\includegraphics[width=0.49\linewidth,trim={0 0 0 0},clip]{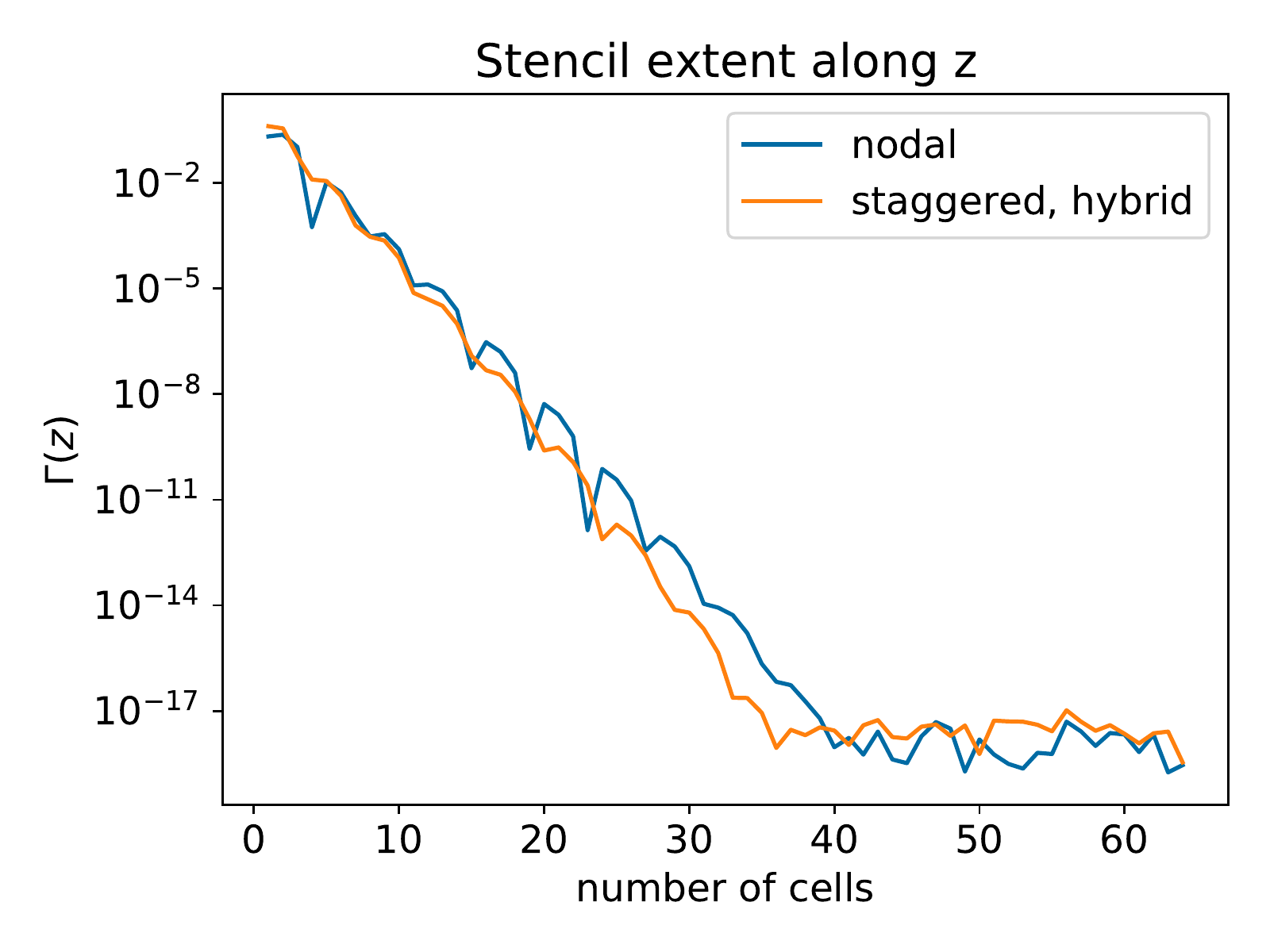}
\caption{\textbf{PSATD Stencils}. Stencil extent of the leading coefficient $\theta_\ctxt^2 \, C$ along $x$, $y$, $z$ for the LWFA test case, computed as prescribed in \eqref{eq_stencil_extent}.}
\label{fig_stencil_lwfa}
\end{figure}

The computational domain is divided in $N_x \times N_y \times N_z = 128 \times 128 \times 2052$ cells and decomposed in 24 subdomains, with $64 \times 64 \times 342$ cells per subdomain. Macro-particles use cubic splines as particle shape factors for current deposition and field gathering and the Vay scheme \citep{Vay2008} for the velocity and position updates. To minimize the number of time steps and speed up the runtime, the simulation is performed using a Lorentz boosted frame of reference in the longitudinal direction, with a Lorentz factor $\gamma = 30$. For stability, the simulation grid follows the plasma with longitudinal Galilean velocity $\vbgal = \vgal \, \hat\zb$ that is then set to $\vgal / c = - \sqrt{1 - 1 / \gamma^2}$, where $c$ denotes the speed of light.
With the Maxwell solver, stencils of order 16 are used in each direction, with the following numbers of ghost cells for the nodal and staggered or hybrid cases, respectively:
\begin{itemize}
\item nodal case: $16$ ghost cells in $(x,y)$ and $16$ ghost cells in $z$;
\item staggered or hybrid case: $8$ ghost cells in $(x,y)$ and $16$ ghost cells in $z$.
\end{itemize}
These choices of ghost cells are based on the measurements of the stencil extents along $x$, $y$ and $z$ shown in Figure~\ref{fig_stencil_lwfa} for this test case, with the curves obtained as prescribed in \eqref{eq_stencil_extent}.

More precisely, we measure the stencil of the leading coefficient $\theta_\ctxt^2 \, C$ in the update equations \eqref{eq_update_B_with_rho_galilean}-\eqref{eq_update_E_with_rho_galilean}.
Figure~\ref{fig_stencil_lwfa} shows that the number of ghost cells can be safely reduced in the transverse directions $(x,y)$ in the staggered or hybrid case, as compared with the nodal case. This is not the case in the longitudinal direction~$z$.
What makes the longitudinal direction $z$ special is the fact that it is the direction of the Galilean coordinate transformation.
The coefficient $\theta_\ctxt^2 \, C$ keeps memory of such coordinate transformation through $\theta_\ctxt^2 = \exp(i \, \vbgal \cdot [\kb]_\ctxt \, \dt)$. 

The time step $\dt$ satisfies $c \, \dt = \dx, \dy < \dz'$, where $\dz'$ denotes the cell size along $z$ in the boosted frame. More precisely, $\dx = \dy \approx 3.125 \, \micron$, $\dz \approx 0.078 \, \micron$, $\dz' = (1 + \beta) \, \gamma \, \dz \approx 4.677 \, \micron$ and $\dt \approx 10.42 \, \text{fs}$.

The choice of a Lorentz boosted frame of reference that travels at a speed close to the speed of light in the direction of the laser makes it possible to simulate the propagation of a laser with a wavelength of a fraction of a micron by using cells that span over a few microns: in the boosted frame of reference the laser beam is elongated by roughly $(1+\beta)\gamma$, while the plasma contracts by roughly $\gamma$. This results in a total speedup of the simulation by $(1+\beta)\gamma^2 \approx 1800$ with $\gamma=30$, compared to the same simulation using a laboratory frame of reference \citep{Vay2007}.

\begin{figure}[!ht]
\includegraphics[width=0.49\linewidth,trim={0 0 0 0},clip]{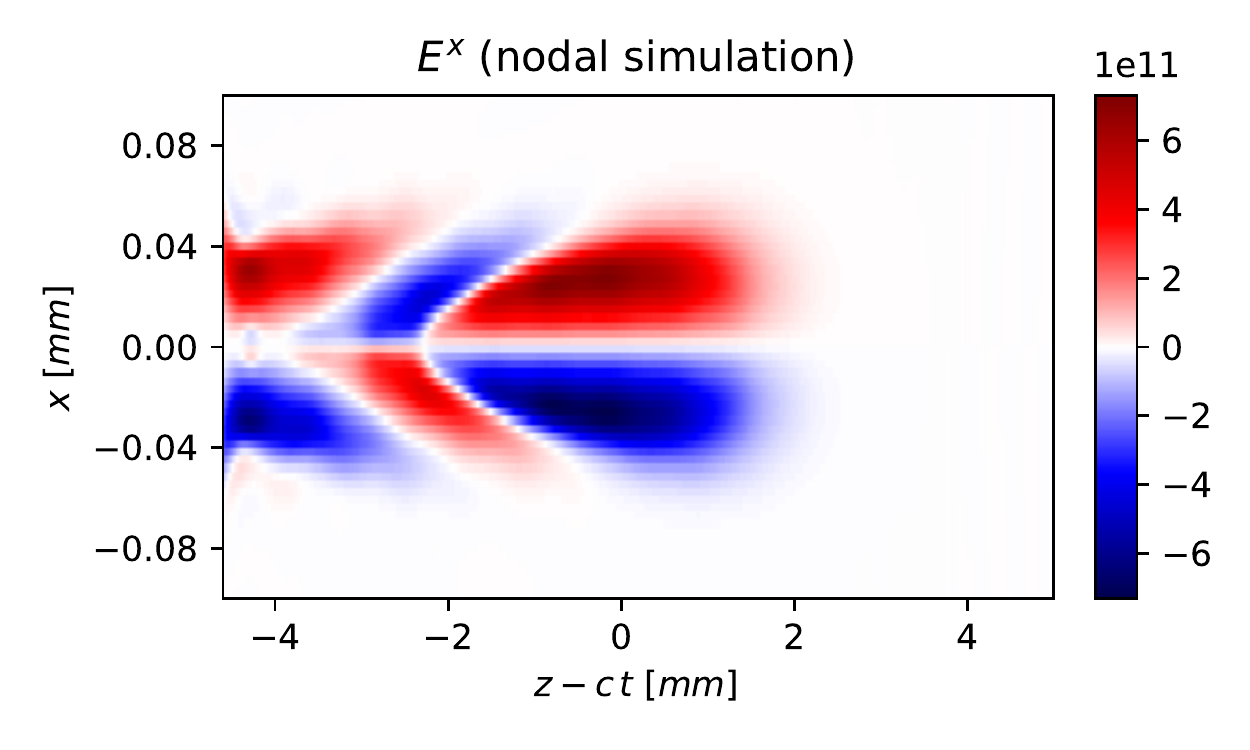}
\includegraphics[width=0.49\linewidth,trim={0 0 0 0},clip]{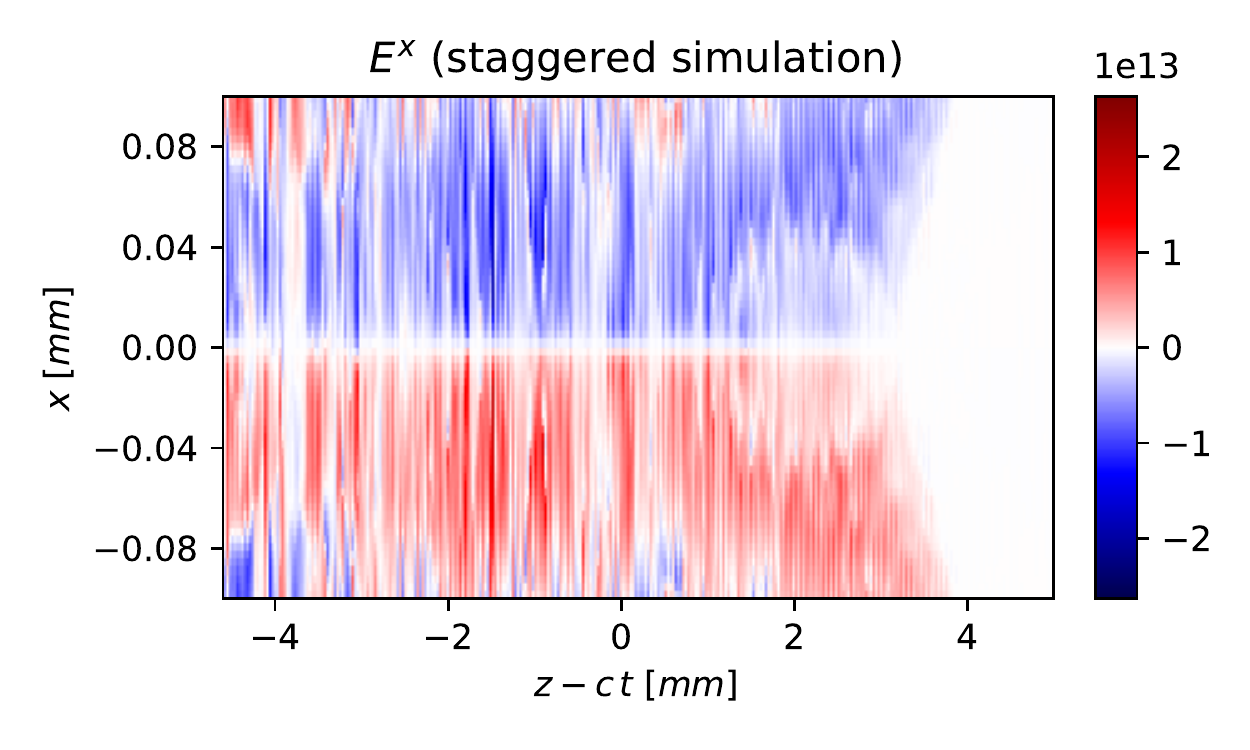} \\
\includegraphics[width=0.49\linewidth,trim={0 0 0 0},clip]{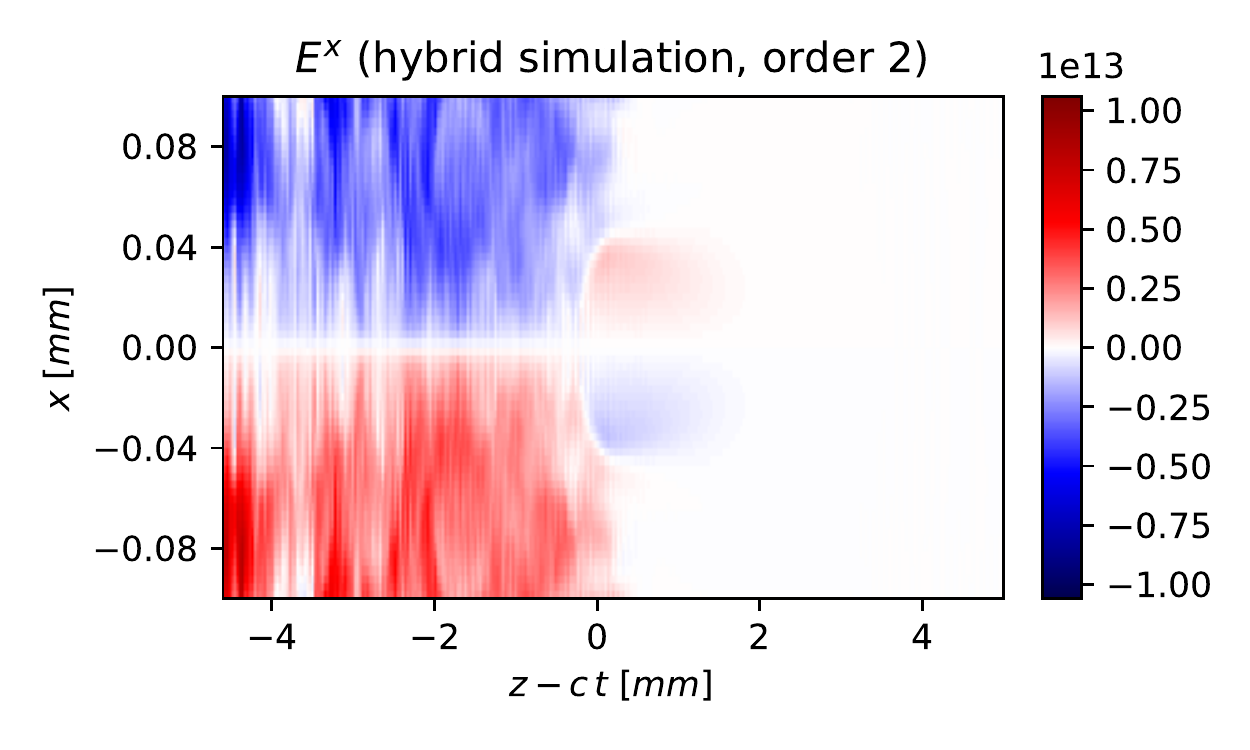}
\includegraphics[width=0.49\linewidth,trim={0 0 0 0},clip]{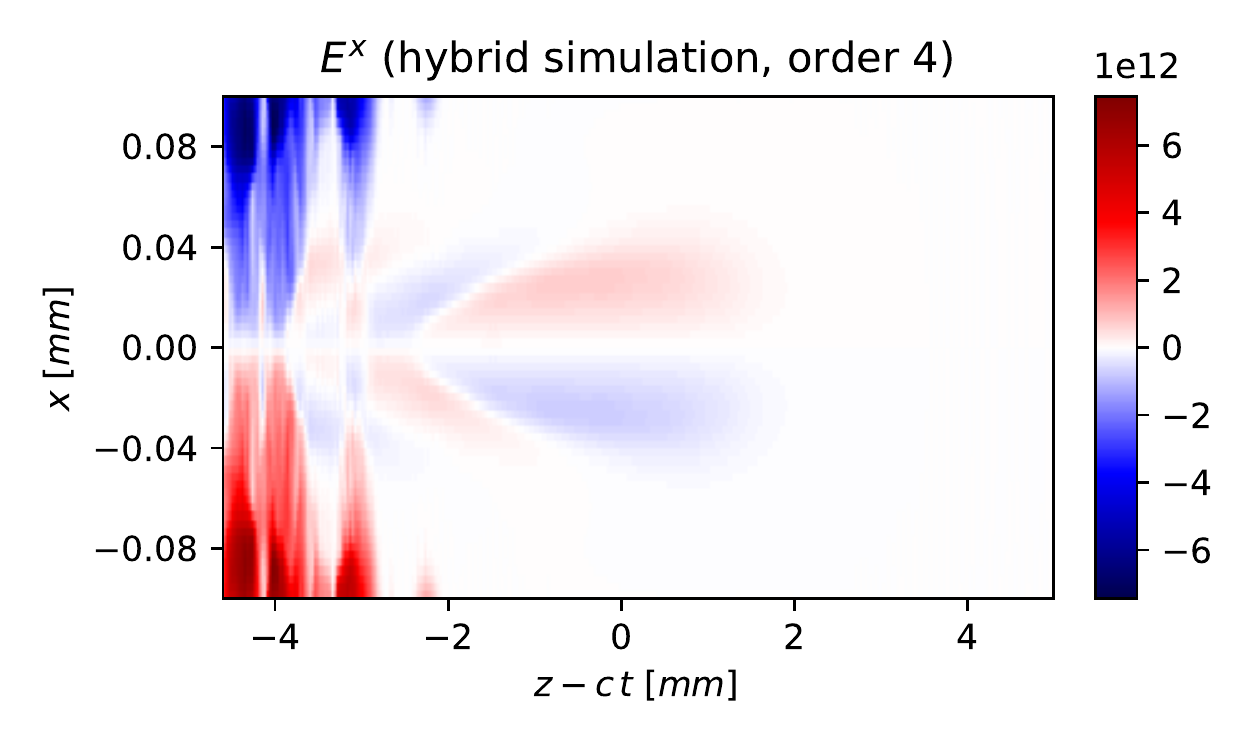} \\
\includegraphics[width=0.49\linewidth,trim={0 0 0 0},clip]{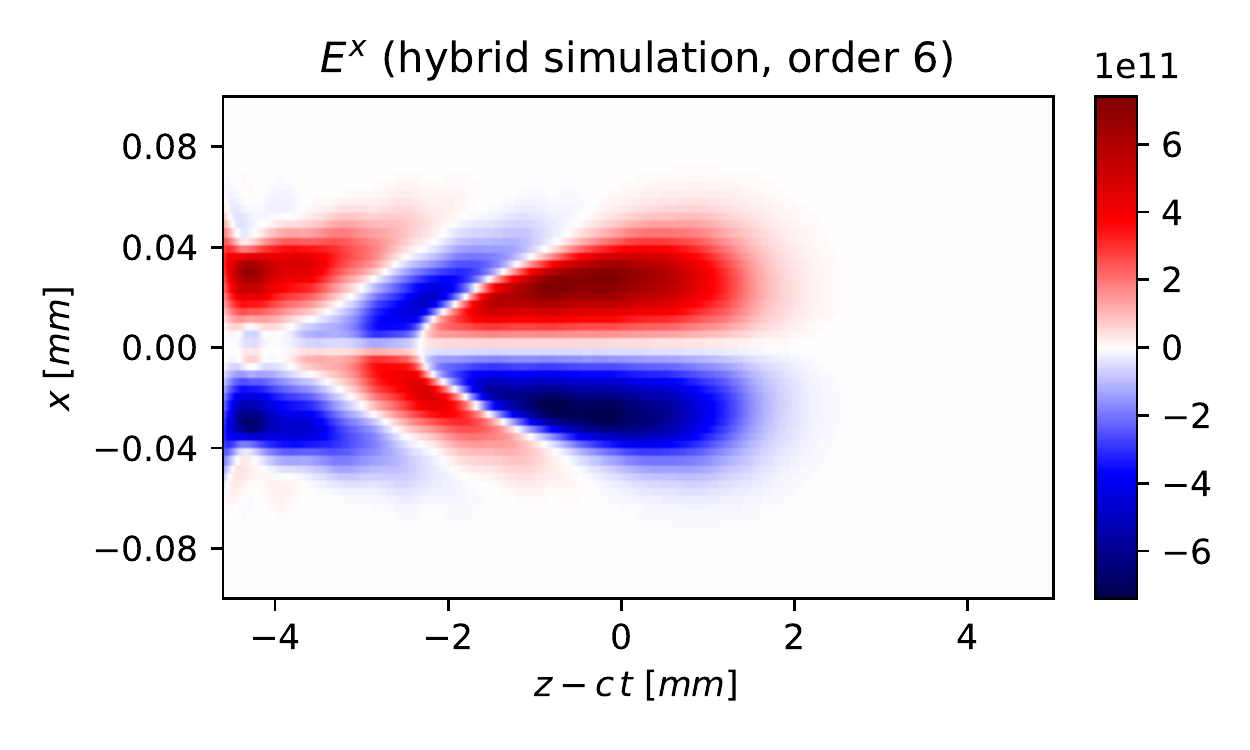}
\hspace{3pt}
\includegraphics[width=0.49\linewidth,trim={0 0 0 0},clip]{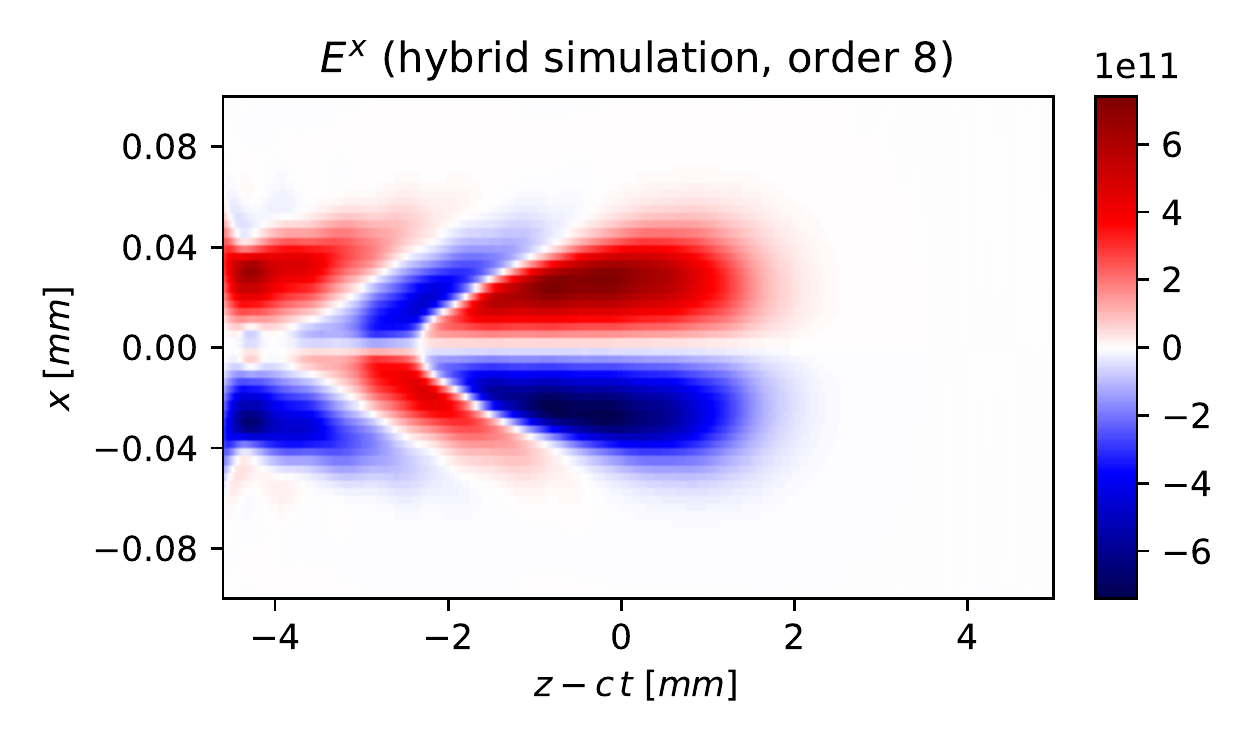}
\caption{\textbf{LWFA Simulations}. Plots of the component $E^x$ of the electric field, at $y=0$, after 1600 iterations, obtained with the standard Galilean PSATD algorithm with a nodal PIC cycle (first row, left column), a staggered PIC cycle (first row, right column), and the hybrid PIC cycle with finite-order centering of fields and currents of order $2m = 2,4,6,8$ in each direction (second to third row, both columns).}
\label{fig_gal_nodal_vs_hybrid}
\end{figure}

Figure~\ref{fig_gal_nodal_vs_hybrid} shows plots of the component $E^x$ of the electric field, at $y=0$, after 1600 iterations, for:
\begin{itemize}
    \item a fully nodal simulation (first row, left column);
    \item a fully staggered simulation (first row, right column);
    \item hybrid simulations with finite-order centering of fields and currents at order $2m=2,4,6,8$ in each direction (second to third row, both columns).
\end{itemize}
While the fully nodal simulation is stable, as expected based on previous work \cite{Lehe2016,Lehe2019}, 
the fully staggered simulation develops a strong numerical Cherenkov instability.
As anticipated in Section~\ref{sec_motivations}, 
the increased stability of the fully nodal case can be recovered with the new hybrid solver, provided that the order of the finite-centering operation is sufficiently high. The required order is as low as $6$ in the present case, based on the results of Figure~\ref{fig_gal_nodal_vs_hybrid}, enabling the hybrid scheme to reach the stability of the fully nodal scheme while using half the number of ghost cells in the transverse directions, thanks to the use of a staggered PSATD Maxwell's solver,
%
resulting in shorter runtimes and smaller computational costs overall.
In fact, the total runtime of the hybrid simulation at order 6 in Figure~\ref{fig_gal_nodal_vs_hybrid} (the lowest order that reproduces the nodal results correctly) is approximately $156 \, \stxt$ on 24 NVIDIA GPUs of the Summit supercomputer (thus, with 1~subdomain per GPU), while the total runtime of the nodal simulation is approximately $289 \, \stxt$, leading to a speed-up of approximately $1.9$.
Table~\ref{table_gal_runtimes} shows the total runtimes of all the simulations shown in Figure~\ref{fig_gal_nodal_vs_hybrid}.

\begin{table}[h]
\centering
\renewcommand{\arraystretch}{1.5}
\setlength{\tabcolsep}{10pt}
\begin{tabular}{c|c|c|c|c|c|c}
\textbf{Simulation} &
nodal &
staggered &
\makecell{hybrid, \\ order 2} &
\makecell{hybrid, \\ order 4} &
\makecell{hybrid, \\ order 6} &
\makecell{hybrid, \\ order 8} \\
\hline
\textbf{Runtime} &
$289 \, \stxt$ &
$157 \, \stxt$ &
$157 \, \stxt$ &
$155 \, \stxt$ &
$156 \, \stxt$ &
$157 \, \stxt$
\end{tabular}
\caption{Runtimes of the nodal, staggered and hybrid simulations shown in Figure~\ref{fig_gal_nodal_vs_hybrid}.}
\label{table_gal_runtimes}
\end{table}


\begin{figure}[!ht]
\centering
\includegraphics[width=0.5\linewidth,trim={0 0 0 0},clip]{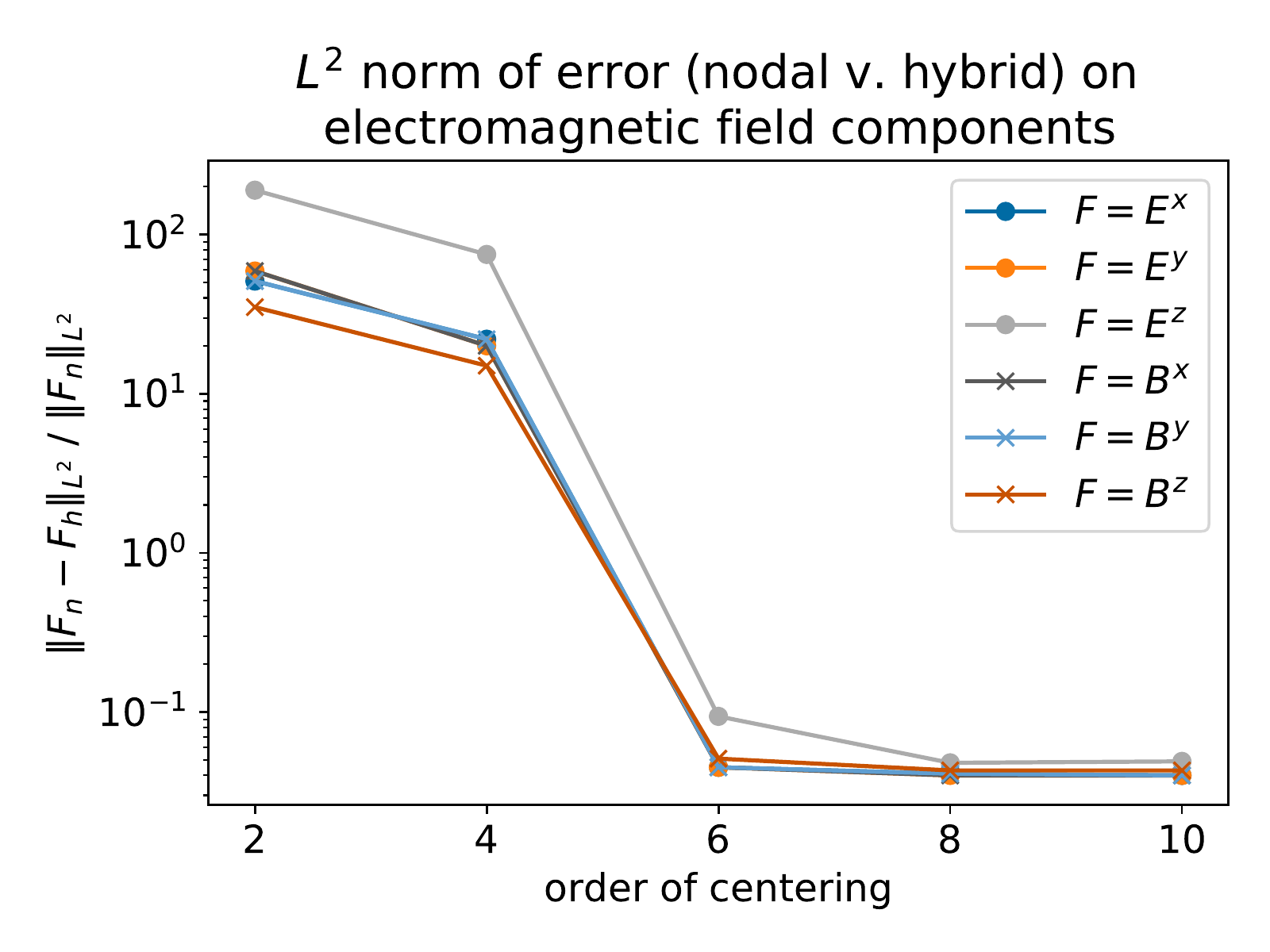}
\caption{\textbf{Convergence}. $L^2$ norm of error, computed as prescribed in~\eqref{L2_norm_error}, for the electromagnetic field components measured from the nodal and hybrid simulations shown in Figure~\ref{fig_gal_nodal_vs_hybrid} (with, in addition, the results from a hybrid simulation at order $10$, confirming that the results have indeed converged).}
\label{fig_L2_gal_interp}
\end{figure}

To give a more quantitative comparison between the nodal and hybrid results, the $L^2$ norm of error was computed for all electromagnetic field components.
More precisely, denoting by $F_\ntxt$ a given electromagnetic field component from a nodal simulation and by $F_{\htxt}$ the corresponding data from a hybrid simulation, the $L^2$ norm of error is given by
\begin{linenomath}
\begin{equation}
\label{L2_norm_error}
\frac{\Vert F_\ntxt - F_{\htxt} \Vert_{L^2}}{\Vert F_\ntxt \Vert_{L^2}} :=
\dfrac{\sqrt{\displaystyle\int \dd^3\xb \, \left[F_\ntxt(\xb) - F_{\htxt}(\xb)\right]^2}}{\sqrt{\displaystyle\int \dd^3\xb \, \left[F_\ntxt(\xb)\right]^2}} \,.
\end{equation}
\end{linenomath}
The errors measured from the nodal and hybrid simulations shown in Figure~\ref{fig_gal_nodal_vs_hybrid} are plotted in Figure~\ref{fig_L2_gal_interp} for $E^x$ as well as all other electromagnetic field components (not shown in Figure~\ref{fig_gal_nodal_vs_hybrid} for brevity).

The dependency of the $L^2$ norm of error of the field components with respect to the order of the centering confirms the dramatic improvement of stability at order 6, which is reinforced further at higher orders.

\subsection{Averaged Galilean PSATD Algorithm: Beam-Driven Plasma Wakefield Acceleration}
\label{sec_avg_test}
This section presents a test of the novel hybrid scheme with the averaged Galilean PSATD algorithm in a Lorentz-boosted frame \citep{Vay2007}, on the numerical simulation of particle beam-driven plasma wakefield acceleration (PWFA), where the electron plasma wave is created by a charged particle beam instead of a laser beam. 

The simulation is performed on a three-dimensional computational domain, parametrized by the Cartesian coordinates $(x,y,z) \in [-200 \, \micron, 200 \, \micron] \times [-200 \, \micron, 200 \, \micron] \times [-220 \, \micron, 10 \, \micron]$.

The plasma is made of electrons and hydrogen ions, injected in the simulation with 2 particles per cell in the transverse directions $(x,y)$ and 1 particle per cell in the longitudinal direction $z$. The plasma profile is constant, with both electron and ion densities equal to $n_0 = 10^{23} \, \mtxt^{-3}$. Moreover, both species are injected assuming zero momentum (cold plasma) in the laboratory frame.

The beam is composed of $10^6$ electrons with total charge $Q = -1 \, \text{nC}$. It is injected with a Gaussian distribution in space with means $\mu_x = \mu_y = 0$ and $\mu_z = -80 \, \micron$ and standard deviations \mbox{$\sigma_x = \sigma_y = 5 \, \micron$} and $\sigma_z = 20.1 \, \micron$, and it follows also a Gaussian momentum distribution (normalized with respect to $mc$) with means $\mu_{u_x} = \mu_{u_y} = 0$ and $\mu_{u_z} = 2000.0$ and standard deviations $\sigma_{u_x} = \sigma_{u_y} = 4.0$ and $\sigma_{u_z} = 20$.

\begin{figure}[!ht]
\includegraphics[width=0.49\linewidth,trim={0 0 0 0},clip]{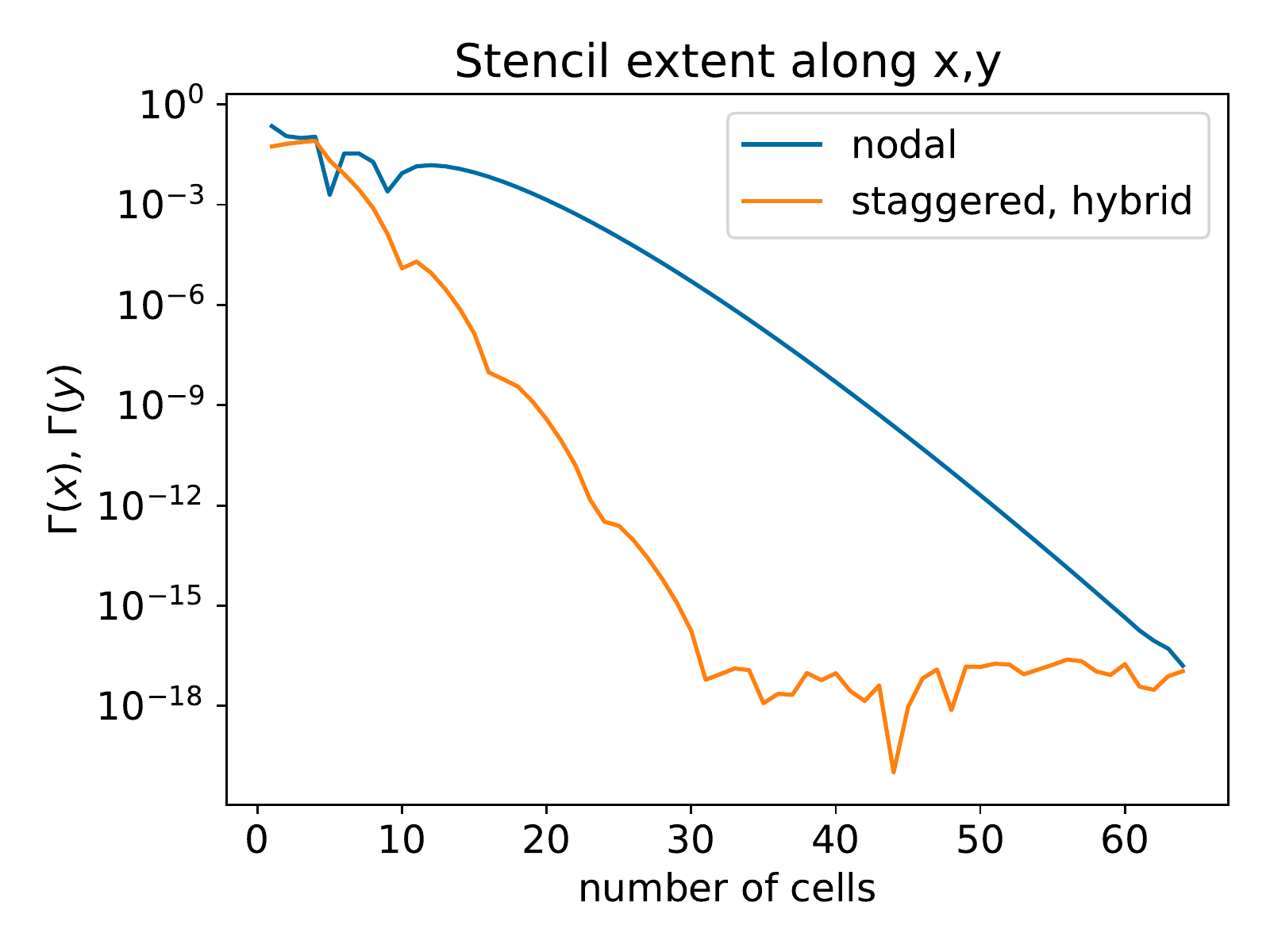}
\includegraphics[width=0.49\linewidth,trim={0 0 0 0},clip]{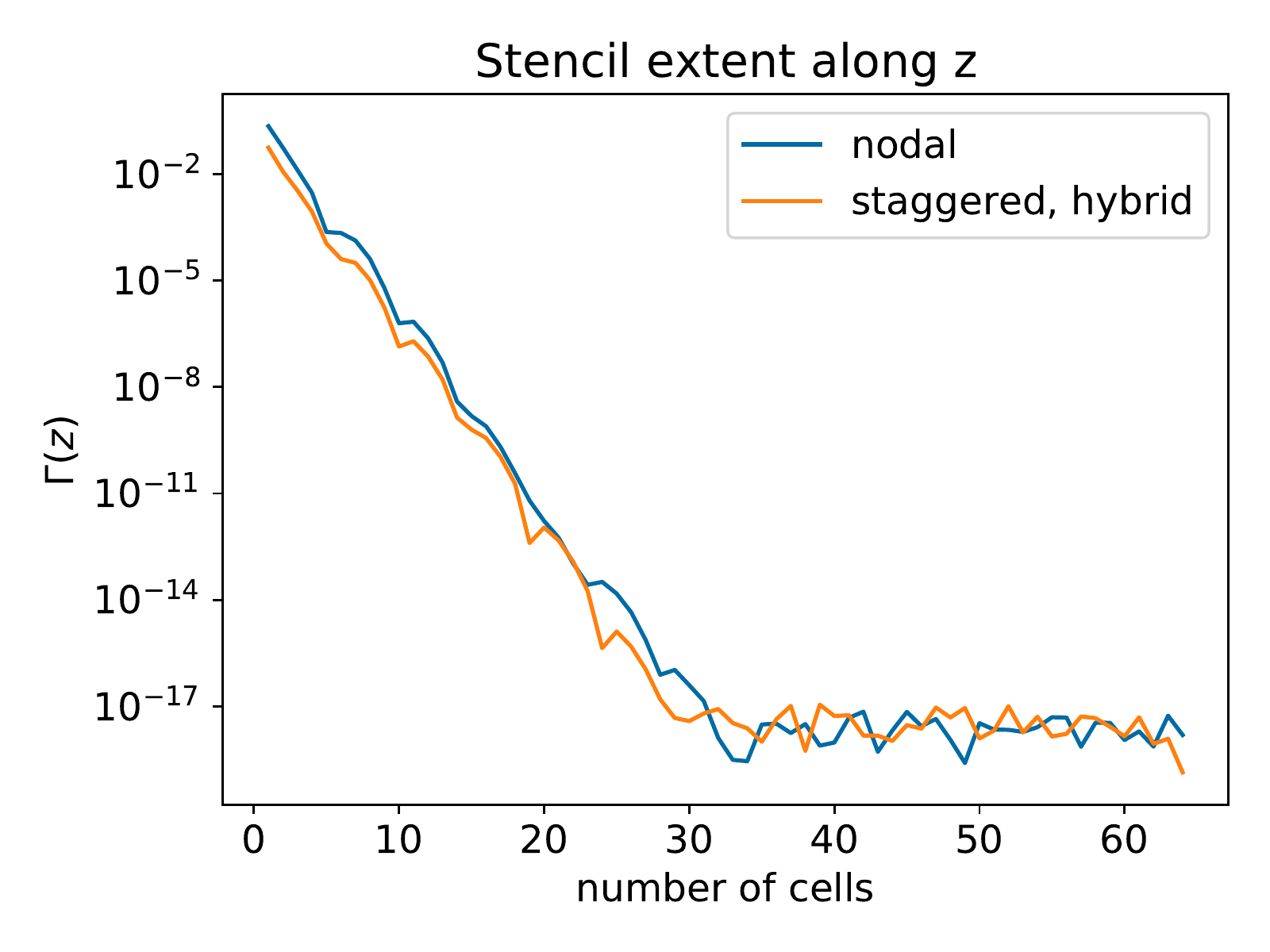}
\caption{\textbf{PSATD Stencils}. Stencil extent of the leading coefficient $\theta_\ctxt^2 \, C$ along $x$, $y$, $z$ for the 3D PWFA test case, computed as prescribed in \eqref{eq_stencil_extent}.}
\label{fig_stencil_pwfa}
\end{figure}

The computational domain is divided in $N_x \times N_y \times N_z = 256 \times 256 \times 256$ cells and decomposed in 24 subdomains, with $128 \times 128 \times 42$ cells per subdomain (except for one single subdomain with $43$ cells along $z$). Macro-particles use cubic splines as particle shape factors for current deposition and field gathering and the Vay scheme \citep{Vay2008} for the velocity and position updates.
To minimize the number of time steps and speed up the runtime, the simulation is performed using a Lorentz boosted frame of reference in the longitudinal direction, with a Lorentz factor $\gamma = 10$.
For stability, the simulation grid follows the plasma with longitudinal Galilean velocity $\vbgal = \vgal \, \hat\zb$ that is then set to $\vgal / c = - \sqrt{1 - 1 / \gamma^2}$, where $c$ denotes the speed of light.
With the Maxwell solver, stencils of order 16 are used in each direction, with the following numbers of ghost cells for the nodal and staggered or hybrid cases, respectively:
\begin{itemize}
\item nodal case: $26$ ghost cells in $(x,y)$ and $16$ ghost cells in $z$;
\item staggered or hybrid case: $8$ ghost cells in $(x,y)$ and $16$ ghost cells in $z$.
\end{itemize}
Here again, these choices of ghost cells are based on the measurements of the stencil extents along $x$, $y$ and $z$ shown in Figure~\ref{fig_stencil_pwfa} for this test case, with the curves obtained as prescribed in \eqref{eq_stencil_extent}.
The same observations made in this regard for the test case presented in Section~\ref{sec_gal_test} hold here.
We measure the stencil of the leading coefficient $\theta_\ctxt^2 \, C$ in the update equations \eqref{eq_update_B_with_rho_galilean}-\eqref{eq_update_E_with_rho_galilean}.
Figure~\ref{fig_stencil_pwfa} shows again that the number of ghost cells can be safely reduced in the transverse directions $(x,y)$ in the staggered or hybrid case, as compared with the nodal case. This is not the case in the longitudinal direction~$z$.
What makes the longitudinal direction $z$ special is the fact that it is the direction of the Galilean coordinate transformation.
The coefficient $\theta_\ctxt^2 \, C$ keeps memory of such coordinate transformation through $\theta_\ctxt^2 = \exp(i \, \vbgal \cdot [\kb]_\ctxt \, \dt)$.

The averaged Galilean PSATD scheme enables large time steps $\dt$ that satisfy $c \, \dt > \dx, \dy$ \cite{Shapoval2021}.
The simulation reported here used $c \, \dt = \dz'/4 \approx 2.87 \dx, \dy$, where $\dz'$ denotes again the cell size along $z$ in the boosted frame. More precisely, $\dx = \dy \approx 1.563 \, \micron$, $\dz \approx 0.8984 \, \micron$, $\dz' = (1 + \beta) \, \gamma \, \dz \approx 17.92 \, \micron$ and $\dt \approx 14.95 \, \text{fs}$.

\begin{figure}[!ht]
\includegraphics[width=0.49\linewidth,trim={0 0 0 0},clip]{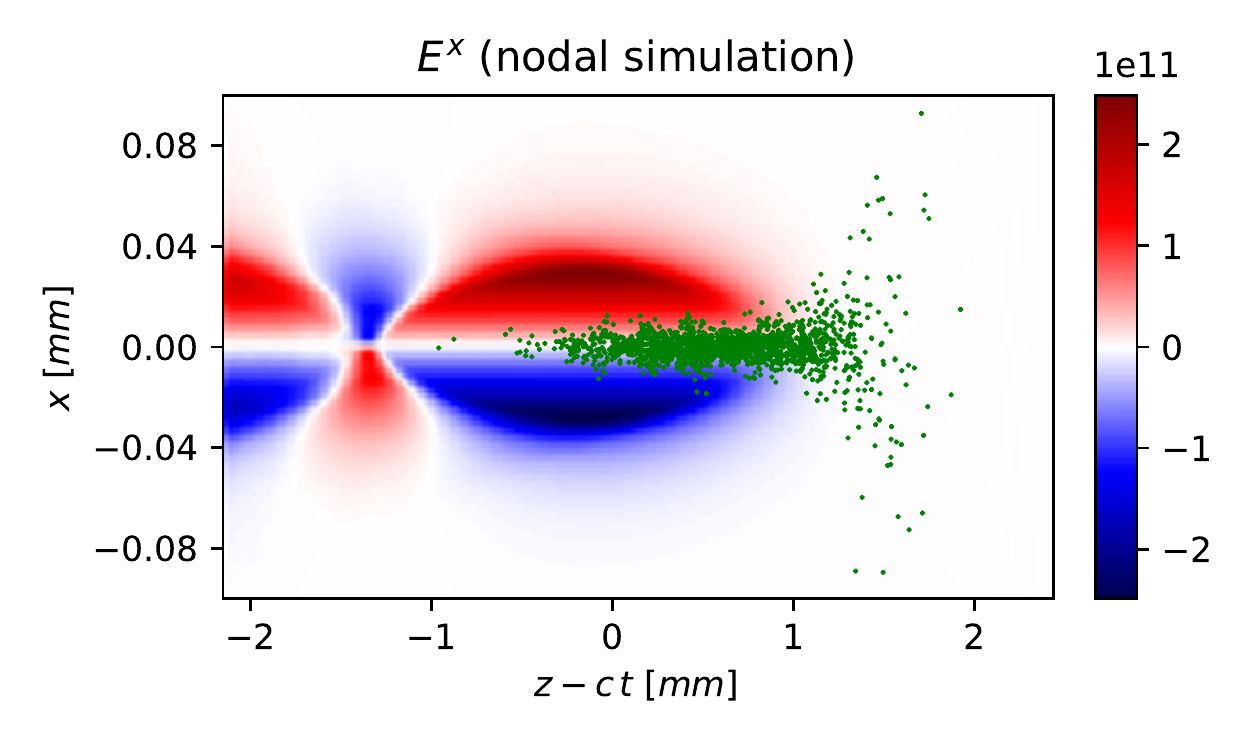}
\includegraphics[width=0.49\linewidth,trim={0 0 0 0},clip]{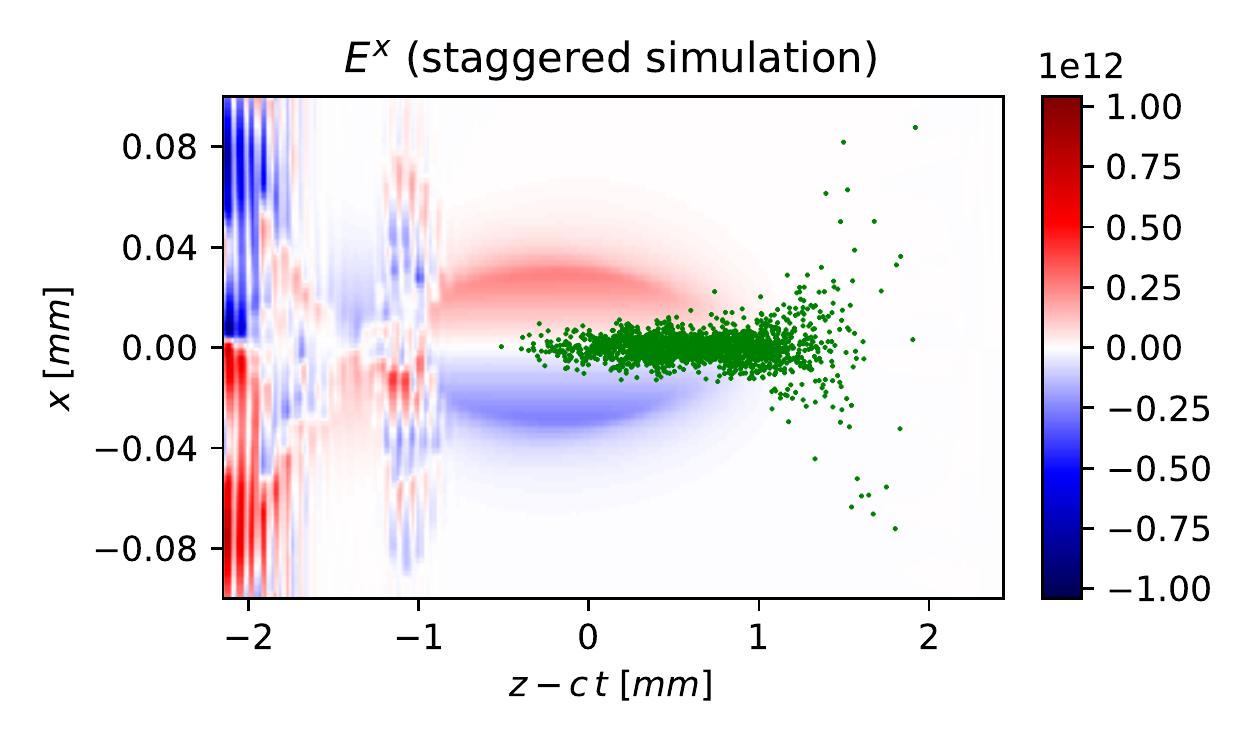} \\
\includegraphics[width=0.49\linewidth,trim={0 0 0 0},clip]{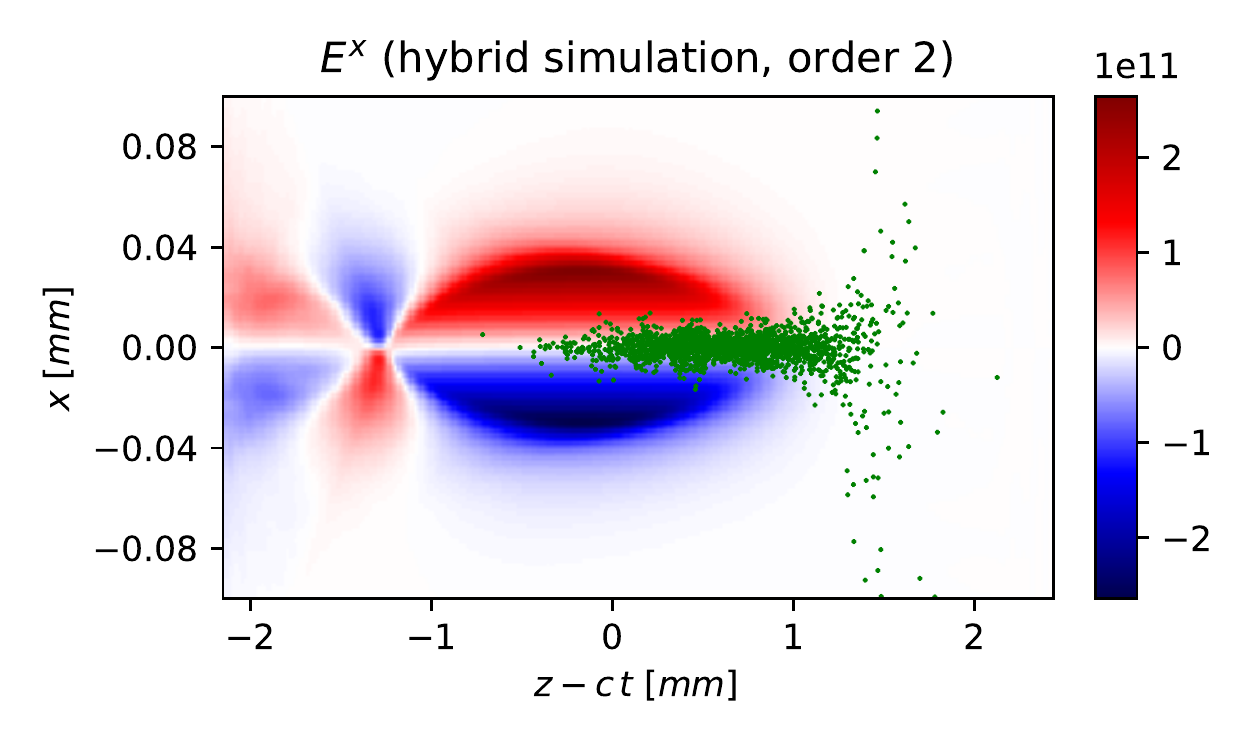}
\includegraphics[width=0.49\linewidth,trim={0 0 0 0},clip]{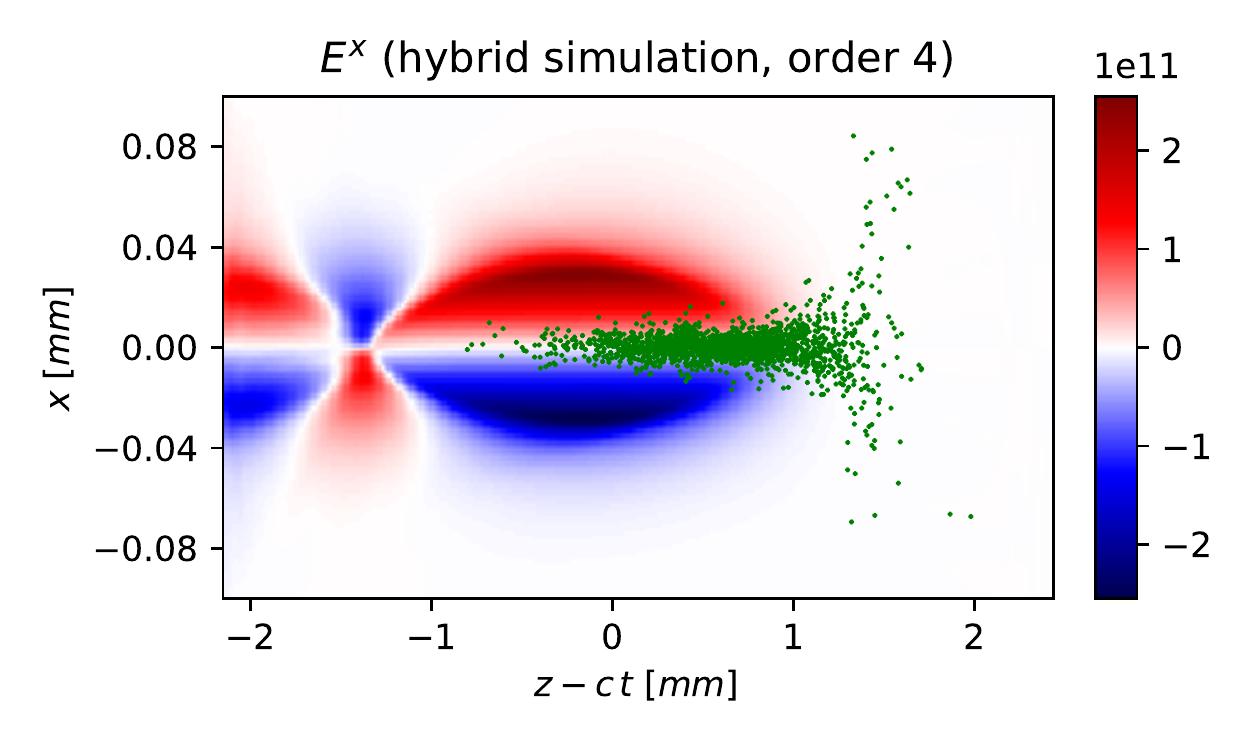} \\
\includegraphics[width=0.49\linewidth,trim={0 0 0 0},clip]{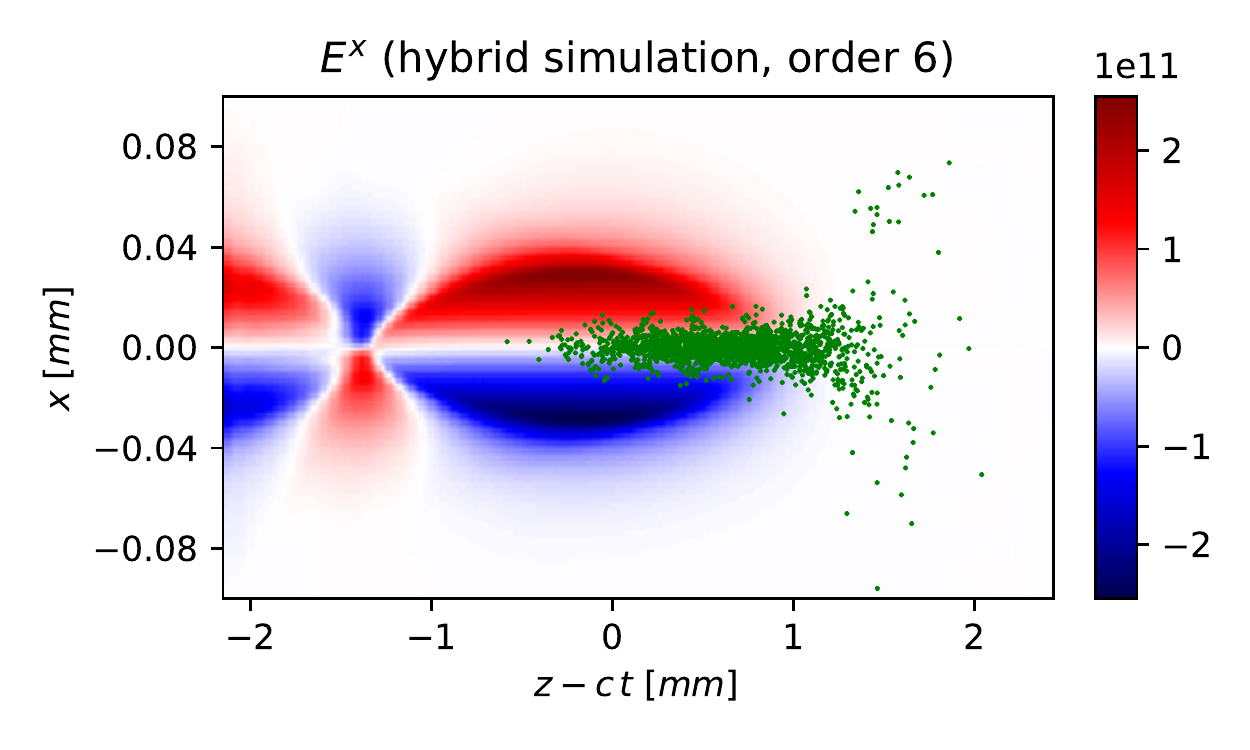}
\hspace{3pt}
\includegraphics[width=0.49\linewidth,trim={0 0 0 0},clip]{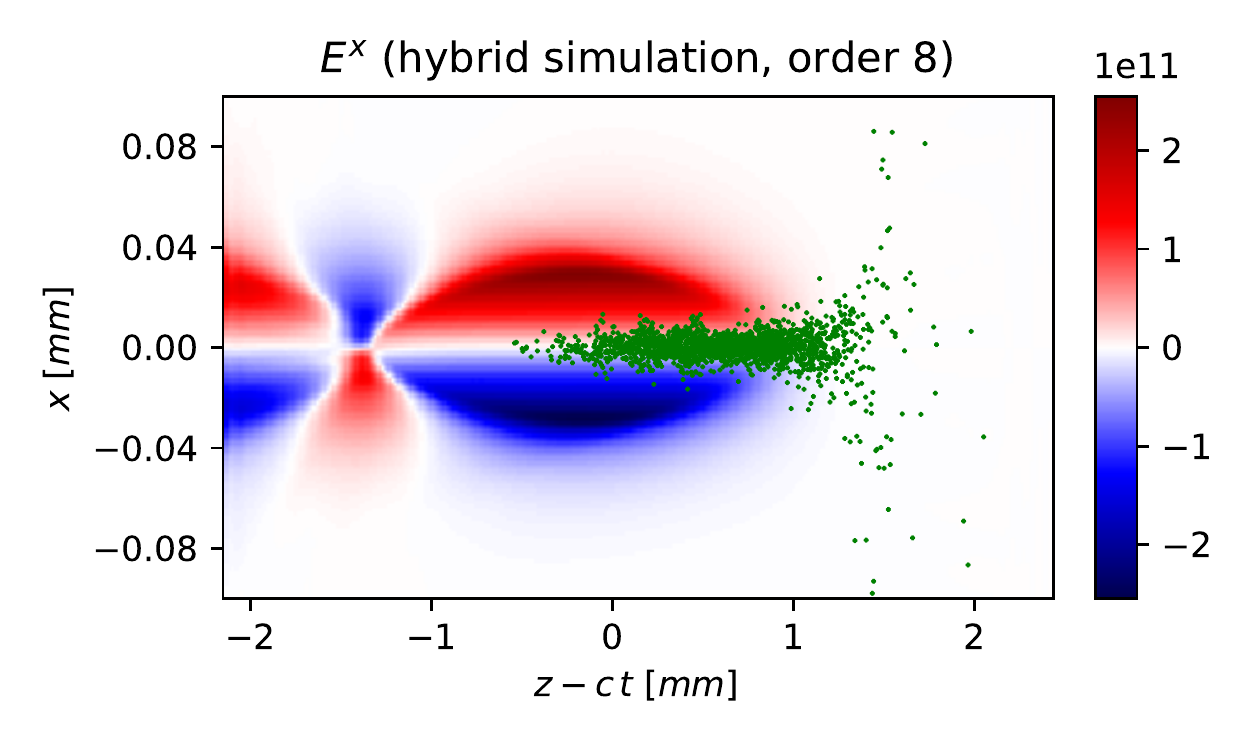}
\caption{\textbf{PWFA Simulations}. Plots of the component $E^x$ of the electric field, at $y=0$, together with a selection of beam particles, after 500 iterations, obtained with the averaged Galilean PSATD algorithm with a nodal PIC cycle (first row,  left column), a staggered PIC cycle (first row, right column), and the hybrid PIC cycle with finite-order centering of fields and currents of order $2m = 2,4,6,8$ in each direction (second to third row, both columns).}
\label{fig_avg_nodal_vs_hybrid}
\end{figure}

Figure \ref{fig_avg_nodal_vs_hybrid} shows plots of the component $E^x$ of the electric field, at $y=0$, together with a selection of beam particles, after 500 iterations, for:
\begin{itemize}
    \item a fully nodal simulation (first row, left column);
    \item a fully staggered simulation (first row, right column);
    \item hybrid simulations with finite-order centering of fields and currents at order $2m = 2,4,6,8$ in each direction (second to third row, both columns).
\end{itemize}

Here again, the fully staggered simulation develops a significant numerical Cherenkov instability (even though slightly weaker than in the LWFA example), which is not present in the nodal case.
In this case, the simulations with the new hybrid scheme are stable for an order of finite centering as low as 2 and reproduce accurately the shape of the field maps for a value as low as 4. 


Thanks to the greater locality of the finite-order stencil of the staggered Maxwell solver used in the hybrid approach, compared to a nodal Maxwell solver, it is possible to use fewer ghost cells between neighboring subdomains, which results in shorter runtimes and smaller computational costs overall.
In fact, the total runtime of the hybrid simulation at order 4 in Figure~\ref{fig_avg_nodal_vs_hybrid} (the lowest order that reproduces the nodal results correctly) is approximately $70 \, \stxt$ on 24 NVIDIA GPUs of the Summit supercomputer (thus, with 1 subdomain per GPU), while the total runtime of the nodal simulation is approximately $142 \, \stxt$, leading to a speed-up of approximately $2$.
Table~\ref{table_avg_runtimes} shows the total runtimes of all the simulations shown in Figure~\ref{fig_avg_nodal_vs_hybrid}.

\begin{table}[h]
\centering
\renewcommand{\arraystretch}{1.5}
\setlength{\tabcolsep}{10pt}
\begin{tabular}{c|c|c|c|c|c|c}
\textbf{Simulation} &
nodal &
staggered &
\makecell{hybrid, \\ order 2} &
\makecell{hybrid, \\ order 4} &
\makecell{hybrid, \\ order 6} &
\makecell{hybrid, \\ order 8} \\
\hline
\textbf{Runtime} &
$142 \, \stxt$ &
$69 \, \stxt$ &
$70 \, \stxt$ &
$70 \, \stxt$ &
$71 \, \stxt$ &
$72 \, \stxt$
\end{tabular}
\caption{Runtimes of the nodal, staggered and hybrid simulations shown in Figure~\ref{fig_avg_nodal_vs_hybrid}.}
\label{table_avg_runtimes}
\end{table}

\begin{figure}[!ht]
\centering
\includegraphics[width=0.5\linewidth,trim={0 0 0 0},clip]{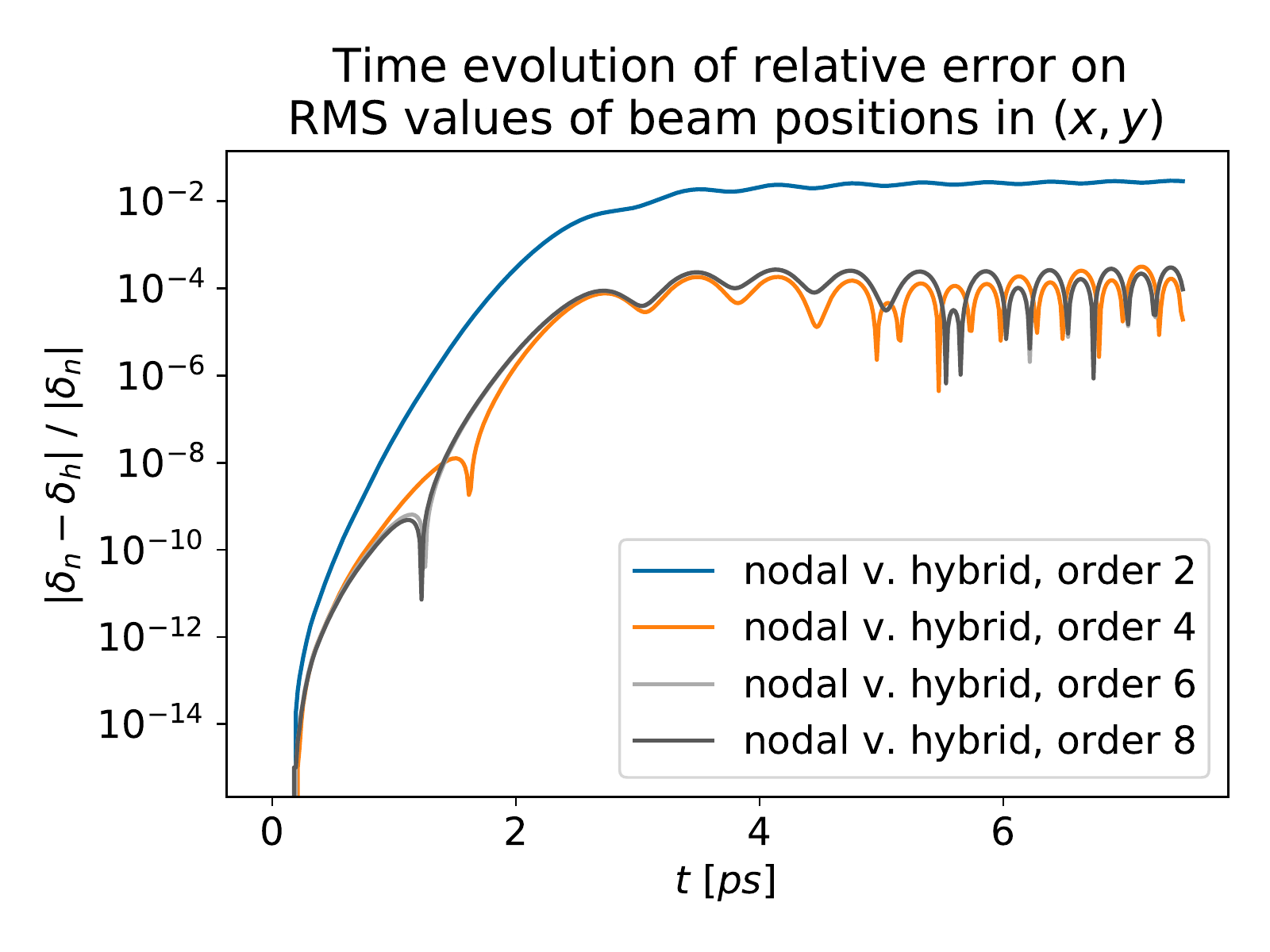}
\caption{\textbf{Convergence}. RMS errors computed as prescribed in~\eqref{eq_RMS_pwfa} for the beam particle positions measured from the nodal and hybrid simulations shown in Figure~\ref{fig_avg_nodal_vs_hybrid}.}
\label{fig_RMS_pwfa}
\end{figure}

Since this test case involves a particle beam, we also compare the root mean square (RMS) values of the beam particle positions in the transverse plane $(x,y)$ between the nodal and hybrid results. More precisely, we measure the averaged RMS quantity
\begin{linenomath}
\begin{equation}
\label{eq_RMS_pwfa}
\delta :=
\frac{1}{2} \sqrt{\dfrac{\sum_p w_p (x_p - \langle x \rangle)^2}{\sum_p w_p}} +
\frac{1}{2} \sqrt{\dfrac{\sum_p w_p (y_p - \langle y \rangle)^2}{\sum_p w_p}} \,,
\end{equation}
\end{linenomath}
and then compute the error $|\delta_\ntxt - \delta_\htxt| / |\delta_\ntxt|$, where we denote by $\delta_\ntxt$ and $\delta_\htxt$ the data from a nodal and hybrid simulation, respectively.
The time evolution of the errors measured from the nodal and hybrid simulations shown in Figure~\ref{fig_avg_nodal_vs_hybrid} are plotted in Figure~\ref{fig_RMS_pwfa}.

These confirm that the hybrid simulations, while stable with a finite centering of order 2, need at least order 4 to be accurate. 

\section{Conclusions}
\label{sec_conclusions}
A novel hybrid PSATD PIC scheme was proposed that combines the advantages of standard nodal and staggered PIC methods.
The novel hybrid scheme employs finite-order interpolation to combine the solution of Maxwell's equations on a staggered grid with the deposition of charges and currents on a nodal grid as well as the gathering of electromagnetic forces from a nodal grid.
The finite-order interpolation proposed to recenter quantities at the nodes or at staggered positions is based on the same coefficients originally introduced by Fornberg~\citep{Fornberg1990} for the high-order approximation of spatial derivatives.

The novel hybrid scheme retains the advantageous properties of staggered Maxwell's solvers (such as lower levels of numerical dispersion, more local stencils resulting in smaller ghost regions for the exchange of fields between parallel subdomains and thus shorter runtimes overall, and better stability at short wavelengths), and avoids at the same time numerical errors coming from low-order interpolation of grid quantities defined at different locations on the grid.
Different classes of PSATD equations (standard PSATD~\citep{Haber1973,Vay2013}, standard Galilean PSATD~\citep{Lehe2016,Kirchen2020}, and averaged Galilean PSATD~\citep{Shapoval2021}) were adapted to the novel hybrid scheme and numerical tests were performed in a variety of physical scenarios, ranging from the modeling of electron-positron pair creation in vacuum to the simulation of laser-driven and particle beam-driven plasma wakefield acceleration.
Further exploration of the properties of the new scheme with regard to, for example, charge conservation, momentum conservation or energy conservation, is planned and will be reported in future publications. 

Though presented here only in the context of PSATD methods, the novel hybrid scheme can be also adapted in a straightforward way to more common FDTD methods, upon which many electromagnetic PIC simulation codes are based.
Therefore, the novel hybrid scheme has the potential to become a useful numerical tool for the simulation of the large variety of physical systems that can be modeled by means of PIC codes, including fusion plasmas, astrophysical plasmas, plasma wakefield particle accelerators, and secondary photon sources driven by ultra-intense lasers. 



\section*{Acknowledgments}
This research used the open-source particle-in-cell code WarpX (\url{https://github.com/ECP-WarpX/WarpX}).
We acknowledge all WarpX contributors.
This research was supported by the Exascale Computing Project (17-SC-20-SC), a collaborative effort of the U.S. Department of Energy Office of Science and the National Nuclear Security Administration.
This research was performed in part under the auspices of the U.S. Department of Energy by Lawrence Berkeley National Laboratory under Contract DE-AC02-05CH11231.
This research was supported by the French National Research Agency (ANR) T-ERC program (Grant No. ANR-18-ERC2-0002).
We also acknowledge the financial support of the Cross-Disciplinary Program on Numerical Simulation of CEA, the French Alternative Energies and Atomic Energy Commission.
This project has received funding from the European Union's Horizon 2020 research and innovation program under grant agreement No. 871072.
This research used resources of the Oak Ridge Leadership Computing Facility, which is a DOE Office of Science User Facility supported under Contract DE-AC05-00OR22725.
An award of computer time (Plasm-In-Silico) was provided by the Innovative and Novel Computational Impact on Theory and Experiment (INCITE) program.
The data that support the findings of this study are available from the corresponding author upon reasonable request.

\appendix

\section{Standard Galilean PSATD: Equations on Staggered Grids}
\label{appendix_1}
This section provides a detailed derivation of the equations for the update of the electromagnetic fields in Fourier space, valid on staggered grids, for the Galilean PSATD algorithm, which had been derived only for nodal grids so far \citep{Lehe2016,Kirchen2020}.

We first recall that the electromagnetic fields $\Eb$ and $\Bb$ are evaluated on a three-dimensional Yee~grid \citep{Yee1966,Cole1997,Cole2002} as follows (please also refer to the schematic illustrated in Section~\ref{sec_motivations}):
\begin{itemize}
\item $E^x$ is evaluated at the \emph{cell nodes} $y_j = j \dy$ and $z_k = k \dz$ along $y$ and $z$ and at the \mbox{\emph{cell centers}} $x_{i+\frac 12} = i \dx + \dx / 2$ along $x$, and it is thus indexed as $E^x_{i+\frac 12,j,k}$;
\item $E^y$ is evaluated at the \emph{cell nodes} $x_i = i \dx$ and $z_k = k \dz$ along $x$ and $z$ and at the \emph{cell centers} $y_{j+\frac 12} = j \dy + \dy / 2$ along $y$, and it is thus indexed as $E^y_{i,j+\frac 12,k}$;
\item $E^z$ is evaluated at the \emph{cell nodes} $x_i = i \dx$ and $y_j = j \dy$ along $x$ and $y$ and at the \emph{cell centers} $z_{k+\frac 12} = k \dz + \dz / 2$ along $z$, and it is thus indexed as $E^z_{i,j,k+\frac 12}$;
\item $B^x$ is evaluated at the \emph{cell nodes} $x_i = i \dx$ along $x$ and at the \emph{cell centers} $y_{j+\frac 12} = j \dy + \dy / 2$ and $z_{k+\frac 12} = k \dz + \dz / 2$ along $y$ and $z$, and it is thus indexed as $B^x_{i,j+\frac 12,k+\frac 12}$;
\item $B^y$ is evaluated at the \emph{cell nodes} $y_j = j \dy$ along $y$ and at the \emph{cell centers} $x_{i+\frac 12} = i \dx + \dx / 2$ and $z_{k+\frac 12} = k \dz + \dz / 2$ along $x$ and $z$, and it is thus indexed as $B^y_{i+\frac 12,j,k+\frac 12}$;
\item $B^z$ is evaluated at the \emph{cell nodes} $z_k = k \dz$ along $z$ and at the \emph{cell centers} $x_{i+\frac 12} = i \dx + \dx / 2$ and $y_{j+\frac 12} = j \dy + \dy / 2$ along $x$ and $y$, and it is thus indexed as $B^z_{i+\frac 12,j+\frac 12,k}$.
\end{itemize}
Moreover, the current density $\Jb$ is evaluated on the same grid as the electric field $\Eb$, while the charge density $\rho$ is evaluated on a fully nodal grid.



\subsection{Finite Differences in Fourier Space}
Let us recall first how to express a one-dimensional finite difference in Fourier space in the \emph{continuum case}.
Given a function $f: \Lambda \ni x \mapsto f(x) \in \Rbb$, we define its Fourier transform \mbox{$\fhat: \Rbb \ni k \mapsto \fhat(k) \in \Cbb$} as
\begin{linenomath}
\begin{equation}
\label{eq_Fourier_continuum}
\fhat(k) = \int \dd x \, f(x) \, e^{-i k x} \,.
\end{equation}
\end{linenomath}
The expression of the finite difference $f(x + a) - f(x - b)$ in Fourier space, with $a, b \in \Rbb$, can be obtained by multiplying the finite difference by $e^{-i k x}$ and integrating over $x$, which yields
\begin{linenomath}
\begin{equation}
\begin{split}
\Fcal \left[f(x + a) - f(x - b)\right] (k) & =
2 \, i \, e^{i k (a-b)/2} \, \sin\left(k \, \frac{a+b}{2}\right) \, \fhat(k) \,,
\end{split}
\end{equation}
\end{linenomath}
where $\Fcal[\cdot](k)$ denotes the Fourier transform of the expression in brackets as a function of $k$.

\subsubsection{Nodal Fields}
We first consider a function $f$ evaluated at $N$ \emph{cell nodes} $x_j = j \dx$ of a one-dimensional periodic grid and thus indexed as $f_j$. We define the Fourier transform of the sequence $\{f_j\}$ as the sequence $\{\fhat_k\}$, where $\fhat_k$ reads
\begin{linenomath}
\begin{equation}
\label{Fourier_discrete}
\fhat_k = \sum_{j=0}^{N-1} f_j \, e^{-i k x_j} \,.
\end{equation}
\end{linenomath}

The periodicity of the grid implies that $f_\ell = f_{\ell-N}$ for any $\ell > N-1$ and that $f_\ell = f_{\ell+N}$ for any $\ell < 0$. As a consequence, the index $j$ in \eqref{Fourier_discrete} must be such that the sum takes into account all $N$ values in the sequence $\{f_j\}$, possibly by periodicity, and can be shifted arbitrarily. In other words, it behaves as the ``mute'' integration variable $x$ in \eqref{eq_Fourier_continuum}.

Let us consider first the \emph{centered finite difference} $f_{j+n} - f_{j-n}$, which for $n > 0$ results in an approximation of the derivative of $f$ at the cell node $x_j$:
\begin{center}
\begin{tikzpicture}[scale=1.]
\draw (-1,1)--(7,1);
\draw (-1,0)--(7,0);
\foreach \x in {-1,...,7} \node at (\x,1) {$\vert$};
\foreach \x in {-1,...,7} \node at (\x,0) {$\vert$};
\node at (0,1+0.5) {$j-n$};
\node at (6,1+0.5) {$j+n$};
\node at (3,0-0.5) {$j$};
\draw[->-=.5] (0,1)--(3,0);
\draw[->-=.5] (6,1)--(3,0);
%
\foreach \x in {-1,...,7} \node at (\x,1) {\color{red}{\small\textbullet}};
\foreach \x in {-1,...,7} \node at (\x,0) {\color{red}{\small\textbullet}};
\end{tikzpicture}
\end{center}

Its expression in Fourier space can be obtained by multiplying the finite difference by $e^{-i k x_j}$ and summing over $j$, which after some algebra yields
\begin{linenomath}
\begin{equation}
\label{centered_finite_difference_Fourier_nodal}
\begin{split}
\sum_{j=0}^{N-1} f_{j+n} \, e^{-i k x_j} & - \sum_{j=0}^{N-1} f_{j-n} \, e^{-i k x_j} 
= 2 \, i \sin(k \, n \, \dx) \, \fhat_k \,.
\end{split}
\end{equation}
\end{linenomath}

Let us now consider the \emph{staggered finite difference} $f_{j+\frac 12+n-\frac 12} - f_{j+\frac 12-n+\frac 12}$, which for $n > 0$ results in an approximation of the derivative of $f$ at the cell center $x_{j+\frac 12}$:
\begin{center}
\begin{tikzpicture}[scale=1.]
\draw (0,1)--(7,1);
\draw (0,0)--(7,0);
\foreach \x in {0,...,7} \node at (\x,1) {$\vert$};
\foreach \x in {0,...,7} \node at (\x,0) {$\vert$};
\node at (1,1+0.5) {$j+\frac 12-n+\frac 12$};
\node at (6,1+0.5) {$j+\frac 12+n-\frac 12$};
\node at (3+0.5,0-0.5) {$j+\frac 12$};
\draw[->-=.5] (1,1)--(3+0.5,0);
\draw[->-=.5] (6,1)--(3+0.5,0);
%
\foreach \x in {0,...,7} \node at (\x,1) {\color{red}{\small\textbullet}};
\foreach \x in {0,...,6} \node at (\x+0.5,0) {\color{blue}{x}};
\end{tikzpicture}
\end{center}

Its expression in Fourier space can be obtained by multiplying the finite difference by $e^{-i k x_j}$ and summing over $j$, which after some algebra yields
\begin{linenomath}
\begin{equation}
\label{staggered_finite_difference_Fourier_nodal}
\begin{split}
\sum_{j=0}^{N-1} f_{j+\frac 12+n-\frac 12} \, e^{-i k x_j} - \sum_{j=0}^{N-1} f_{j+\frac 12-n+\frac 12} \, e^{-i k x_j} 
& = e^{i k \dx / 2} \, 2 \, i \sin(k \, (n-1/2) \, \dx) \, \fhat_k \,.
\end{split}
\end{equation}
\end{linenomath}

Note that the shift factor $e^{i k \dx / 2}$ in \eqref{staggered_finite_difference_Fourier_nodal} is the mathematical consequence of the fact that $f$ is evaluated at the cell nodes but we are looking for an approximation of its derivative at a cell center.

\subsubsection{Cell-centered Fields}
We now consider a function $f$ evaluated at $N$ \emph{cell centers} \mbox{$x_{j+\frac 12} = j \dx + \dx / 2$} of a one-dimensional periodic grid and thus indexed as $f_{j+\frac 12}$.
We define the Fourier transform of the sequence $\{f_{j+\frac 12}\}$ as the sequence $\{\fhat_k\}$, where $\fhat_k$ reads
\begin{linenomath}
\begin{equation}
\label{Fourier_discrete_staggered}
\fhat_k = \sum_{j=0}^{N-1} f_{j+\frac 12} \, e^{-i k x_{j+\frac 12}} \,.
\end{equation}
\end{linenomath}

The \emph{centered finite difference} $f_{j+\frac 12+n} - f_{j+\frac 12-n}$, which for $n > 0$ results in an approximation of the derivative of $f$ at the cell center $x_{j+\frac 12}$, can be expressed in Fourier space as the corresponding nodal result \eqref{centered_finite_difference_Fourier_nodal}.

Similarly, the \emph{staggered finite difference} $f_{j+n-\frac 12} - f_{j-n+\frac 12}$, which for $n > 0$ results in an approximation of the derivative of $f$ at the cell node $x_j$, can be expressed in Fourier space as the corresponding nodal result \eqref{staggered_finite_difference_Fourier_nodal}, but with an inverse shift factor $e^{-i k \dx / 2}$.

\subsubsection{Summary}
Denoting by $\alpha_{m,n}^\ctxt$ and $\alpha_{m,n}^\stxt$ the centered and staggered Fornberg coefficients \citep{Fornberg1990}
\begin{linenomath}
\begin{subequations}
\begin{align}
& \alpha_{m,n}^\ctxt := (-1)^{n+1} \frac{2 (m!)^2}{(m-n)! \, (m+n)!} \,, \\[5pt]
& \alpha_{m,n}^\stxt := (-1)^{n+1} \left[\frac{(2m)!}{2^{2m} m!}\right]^2 \frac{4}{(2n-1) (m-n)! \, (m+n-1)!} \,,
\end{align}
\end{subequations}
\end{linenomath}
(introduced in Section~\ref{sec_algo}), a finite-order \emph{centered finite-difference} approximation of the \emph{derivative of a nodal field at a cell node} is expressed in Fourier space by means of~\eqref{centered_finite_difference_Fourier_nodal} and reads
\begin{linenomath}
\begin{equation}
\label{centered_finite_difference_nodal}
\Fcal \left[
\sum_{n=1}^{m} \alpha_{m,n}^\ctxt
\frac{f_{j+n} - f_{j-n}}{2 \, n \, \dx}
\right](k)
= i \left(\sum_{n=1}^{m} \alpha_{m,n}^\ctxt
\frac{\sin(k \, n \, \dx)}{n \, \dx}\right) \fhat_k
=: i \, [k]_\ctxt \, \fhat_k \,.
\end{equation}
\end{linenomath}

A finite-order \emph{staggered finite-difference} approximation of the \emph{derivative of a nodal field at a cell center} is expressed in Fourier space by means of \eqref{staggered_finite_difference_Fourier_nodal} and reads
\begin{linenomath}
\begin{equation}
\label{staggered_finite_difference_nodal}
\Fcal \left[
\sum_{n=1}^{m} \alpha_{m,n}^\stxt
\frac{f_{j+\frac 12+n-\frac 12} - f_{j+\frac 12-n+\frac 12}}{2 \, (n-1/2) \, \dx}
\right](k)
= i \, e^{i k \dx / 2} \left(\sum_{n=1}^{m} \alpha_{m,n}^\stxt
\frac{\sin(k \, (n-1/2) \, \dx)}{(n-1/2) \, \dx}\right) \fhat_k \,. 
\end{equation}
\end{linenomath}

Similarly, a finite-order \emph{centered finite-difference} approximation of the \emph{derivative of a cell-centered field at a cell center} is expressed in Fourier space by means of \eqref{centered_finite_difference_nodal},
and a finite-order \emph{staggered finite-difference} approximation of the \emph{derivative of a cell-centered field at a cell node} is expressed in Fourier space by means of \eqref{staggered_finite_difference_nodal}, but with an inverse shift factor $e^{-i k \dx / 2}$.
Finally, for later convenience, we introduce also the staggered modified wave numbers
\begin{linenomath}
\begin{equation}
[k]_\stxt := \sum_{n=1}^{m} \alpha_{m,n}^\stxt \frac{\sin(k \, (n-1/2) \, \dx)}{(n-1/2) \, \dx} \,.
\end{equation}
\end{linenomath}

\subsection{Faraday's Law}
In the Galilean coordinates $\xb = \xb' - \vbgal t$, Faraday's law reads
\begin{linenomath}
\begin{equation}
\label{eq_Faraday_app}
\left(\frac{\de}{\de t} - \vbgal \cdot \Nabla\right) \Bb = - \Nabla \times \Eb \,,
\end{equation}
\end{linenomath}
where $\Nabla$ denotes spatial derivatives with respect to the Galilean coordinates $\xb$ on the spatial grids where $\Eb$ and $\Bb$ are defined, respectively.

The component of \eqref{eq_Faraday_app} along $x$ reads
\begin{linenomath}
\begin{equation}
\label{eq_Faraday_x}
\frac{\de B^x}{\de t} - \vgal^x \frac{\de B^x}{\de x} - \vgal^y \frac{\de B^x}{\de y} - \vgal^z \frac{\de B^x}{\de z} = - \left(\frac{\de E^z}{\de y} - \frac{\de E^y}{\de z}\right) \,.
\end{equation}
\end{linenomath}

At finite order $2m$, \eqref{eq_Faraday_x} reads
\begin{linenomath}
\begin{equation}
\label{eq_Faraday_x_ijk}
\begin{split}
\frac{\de B^x_{i,j+\frac 12,k+\frac 12}}{\de t}
& - \vgal^x \sum_{n=1}^{m} \alpha_{m,n}^\ctxt \frac{B^x_{i+n,j+\frac 12,k+\frac 12} - B^x_{i-n,j+\frac 12,k+\frac 12}}{2n \, \Delta x} \\[5pt]
& - \vgal^y \sum_{n=1}^{m} \alpha_{m,n}^\ctxt \frac{B^x_{i,j+\frac 12+n,k+\frac 12} - B^x_{i,j+\frac 12-n,k+\frac 12}}{2n \, \Delta y} \\[5pt]
& - \vgal^z \sum_{n=1}^{m} \alpha_{m,n}^\ctxt \frac{B^x_{i,j+\frac 12,k+\frac 12+n} - B^x_{i,j+\frac 12,k+\frac 12-n}}{2n \, \Delta z} \\[5pt]
& = - \left(\sum_{n=1}^{m} \alpha_{m,n}^\stxt \frac{E^z_{i,j+\frac 12+n-\frac 12,k+\frac 12} - E^z_{i,j+\frac 12-n+\frac 12,k+\frac 12}}{2 \, (n-1/2) \, \Delta y} \right. \\[5pt]
& \qquad \left.
- \sum_{n=1}^{m} \alpha_{m,n}^\stxt \frac{E^y_{i,j+\frac 12,k+\frac 12+n-\frac 12} - E^y_{i,j+1/2,k+\frac 12-n+\frac 12}}{2 \, (n-1/2) \, \Delta z}\right) \,.
\end{split}
\end{equation}
\end{linenomath}

The expression of \eqref{eq_Faraday_x_ijk} in Fourier space can be obtained by multiplying both sides of the equation by $e^{-i k^x x_i} e^{-i k^y y_{j+\frac 12}} e^{-i k^z z_{k+\frac 12}}$ and sum over $i,j,k=0,\dots,N-1$. The factor $e^{-i k^y y_{j+\frac 12}}$ brings an additional factor $e^{-i k^y \dy / 2}$ with respect to the factor $e^{-i k^y y_j}$ that is needed to recover the Fourier transform of $E^z$ along $y$ (where $E^z$ is nodal). Similarly, the factor $e^{-i k^z z_{k+\frac 12}}$ brings an additional factor $e^{-i k^z \dz / 2}$ with respect to the factor $e^{-i k^z z_k}$ that is needed to recover the Fourier transform of $E^y$ along $z$ (where $E^y$ is nodal). As a result, the Fourier expression of \eqref{eq_Faraday_x_ijk} reads
\begin{linenomath}
\begin{equation}
\label{eq_Faraday_x_Fourier}
\left(\frac{\de}{\de t} - i \, \vbgal \cdot [\kb]_\ctxt\right) \wh{B^x}
= - i \, \left([k^y]_\stxt \, \wh{E^z} - [k^z]_\stxt \, \wh{E^y}\right) \,.
\end{equation}
\end{linenomath}

The components of \eqref{eq_Faraday_app} along $y$ and $z$ can be computed in a similar way, eventually resulting in the following expression of Faraday's law in Fourier space:
\begin{linenomath}
\begin{equation}
\label{eq_Faraday_Fourier}
\left(\frac{\de}{\de t} - i \, \vbgal \cdot [\kb]_\ctxt\right) \wh\Bb
= - i \, [\kb]_\stxt \times \wh\Eb \,.
\end{equation}
\end{linenomath}

\subsection{Ampère-Maxwell's Law}

In the Galilean coordinates $\xb = \xb' - \vbgal t$, Ampère-Maxwell's law reads
\begin{linenomath}
\begin{equation}
\label{eq_AmpereMaxwell_app}
\frac{1}{c^2}\left(\frac{\de}{\de t} - \vbgal \cdot \Nabla\right) \Eb = \Nabla \times \Bb - \mu_0 \Jb \,,
\end{equation}
\end{linenomath}
where $\Nabla$ denotes spatial derivatives with respect to the Galilean coordinates $\xb$ on the spatial grids where $\Bb$ and $\Eb$ are defined, respectively.

The component of \eqref{eq_AmpereMaxwell_app} along $x$ reads
\begin{linenomath}
\begin{equation}
\label{eq_AmpereMaxwell_x}
\frac{1}{c^2}\left(\frac{\de E^x}{\de t} - \vgal^x \frac{\de E^x}{\de x} - \vgal^y \frac{\de E^x}{\de y} - \vgal^z \frac{\de E^x}{\de z}\right) = \left(\frac{\de B^z}{\de y} - \frac{\de B^y}{\de z}\right) - \mu_0 J^x \,.
\end{equation}
\end{linenomath}

At finite order $2m$, \eqref{eq_AmpereMaxwell_x} reads
\begin{linenomath}
\begin{equation}
\label{eq_AmpereMaxwell_x_ijk}
\begin{split}
\frac{1}{c^2}\Bigg(
\frac{\de E^x_{i+\frac 12,j,k}}{\de t}
& - \vgal^x \sum_{n=1}^{m} \alpha_{m,n}^\ctxt \frac{E^x_{i+\frac 12+n,j,k} - E^x_{i+\frac 12-n,j,k}}{2n \, \Delta x} \\[5pt]
& - \vgal^y \sum_{n=1}^{m} \alpha_{m,n}^\ctxt \frac{E^x_{i+\frac 12,j+n,k} - E^x_{i+\frac 12,j-n,k}}{2n \, \Delta y} \\[5pt]
& - \vgal^z \sum_{n=1}^{m} \alpha_{m,n}^\ctxt \frac{E^x_{i+\frac 12,j,k+n} - E^x_{i+\frac 12,j,k-n}}{2n \, \Delta z}\Bigg) \\[5pt]
& = \left(\sum_{n=1}^{m} \alpha_{m,n}^\stxt \frac{B^z_{i+\frac 12,j+n-\frac 12,k} - B^z_{i+\frac 12,j-n+\frac 12,k}}{2 \, (n-1/2) \, \Delta y} \right. \\[5pt]
& \quad \left. - \sum_{n=1}^{m} \alpha_{m,n}^\stxt \frac{B^y_{i+\frac 12,j,k+n-\frac 12} - B^y_{i+\frac 12,j,k-n+\frac 12}}{2 \, (n-1/2) \, \Delta z}\right)
- \mu_0 J^x_{i+\frac 12,j,k} \,.
\end{split}
\end{equation}
\end{linenomath}

The expression of \eqref{eq_AmpereMaxwell_x_ijk} in Fourier space can be obtained by multiplying both sides of the equation by $e^{-i k^x x_{i+\frac 12}} e^{-i k^y y_j} e^{-i k^z z_k}$ and sum over $i,j,k=0,\dots,N-1$. The factor $e^{-i k^y y_j}$ brings an additional factor $e^{i k^y \dy / 2}$ with respect to the factor $e^{-i k^y y_{j+\frac 12}}$ that is needed to recover the Fourier transform of $B^z$ along $y$ (where $B^z$ is cell-centered). Similarly, the factor $e^{-i k^z z_k}$ brings an additional factor $e^{i k^z \dz / 2}$ with respect to the factor $e^{-i k^z z_{k+\frac 12}}$ that is needed to recover the Fourier transform of $B^y$ along $z$ (where $B^y$ is cell-centered). As a result, the Fourier expression of \eqref{eq_AmpereMaxwell_x_ijk} reads
\begin{linenomath}
\begin{equation}
\label{eq_AmpereMaxwell_x_Fourier}
\frac{1}{c^2}\left(\frac{\de}{\de t} - i \, \vbgal \cdot [\kb]_\ctxt\right) \wh{E^x}
= i \, \left([k^y]_\stxt \, \wh{B^z} - [k^z]_\stxt \, \wh{B^y}\right) - \mu_0 \wh{J^x} \,.
\end{equation}
\end{linenomath}

The components of \eqref{eq_AmpereMaxwell_app} along $y$ and $z$ can be computed in a similar way, eventually resulting in the following expression of Ampère-Maxwell's law in Fourier space:
\begin{linenomath}
\begin{equation}
\label{eq_AmpereMaxwell_Fourier}
\frac{1}{c^2}\left(\frac{\de}{\de t} - i \, \vbgal \cdot [\kb]_\ctxt\right) \wh{\Eb}
= i \, [\kb]_\stxt \times \wh{\Bb} - \mu_0 \wh{\Jb} \,.
\end{equation}
\end{linenomath}

\subsection{Continuity Equation}
In the Galilean coordinates $\xb = \xb' - \vbgal t$, the continuity equation reads
\begin{linenomath}
\begin{equation}
\label{eq_continuity}
\left(\frac{\de}{\de t} - \vbgal \cdot \Nabla\right) \rho + \Nabla \cdot \Jb = 0 \,,
\end{equation}
\end{linenomath}
where $\Nabla$ denotes spatial derivatives with respect to the Galilean coordinates $\xb$ on the spatial grids where $\rho$ and $\Jb$ are defined, respectively.

At finite order $2m$, \eqref{eq_continuity} reads
\begin{linenomath}
\begin{equation}
\label{eq_continuity_ijk}
\begin{split}
\frac{\de \rho_{i,j,k}}{\de t}
& - \vgal^x \sum_{n=1}^{m} \alpha_{m,n}^\ctxt \frac{\rho_{i+n,j,k} - \rho_{i-n,j,k}}{2n \, \dx} \\[5pt]
& - \vgal^y \sum_{n=1}^{m} \alpha_{m,n}^\ctxt \frac{\rho_{i,j+n,k} - \rho_{i,j-n,k}}{2n \, \dy}
- \vgal^z \sum_{n=1}^{m} \alpha_{m,n}^\ctxt \frac{\rho_{i,j,k+n} - \rho_{i,j,k-n}}{2n \, \dz} \\[5pt]
& = \sum_{n=1}^{m} \alpha_{m,n}^\stxt \frac{J^x_{i+n-\frac 12,j,k} - J^x_{i-n+\frac 12,j,k}}{2 \, (n-1/2) \, \dx} \\[5pt]
& + \sum_{n=1}^{m} \alpha_{m,n}^\stxt \frac{J^y_{i,j+n-\frac 12,k} - J^y_{i,j-n+\frac 12,k}}{2 \, (n-1/2) \, \dy}
+ \sum_{n=1}^{m} \alpha_{m,n}^\stxt \frac{J^y_{i,j,k+n-\frac 12} - J^y_{i,j,k-n+\frac 12}}{2 \, (n-1/2) \, \dz} \,.
\end{split}
\end{equation}
\end{linenomath}

The expression of \eqref{eq_continuity_ijk} in Fourier space can be obtained by multiplying both sides of the equation by $e^{-i k^x x_i} e^{-i k^y y_j} e^{-i k^z z_k}$ and sum over $i,j,k=0,\dots,N-1$. The factor $e^{-i k^x x_i}$ brings an additional factor $e^{i k^x \dx / 2}$ with respect to the factor $e^{-i k^x x_{i+\frac 12}}$ that is needed to recover the Fourier transform of $J^x$ along $x$ (where $J^x$ is cell-centered).
The same argument applies to $J^y$ and $J^z$ with circular permutation of the indices.
As a result, the Fourier expression of \eqref{eq_continuity_ijk} reads
\begin{linenomath}
\begin{equation}
\label{eq_continuity_Fourier}
\left(\frac{\de}{\de t} - i \, \vbgal \cdot [\kb]_{\ctxt}\right) \wh\rho + i \, [\kb]_\stxt \cdot \wh\Jb = 0 \,.
\end{equation}
\end{linenomath}

\subsection{Update Equations}
We now combine equations \eqref{eq_Faraday_Fourier} and \eqref{eq_AmpereMaxwell_Fourier},
\begin{linenomath}
\begin{align}
\label{eq_Faraday_Fourier2}
\left(\frac{\de}{\de t} - i \, \vbgal \cdot [\kb]_\ctxt\right) \wh{\Bb}
& = - i \, [\kb]_{\stxt} \times \wh{\Eb} \,, \\[5pt]
\label{eq_AmpereMaxwell_Fourier2}
\frac{1}{c^2}\left(\frac{\de}{\de t} - i \, \vbgal \cdot [\kb]_\ctxt\right) \wh{\Eb}
& = i \, [\kb]_{\stxt} \times \wh{\Bb} - \mu_0 \, \wh{\Jb} \,,
\end{align}
\end{linenomath}
together with the continuity equation \eqref{eq_continuity_Fourier}, in order to obtain the finite-order update equations for $\wh\Eb$ and $\wh\Bb$ similar to equations (4a)-(4b) of \citep{Kirchen2020}.
The notation introduced in Section~\ref{sec_algo} is used for the rest of the derivation, in particular for the frequencies
$\Omega_\ctxt := \vbgal \cdot [\kb]_\ctxt$,
$\omega_\ctxt := c \, [k]_\ctxt$ and
$\omega_\stxt := c \, [k]_\stxt$,
where $[k]_\ctxt$ and $[k]_\stxt$ denote the magnitudes of the centered and staggered modified wave vectors $[\kb]_\ctxt$ and $[\kb]_\stxt$, respectively.

Taking the time derivative of \eqref{eq_Faraday_Fourier2} yields
\begin{linenomath}
\begin{equation}
\begin{split}
\frac{\de^2 \wh\Bb}{\de t^2} - i \, \Omega_\ctxt \frac{\de \wh\Bb}{\de t}
& = - i \, [\kb]_{\stxt} \times \frac{\de \wh\Eb}{\de t} \\[5pt]
& = i \, \Omega_\ctxt \frac{\de \wh\Bb}{\de t}
+ \Omega_\ctxt^2 \, \wh\Bb
- \omega_{\stxt}^2 \, \wh{\Bb} 
+ c^2 \, ([\kb]_{\stxt} \cdot \wh\Bb) \, [\kb]_{\stxt}
+ i \, c^2 \mu_0 \, [\kb]_{\stxt} \times \wh{\Jb} \,,
\end{split}
\end{equation}
\end{linenomath}
which, by setting $[\kb]_{\stxt} \cdot \wh\Bb = 0$ thanks to magnetic Gauss' law, reads
\begin{linenomath}
\begin{equation}
\label{eq_FAM1_o2_new}
\left(\frac{\de}{\de t} - i \, \Omega_\ctxt\right)^2 \wh\Bb
+ \omega_\stxt^2 \, \wh{\Bb}
= \frac{i}{\eps_0} \, [\kb]_{\stxt} \times \wh{\Jb} \,.
\end{equation}
\end{linenomath}

Taking now the time derivative of \eqref{eq_AmpereMaxwell_Fourier2} yields
\begin{linenomath}
\begin{equation}
\begin{split}
\frac{\de^2 \wh\Eb}{\de t^2} - i \, \Omega_\ctxt \frac{\de \wh\Eb}{\de t}
& = i \, c^2 \, [\kb]_{\stxt} \times \frac{\de \wh{\Bb}}{\de t} - c^2 \, \mu_0 \frac{\de \wh\Jb}{\de t} \\[5pt]
& = i \, \Omega_\ctxt \frac{\de \wh\Eb}{\de t}
+ \Omega_\ctxt^2 \, \wh\Eb + i \, c^2 \mu_0 \, \Omega_\ctxt \, \wh\Jb 
+ c^2 \, ([\kb]_{\stxt} \cdot \wh{\Eb}) \, [\kb]_{\stxt}
- \omega_{\stxt}^2 \, \wh{\Eb}
- c^2 \mu_0 \frac{\de \wh\Jb}{\de t} \,,
\end{split}
\end{equation}
\end{linenomath}
which, by setting $[\kb]_{\stxt} \cdot \wh{\Eb}=- i \, \wh\rho / \eps_0$ thanks to Gauss' law and $\de \wh\Jb / \de t = 0$ (from the assumption that $\wh\Jb$ is constant over a single time step), reads
\begin{linenomath}
\begin{equation}
\label{eq_FAM2_o2_new}
\left(\frac{\de}{\de t} - i \, \Omega_\ctxt\right)^2 \wh\Eb
+ \omega_\stxt^2 \, \wh\Eb
= \frac{i}{\eps_0} \, \Omega_\ctxt \, \wh\Jb
- i \, \frac{c^2}{\eps_0} \wh\rho \, [\kb]_{\stxt} \,.
\end{equation}
\end{linenomath}

Before integrating \eqref{eq_FAM1_o2_new} and \eqref{eq_FAM2_o2_new} over one time step, we integrate the continuity equation \eqref{eq_continuity_Fourier}. In terms of the variable $\Omega_\ctxt$, this reads
\begin{linenomath}
\begin{equation}
\label{eq_continuity_Fourier_new}
\left(\frac{\de}{\de t} - i \, \Omega_\ctxt\right) \wh\rho + i \, [\kb]_{\stxt} \cdot \wh\Jb = 0 \,.
\end{equation}
\end{linenomath}
This equation is of the general form
\begin{linenomath}
\begin{equation}
\label{eq_whrho_general}
\frac{\de \wh\rho}{\de t} + \alpha \, \wh\rho + \beta = 0 \,,
\end{equation}
\end{linenomath}
with $\alpha = - i \, \Omega_\ctxt$ and $\beta = i \, [\kb]_{\stxt} \cdot \wh\Jb$.
Integrating \eqref{eq_whrho_general} between $n \dt$ and $t$ yields
\begin{linenomath}
\begin{equation}
\wh\rho(t) = \kappa \, e^{- \alpha (t - n \dt)} - \frac{\beta}{\alpha} \,.
\end{equation}
\end{linenomath}
The constant $\kappa$ can be determined by setting $t = n \dt$ and $\wh\rho(n \dt) = \wh\rho^{\, n}$:
\begin{linenomath}
\begin{equation}
\label{eq_kappa}
\kappa = \wh\rho^{\, n} + \frac{\beta}{\alpha} \,. 
\end{equation}
\end{linenomath}
Moreover, the constraint $\wh\rho((n+1) \dt) = \wh\rho^{\, n+1}$ requires that
\begin{linenomath}
\begin{equation}
\kappa \, e^{- \alpha \dt} - \frac{\beta}{\alpha} = \wh\rho^{\, n+1} \,,
\end{equation}
\end{linenomath}
which, thanks to \eqref{eq_kappa}, yields
\begin{linenomath}
\begin{equation}
\label{eq_beta}
\frac{\beta}{\alpha} = \frac{\wh\rho^{\, n+1} - \wh\rho^{\, n} \, e^{- \alpha \dt}}{e^{-\alpha \dt} - 1} \,.
\end{equation}
\end{linenomath}
Inserting the values of $\alpha$ and $\beta$ finally yields
\begin{linenomath}
\begin{equation}
\label{eq_rho_t}
\wh\rho(t) = \frac{\wh\rho^{\, n+1} - \wh\rho^{\, n}}{e^{i \Omega_\ctxt \dt} - 1} \, e^{i \Omega_\ctxt (t - n \dt)}
- \frac{\wh\rho^{\, n+1} - \wh\rho^{\, n} e^{i \Omega_\ctxt \dt}}{e^{i \Omega_\ctxt \dt} - 1} \,.
\end{equation}
\end{linenomath}
We remark that \eqref{eq_rho_t} corresponds to equation (9) of \citep{Lehe2016}.
We now insert the solution \eqref{eq_rho_t} into \eqref{eq_FAM2_o2_new} and obtain the new system
\begin{linenomath}
\begin{align}
\label{eq_FAM1_o2_ready}
\left(\frac{\de}{\de t} - i \, \Omega_\ctxt\right)^2 \wh\Bb + \omega_\stxt^2 \, \wh{\Bb}
& = \frac{i}{\eps_0} \, [\kb]_{\stxt} \times \wh{\Jb} \,, \\[5pt]
\label{eq_FAM2_o2_ready}
\begin{split}
\left(\frac{\de}{\de t} - i \, \Omega_\ctxt\right)^2 \wh\Eb + \omega_\stxt^2 \wh\Eb
& = \frac{i}{\eps_0} \, \Omega_\ctxt \, \wh\Jb
+ i \, \frac{c^2}{\eps_0} \frac{\wh\rho^{\, n+1} - \wh\rho^{\, n} e^{i \Omega_\ctxt \dt}}{e^{i \Omega_\ctxt \dt} - 1} \, [\kb]_{\stxt} \\[5pt]
& \quad - i \, \frac{c^2}{\eps_0} \frac{\wh\rho^{\, n+1} - \wh\rho^{\, n}}{e^{i \Omega_\ctxt \dt} - 1} \, [\kb]_{\stxt} \, e^{i \Omega_\ctxt (t - n \dt)}
\,.
\end{split}
\end{align}
\end{linenomath}
Equations \eqref{eq_FAM1_o2_ready} and \eqref{eq_FAM2_o2_ready} correspond to equations (A1a) and (A1b) of \citep{Lehe2016}, respectively.
Both equations can be cast into the general form
\begin{linenomath}
\begin{equation}
\label{eq_EB_general}
\left(\frac{\de}{\de t} - i \, \Omega_\ctxt\right)^2 f + \omega_\stxt^2 \, f = \alpha + \beta \, e^{i \Omega_\ctxt (t - n \dt)} \,,
\end{equation}
where the coefficients $\alpha$ and $\beta$ are given by
\begin{equation}
\label{eq_alphabeta_B}
\alpha = \frac{i}{\eps_0} \, [\kb]_{\stxt} \times \wh{\Jb} \,, \quad
\beta = 0 \,,
\end{equation}
\end{linenomath}
in the case of equation \eqref{eq_FAM1_o2_ready}, and
\begin{linenomath}
\begin{equation}
\label{eq_alphabeta_E}
\alpha = \frac{i}{\eps_0} \, \Omega_\ctxt \, \wh\Jb
+ i \, \frac{c^2}{\eps_0} \frac{\wh\rho^{\, n+1} - \wh\rho^{\, n} e^{i \Omega_\ctxt \dt}}{e^{i \Omega_\ctxt \dt} - 1} \, [\kb]_{\stxt} \,, \quad
\beta = - i \, \frac{c^2}{\eps_0} \frac{\wh\rho^{\, n+1} - \wh\rho^{\, n}}{e^{i \Omega_\ctxt \dt} - 1} \, [\kb]_{\stxt} \,,
\end{equation}
\end{linenomath}
in the case of equation \eqref{eq_FAM2_o2_ready}. A solution $f_0$ of the homogeneous equation associated to \eqref{eq_EB_general} is
\begin{linenomath}
\begin{equation}
f_0(t) = \kappa_1 \cos(\omega_\stxt (t - n \dt)) \, e^{i \Omega_\ctxt (t - n \dt)} + \kappa_2 \sin(\omega_\stxt (t - n \dt)) \, e^{i \Omega_\ctxt (t - n \dt)} \,.
\end{equation}
\end{linenomath}
Moreover, a specific solution $\bar f$ of \eqref{eq_EB_general} is
\begin{linenomath}
\begin{equation}
\bar f(t) = \frac{\alpha}{\omega_\stxt^2 - \Omega_\ctxt^2}
+ \frac{\beta}{\omega_\stxt^2} \, e^{i \Omega_\ctxt (t - n \dt)} \,.
\end{equation}
\end{linenomath}
The general solution $f$ of \eqref{eq_EB_general} is then obtained by combining $f_0$ and $\bar f$, which yields
\begin{linenomath}
\begin{equation}
\label{eq_EB_solution}
\begin{split}
f(t) & = \kappa_1 \cos(\omega_\stxt (t - n \dt)) \, e^{i \Omega_\ctxt (t - n \dt)} + \kappa_2 \sin(\omega_\stxt (t - n \dt)) \, e^{i \Omega_\ctxt (t - n \dt)} \\[5pt]
& \quad + \frac{\alpha}{\omega_\stxt^2 - \Omega_\ctxt^2}
+ \frac{\beta}{\omega_\stxt^2} \, e^{i \Omega_\ctxt (t - n \dt)} \,.
\end{split}
\end{equation}
\end{linenomath}
The integration constants $\kappa_1$ and $\kappa_2$ can be determined by the initial conditions $f(n \dt)$ and $\de f / \de t (n \dt)$:
\begin{linenomath}
\begin{align}
\kappa_1 & = f(n \dt) - \frac{\alpha}{\omega_\stxt^2 - \Omega_\ctxt^2}
- \frac{\beta}{\omega_\stxt^2} \,, \\[5pt]
\kappa_2 & = \frac{1}{\omega_\stxt} \left(\frac{\de f}{\de t}(n \dt) - i \, \Omega_\ctxt \, f(n \dt) + i \, \Omega_\ctxt \, \frac{\alpha}{\omega_\stxt^2 - \Omega_\ctxt^2}\right) \,.
\end{align}
\end{linenomath}
Evaluating the solution \eqref{eq_EB_solution} at $t = (n + 1) \dt$ finally yields
\begin{linenomath}
\begin{equation}
f((n+1) \dt) = \kappa_1 \cos(\omega_\stxt \dt) e^{i \Omega_\ctxt \dt} + \kappa_2 \sin(\omega_\stxt \dt) e^{i \Omega_\ctxt \dt}
+ \frac{\alpha}{\omega_\stxt^2 - \Omega_\ctxt^2}
+ \frac{\beta}{\omega_\stxt^2} \, e^{i \Omega_\ctxt \dt} \,,
\end{equation}
\end{linenomath}
which can be rewritten as
\begin{linenomath}
\begin{equation}
\label{eq_EB_solution_tnp1}
f((n+1) \dt) = \theta_\ctxt^2 C f(n \dt) + \theta_\ctxt \frac{\chi_1}{\omega_\ctxt^2} \, \alpha
+ \theta_\ctxt^2 \frac{(1 - C)}{\omega_\stxt^2} \, \beta
+ \theta_\ctxt^2 \frac{S}{\omega_\stxt} \left(\frac{\de f}{\de t}(n \dt) - i \, \Omega_\ctxt \, f(n \dt)\right) \,,
\end{equation}
\end{linenomath}
where $C$, $S$, $\theta_\ctxt$, $\theta_\ctxt^*$ and $\chi_1$ are defined as in~Section~\ref{sec_algo}. 
Equation \eqref{eq_EB_solution_tnp1} corresponds to equation (A7) of~\citep{Lehe2016}, with $\chi_1$ defined in \eqref{eq_chi1} replacing the corresponding definition in equation (12c) of~\citep{Lehe2016}.
By inserting \eqref{eq_alphabeta_B}-\eqref{eq_alphabeta_E} in \eqref{eq_EB_solution_tnp1} we then obtain \eqref{eq_update_B_with_rho_galilean}-\eqref{eq_update_E_with_rho_galilean}, after some algebra.

\section{Standard Galilean PSATD: Vacuum Dispersion Relation}
\label{appendix_2}
This section presents the derivation of the dispersion relation for the update equations \eqref{eq_update_B_with_rho_galilean}-\eqref{eq_update_E_with_rho_galilean} in vacuum:
\begin{linenomath}
\begin{subequations}
\begin{align}
\label{eq_update_B_with_rho_galilean_disprel}
\wh\Bb^{n+1} & = \theta_\ctxt^2 C \wh\Bb^{n}
- i \, \theta_\ctxt^2 \frac{S}{\omega_\stxt} \, [\kb]_{\stxt} \times \wh\Eb^{n} \,, \\[5pt]
\label{eq_update_E_with_rho_galilean_disprel}
\wh\Eb^{n+1} & = \theta_\ctxt^2 C \wh\Eb^{n}
+ i \, c^2 \, \theta_\ctxt^2 \frac{S}{\omega_\stxt} \, [\kb]_{\stxt} \times \wh\Bb^{n} \,.
\end{align}
\end{subequations}
\end{linenomath}
For this purpose it is useful to first rewrite \eqref{eq_update_B_with_rho_galilean_disprel}-\eqref{eq_update_E_with_rho_galilean_disprel} by taking advantage of the fact that these equations result from the analytical integration of Maxwell's equations and are therefore time-reversible.
More precisely, we can rewrite \eqref{eq_update_B_with_rho_galilean_disprel}-\eqref{eq_update_E_with_rho_galilean_disprel} by performing the following time-reversal operations: interchange $n$ and $n+1$, change sign to the magnetic field components, and change sign to the Galilean velocity (that is, replace $\theta_\ctxt$ with $\theta_\ctxt^*$). This results in the following equations:
\begin{linenomath}
\begin{subequations}
\begin{align}
\label{eq_update_B_with_rho_galilean_disprel_tr}
\wh\Bb^{n} & = \theta_\ctxt^{*2} C \wh\Bb^{n+1}
+ i \, \theta_\ctxt^{*2} \frac{S}{\omega_\stxt} \, [\kb]_{\stxt} \times \wh\Eb^{n+1} \,, \\[5pt]
\label{eq_update_E_with_rho_galilean_disprel_tr}
\wh\Eb^{n} & = \theta_\ctxt^{*2} C \wh\Eb^{n+1}
- i \, c^2 \, \theta_\ctxt^{*2} \frac{S}{\omega_\stxt} \, [\kb]_{\stxt} \times \wh\Bb^{n+1} \,.
\end{align}
\end{subequations}
\end{linenomath}
Subtracting \eqref{eq_update_B_with_rho_galilean_disprel_tr}-\eqref{eq_update_E_with_rho_galilean_disprel_tr} multiplied by $\theta_\ctxt$ from \eqref{eq_update_B_with_rho_galilean_disprel}-\eqref{eq_update_E_with_rho_galilean_disprel} multiplied by $\theta_\ctxt^*$ then yields
\begin{linenomath}
\begin{subequations}
\begin{align}
\label{eq_update_B_with_rho_galilean_disprel_ts}
(1 + C) \left(\theta_\ctxt^* \wh\Bb^{n+1} - \theta_\ctxt \wh\Bb^{n}\right) & = 
- i \, \frac{S}{\omega_\stxt} \, [\kb]_{\stxt} \times \left(\theta_\ctxt^* \wh\Eb^{n+1} + \theta_\ctxt \wh\Eb^{n}\right) \,, \\[5pt]
\label{eq_update_E_with_rho_galilean_disprel_ts}
(1 + C) \left(\theta_\ctxt^* \wh\Eb^{n+1} - \theta_\ctxt \wh\Eb^{n}\right) & = 
i \, c^2 \, \frac{S}{\omega_\stxt} \, [\kb]_{\stxt} \times \left(\theta_\ctxt^* \wh\Bb^{n+1} + \theta_\ctxt \wh\Bb^{n}\right) \,.
\end{align}
\end{subequations}
\end{linenomath}
We now look for electromagnetic modes of the form
$\wh\Bb^n = \wh\Bb \, e^{-i (\omega - \Omega) n \dt}$ and
$\wh\Eb^n = \wh\Eb \, e^{-i (\omega - \Omega) n \dt}$,
with $\Omega := \vbgal \cdot \kb$, and the same for $n+1$.
Equations \eqref{eq_update_B_with_rho_galilean_disprel_ts}-\eqref{eq_update_E_with_rho_galilean_disprel_ts} then yield
\begin{linenomath}
\begin{subequations}
\begin{align}
\label{eq_update_B_with_rho_galilean_disprel_mode}
(1 + C) \, s_\omega \, \wh\Bb & = 
\frac{S}{\omega_\stxt} \, c_\omega \, [\kb]_{\stxt} \times \wh\Eb \,, \\[5pt]
\label{eq_update_E_with_rho_galilean_disprel_mode}
(1 + C) \, s_\omega \, \wh\Eb & = 
- c^2 \, \frac{S}{\omega_\stxt} \, c_\omega \, [\kb]_{\stxt} \times \wh\Bb \,,
\end{align}
\end{subequations}
\end{linenomath}
where $s_\omega := \sin((\omega - \delta \Omega) \dt/2)$, $c_\omega := \cos((\omega - \delta \Omega) \dt/2)$, with $\delta \Omega := \Omega - \Omega_\ctxt$. The two equations can be decoupled by taking the cross product with $[\kb]_\stxt$, which finally yields
\begin{linenomath}
\begin{equation}
\tan^2\left(\frac{(\omega - \delta \Omega) \dt}{2}\right)
= \frac{S^2}{(1 + C)^2}
= \tan^2\left(\frac{\omega_\stxt \dt}{2}\right) \,,
\end{equation}
\end{linenomath}
leading to the vacuum dispersion relation
\begin{linenomath}
\begin{equation}
\label{vacuum_disprel_gal}
\omega = \omega_\stxt + \delta \Omega = c \, [k]_\stxt + \vbgal \cdot (\kb - [\kb]_\ctxt) \,.
\end{equation}
\end{linenomath}
In the case of the standard PSATD equations, $\vbgal = 0$, $\Omega = \Omega_\ctxt = 0$, and the vacuum dispersion relation reads
\begin{linenomath}
\begin{equation}
\label{vacuum_disprel_psatd}
\omega = \omega_\stxt = c \, [k]_\stxt \,.
\end{equation}
\end{linenomath}
As expected for consistency, the vacuum dispersion relation \eqref{vacuum_disprel_gal} corresponds to equation (12) of \citep{Kirchen2020}, when considering the purely nodal case where all staggered quantities are replaced by the corresponding centered quantities.

\section{Standard Galilean PSATD: Staggering, Finite-Order Centering, and NCI}
\label{appendix_3}
This section presents a few heuristic arguments that help understand the role of staggering and finite-order centering in relation to NCI mitigation.
In order to fix the ideas, in the following we refer to the dispersion analysis presented in~\citep{Lehe2016}, for the two-dimensional case on nodal grids, at infinite spectral order (in which case the finite-order modified wave vectors $[\kb]_\ctxt$ and $[\kb]_\stxt$ coincide and are both simply equal to $\kb$).

As mentioned in Section~\ref{sec_motivations} as well as observed with the test cases presented in Sections~\ref{sec_gal_test}~and~\ref{sec_avg_test}, fully nodal PIC cycles mitigate NCI more effectively than fully staggered PIC cycles.
With regard to the derivation of the dispersion analysis presented in~\citep{Lehe2016}, the net effect of staggering a given grid quantity $F$ (that is, a component of the electromagnetic fields $\Eb$ and $\Bb$ or the current density $\Jb$) is that its Fourier transform $\Fhat$ gets multiplied by a coefficient $\zeta_F$, which depends on the staggering of the grid quantity with a certain functional dependency. More precisely, $\Fhat$ is replaced by $\zeta_F \Fhat$, where $\zeta_F$ is of the form $\zeta_F = (-1)^{\lb \cdot \epsb}$, where $\lb$ represents the vector of alias numbers associated with different Brillouin zones of the $\kb$ space and the components of the vector $\epsb$ can be either $0$ or $1$, depending on whether the grid quantity $F$ is nodal or staggered in that direction, respectively.
We remark that the staggering that matters here is the one of the grids used for current deposition and field gathering in the PIC cycle.

The effect, in terms of NCI mitigation, of the new coefficient $\zeta_F$, which appears in front of the Fourier transforms of the various grid quantities, is explained briefly in the following.
By taking the ultra-relativistic and low-density limits of the final dispersion relation derived in~\citep{Lehe2016}, for the optimal case where the Galilean velocity matches the plasma velocity ($\vbgal = \vb_0 = v_0 \ub_z$), and considering a small frequency perturbation $\delta\omega$ around the relativistic plasma mode $k_z v_0$, the dispersion relation reduces to the simpler form
\begin{linenomath}
\begin{equation}
\label{eq_lim_disprel_stagg}
\delta\omega^2 \propto \sum_{\lb} g(\kb_{\lb})
\left(\left[1 - \left(\frac{k^z}{k}\right)^2\right] \zeta_{E^x}
+ \left(\frac{k^z}{k}\right)^2 \zeta_{J^z} \, \zeta_{E^x}
- \zeta_{J^z} \, \zeta_{B^y}\right)\,.
\end{equation}
\end{linenomath}
The key observation here is that the right hand side of~\eqref{eq_lim_disprel_stagg} happens to vanish in the nodal case, where $\epsb = (0,0)$ and $\zeta_F = 1$ for all grid quantities, while it is non-zero and can be negative for some alias numbers in the staggered case, where $\zeta_F \neq 1$ for some grid quantities, eventually leading to instability.
In other words, the instability observed in the case of fully staggered PIC cycles is driven by the fact that $\zeta_F$ can deviate from unity and possibly become negative for some of the grid quantities.
This information is key in helping us understand what happens in the case of a hybrid PIC cycle, and how finite-order centering improves stability with respect to fully staggered simulations.

As mentioned above, the staggering that matters here, in the sense of being responsible for the terms $(-1)^{\lb \cdot \epsb}$ appearing in front of the Fourier transforms of the various grid quantities, is the staggering of the grids used for current deposition and field gathering in the PIC cycle.
In the case of a hybrid PIC cycle, the current is deposited on a nodal grid and the electromagnetic forces are gathered from a nodal grid.
This observation alone could lead us to the erroneous conclusion that the hybrid PIC cycle, per se, should guarantee full stability, as in the nodal case, irrespective of the order $2m$ of the finite centering of fields and currents.
However, this is not the case, as we show in the following.

In order to understand what role the order of the finite centering of fields and currents plays in terms of NCI mitigation, it is necessary to express the centering of a given grid quantity $F$, from a nodal grid to a staggered grid (or vice versa), in Fourier space.

\begin{figure}[!ht]
\centering
\includegraphics[width=0.6\linewidth,trim={0 0 0 0},clip]{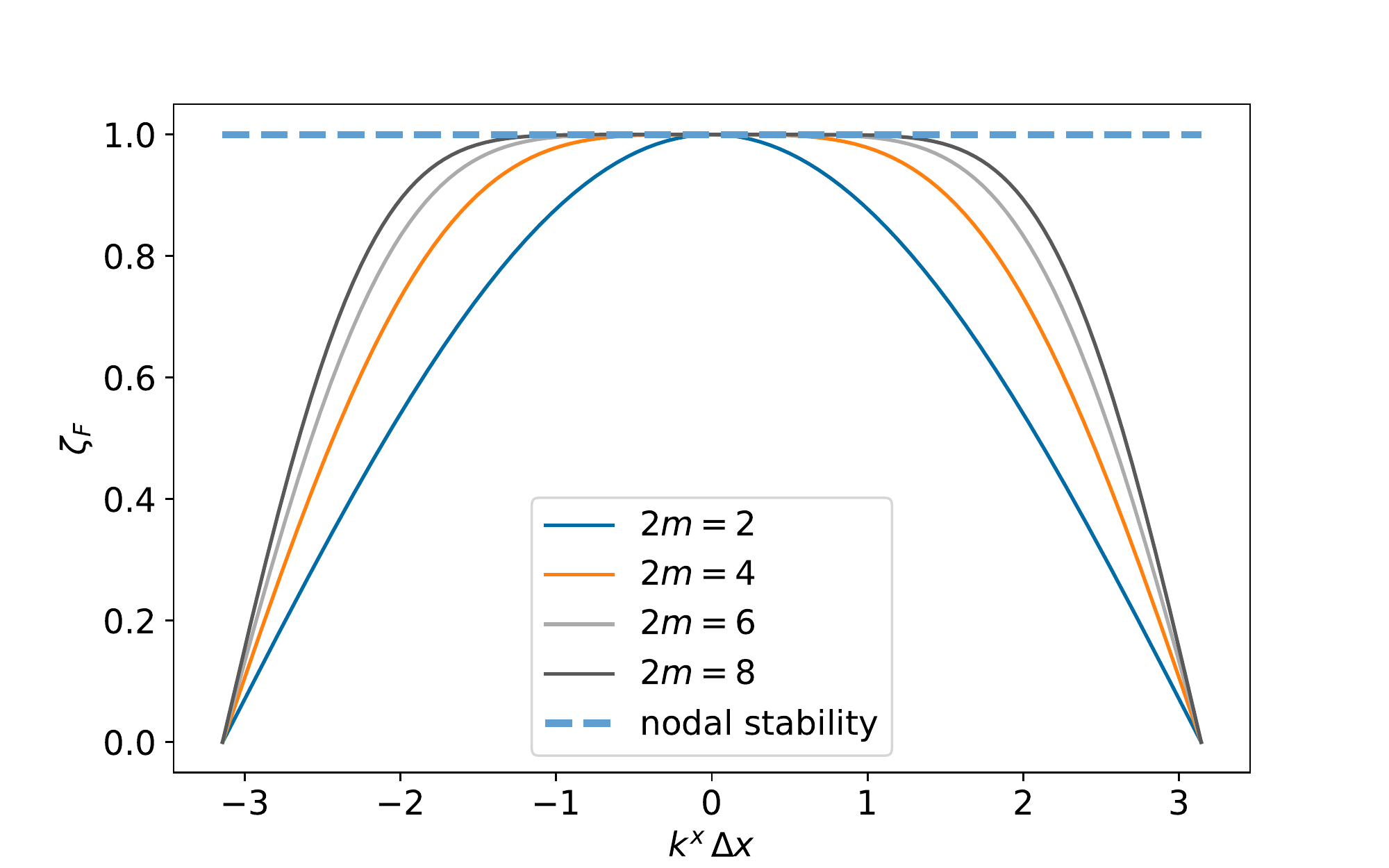}
\caption{\textbf{Finite-order Centering and NCI}. $\zeta_F$ as a function of $k^x \dx$: the higher the order $2m$ of the finite centering, the smaller the deviation of $\zeta_F$ from unity (or, in other words, the smaller the region in units of $k^x \dx$ where $0 < \zeta_F < 1$).}
\label{fig_disprel_zeta}
\end{figure}

For this purpose, let us consider, for instance, a staggered finite sum of the form $F_{j-n+\frac 12} + F_{j+n-\frac 12}$, for a grid quantity $F$ collocated on the cell centers of a one-dimensional periodic grid.
By means of the same type of algebraic manipulations described in \ref{appendix_1}, its expression in Fourier space can be obtained by multiplying the finite sum by $e^{-i k x_{j+\frac 12}}$ and summing over $j = 0, \dots, N-1$, which yields
\begin{linenomath}
\begin{equation}
\label{staggered_finite_sum_Fourier_staggered}
\sum_{j=0}^{N-1} F_{j+n-\frac 12} \, e^{-i k x_{j+\frac 12}}
+ \sum_{j=0}^{N-1} F_{j-n+\frac 12} \, e^{-i k x_{j+\frac 12}}
= e^{-i k \dx / 2} \, 2 \cos(k \, (n-1/2) \, \dx) \, \Fhat_k \,.
\end{equation}
\end{linenomath}
Therefore, the expression in Fourier space of a finite-order interpolation of the form
\begin{linenomath}
\begin{equation}
F^\ntxt_j = \sum_{n=1}^{m} \alpha_{m,n}^\stxt \frac{F^\stxt_{j+n-1/2} + F^\stxt_{j-n+1/2}}{2} \,,
\end{equation}
\end{linenomath}
which is used for the centering of fields and currents as described in Section~\ref{sec_interpolation} (with a slight difference in the notation used here, which emphasizes the fact that the function on the right hand side is collocated by definition on the cell centers of the grid, while the function on the left hand side, which represents the result of the interpolation, is collocated by definition on the cell nodes), can be obtained by multiplying both sides of the equation by $e^{-i k x_j}$ and summing over $j = 0, \dots, N-1$, which after some algebra yields
\begin{linenomath}
\begin{equation}
\begin{split}
\Fhat^\ntxt_k
& = \sum_{j=0}^{N-1} F^\ntxt_j \, e^{-i k x_j} =
\sum_{n=1}^{m} \frac{\alpha_{m,n}^\stxt}{2} \left(
\sum_{j=0}^{N-1} F^\stxt_{j-n+\frac 12} \, e^{-i k x_j}
+ \sum_{j=0}^{N-1} F^\stxt_{j+n-\frac 12}  \, e^{-i k x_j}\right) \\[5pt]
& = \sum_{n=1}^{m} \frac{\alpha_{m,n}^\stxt}{2} \, e^{i k \dx / 2} \left(
\sum_{j=0}^{N-1} F^\stxt_{j-n+\frac 12} \, e^{-i k x_{j+\frac 12}}
+ \sum_{j=0}^{N-1} F^\stxt_{j+n-\frac 12}  \, e^{-i k x_{j+\frac 12}}
\right) \\[5pt]
& = \sum_{n=1}^{m} \alpha_{m,n}^\stxt \cos(k \, (n-1/2) \, \dx) \, \Fhat^\stxt_k \,.
\end{split}
\end{equation}
\end{linenomath}
As a consequence, the net effect of the finite-order centering of a given grid quantity $F$ (that is, a component of the electromagnetic fields $\Eb$ and $\Bb$ or the current density $\Jb$) is that its Fourier transform $\Fhat$ gets multiplied by a coefficient $\zeta_F$, which this time reads
\begin{linenomath}
\begin{equation}
\zeta_F = \sum_{n=1}^{m} \alpha_{m,n}^\stxt \cos(k^x \, (n-1/2) \, \dx) \,,
\end{equation}
\end{linenomath}
considering centering only along one direction, say $x$, for simplicity (in the general case, different grid quantities will be centered along different directions, depending on their staggering).
As mentioned in the discussion following~\eqref{eq_lim_disprel_stagg}, the deviation of $\zeta_F$ from unity is the factor that determines how well NCI is mitigated in this particular case.
A plot of $\zeta_F$ as a function of $k^x \dx$, as illustrated in Figure~\ref{fig_disprel_zeta}, shows that the higher the order $2m$ of the finite centering, the smaller the deviation of $\zeta_F$ from unity (or, in other words, the smaller the region in units of $k^x \dx$ where $0 < \zeta_F < 1$), which confirms the stability patterns observed with the test cases presented in Sections~\ref{sec_gal_test} and~\ref{sec_avg_test}.

\clearpage
\bibliography{references}

\end{document}